\documentclass{article}

\usepackage{arxiv}

\usepackage[utf8]{inputenc} %
\usepackage[T1]{fontenc}    %
\usepackage[hidelinks]{hyperref}      %
\usepackage{url}            %
\usepackage{booktabs}       %
\usepackage{amsfonts}       %
\usepackage{nicefrac}       %
\usepackage{microtype}      %
\usepackage{graphicx}
\usepackage{doi}
\usepackage{latexsym}
\usepackage{colortbl}
\usepackage{pifont}
\usepackage{wrapfig}
\usepackage{svg}
\usepackage{adjustbox}

\usepackage{array}
\usepackage{makecell} %
\usepackage{multirow} %
\usepackage{color}
\usepackage{rotating}
\usepackage{tabularray}

\usepackage{tikz}
\usepackage{pgfplots}
\usepackage{pgfplotstable}
\usepackage{floatrow}

\newfloatcommand{capbtabbox}{table}[][\FBwidth]

\pgfplotsset{compat=1.18}

\usepackage[most]{tcolorbox}
\usepackage{enumitem,kantlipsum}
\newcommand{\customlabel}[1]{\small{\textbf{C#1.}}}

\usepackage{mathastext} %

\usepackage[
  backend=biber,
  style=numeric,
  sorting=ynt,
  maxnames=20,
  minnames=2
]{biblatex}
\definecolor{darkgreen}{RGB}{0,100,0}
\definecolor{darkorange}{RGB}{255,140,0}
\definecolor{tgreen}{HTML}{CDEB8B}
\definecolor{tred}{HTML}{CCE5FF}
\newcommand{\yes}[1][\ding{51}]{\cellcolor{tgreen}\textcolor{black}{#1}}%
\newcommand{\no}{\cellcolor{tred}\textcolor{black}{\ding{55}}}%
\newcommand{\question}{\cellcolor{yellow!25}\textcolor{black}{?}}%
\usepackage{xparse}
\NewDocumentCommand{\heatgreen}{O{} m O{}}{%
  \pgfmathsetmacro{\saturation}{0.05 + 0.4 * #2 * 0.01}
  \pgfmathsetmacro{\brightness}{1 - 0.4 * #2 * 0.01}
  \xdef\tempcolor{0.33, \saturation, \brightness}
  \cellcolor[hsb]{\tempcolor}%
  \ifthenelse{\equal{#1}{bold}}{%
    \textbf{#2\ifthenelse{\equal{#3}{}}{}{\scriptsize{$\pm$ #3}}}%
  }{%
    \ifthenelse{\equal{#1}{underline}}{%
      \underline{#2\ifthenelse{\equal{#3}{}}{}{\scriptsize{$\pm$ #3}}}%
    }{%
      #2\ifthenelse{\equal{#3}{}}{}{\scriptsize{$\pm$ #3}}%
    }%
  }%
}
\NewDocumentCommand{\heatblue}{O{} m O{}}{%
  \pgfmathsetmacro{\saturation}{0.05 + 0.4 * #2 * 0.01}
  \pgfmathsetmacro{\brightness}{1 - 0.4 * #2 * 0.01}
  \xdef\tempcolor{0.6, \saturation, \brightness}
  \cellcolor[hsb]{\tempcolor}%
  \ifthenelse{\equal{#1}{bold}}{%
    \textbf{#2\ifthenelse{\equal{#3}{}}{}{\scriptsize{$\pm$ #3}}}%
  }{%
    \ifthenelse{\equal{#1}{underline}}{%
      \underline{#2\ifthenelse{\equal{#3}{}}{}{\scriptsize{$\pm$ #3}}}%
    }{%
      #2\ifthenelse{\equal{#3}{}}{}{\scriptsize{$\pm$ #3}}%
    }%
  }%
}
\NewDocumentCommand{\heatred}{O{} m O{}}{%
  \pgfmathsetmacro{\saturation}{0.05 + 0.4 * #2 * 0.01}
  \pgfmathsetmacro{\brightness}{1 - 0.4 * #2 * 0.01}
  \xdef\tempcolor{0.8, \saturation, \brightness}
  \cellcolor[hsb]{\tempcolor}%
  \ifthenelse{\equal{#1}{bold}}{%
    \textbf{#2\ifthenelse{\equal{#3}{}}{}{\scriptsize{$\pm$ #3}}}%
  }{%
    \ifthenelse{\equal{#1}{underline}}{%
      \underline{#2\ifthenelse{\equal{#3}{}}{}{\scriptsize{$\pm$ #3}}}%
    }{%
      #2\ifthenelse{\equal{#3}{}}{}{\scriptsize{$\pm$ #3}}%
    }%
  }%
}

\newcommand{\dirarrow}{
  \includegraphics[scale=0.1]{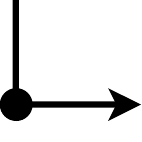}
}

\newcommand{\datalink}[2]{\href{#1}{#2}}

\usepackage{glossaries}
\usepackage{glossaries-extra}
\setabbreviationstyle[acronym]{long-short}
\glsdisablehyper

\newacronym{oc}{OC}{organic computing}
\newacronym{dl}{DL}{deep learning}
\newacronym{ssl}{SSL}{self-supervised learning}
\newacronym{sl}{SL}{supervised learning}
\newacronym{ml}{ML}{machine learning}
\newacronym{tl}{TL}{transfer learning}
\newacronym{ws}{WS}{weak supervision}
\newacronym{al}{AL}{active learning}
\newacronym{dal}{DAL}{deep active learning}
\newacronym{mlp}{MLP}{multi-layer perceptron}
\newacronym{dnn}{DNN}{deep neural network}
\newacronym{sota}{SotA}{state-of-the-art}
\newacronym{cv}{CV}{computer vision}
\newacronym{plm}{PLMs}{pretrained language models}
\newacronym{cnn}{CNN}{convolutional neural network}
\newacronym{rnn}{RNN}{recurrent neural network}
\newacronym{ast}{AST}{audio spectrogram transformer}
\newacronym{w2v2}{W2V2}{Wav2Vec 2.0}
\newacronym{fsl}{FSL}{Few-Shot Learning}
\newacronym{rl}{RL}{Reinforcement Learning}
\newacronym{hf}{HF}{hugging face}
\newacronym{vit}{ViT}{vision transformer}
\newacronym{lora}{LoRA}{Low-Rank Adaptation}
\newacronym{llm}{LLM}{large language model}

\newacronym{pam}{PAM}{passive acoustic monitoring}
\newacronym{sed}{SED}{sound event detection}
\newacronym{bsc}{BSC}{bird sound classification}
\newacronym{snr}{SNR}{signal-to-noise ratio}
\newacronym{stft}{STFT}{Short-Time Fourier Transform}

\newacronym{esc}{ESC}{ESC-50}
\newacronym{as}{AS}{AudioSet}
\newacronym{ac}{AC}{AudioCaps}
\newacronym{vggs}{VGGS}{VGGSound}
\newacronym{fsd}{FSD}{FSD50k}
\newacronym{wtk}{WTK}{Watkins Marine Mammal - Best of Cuts}
\newacronym{bat}{BAT}{Bats}
\newacronym{cbi}{CBI}{Cornell Bird Identification}
\newacronym{dog}{DOG}{Dogs}
\newacronym{hum}{HUM}{HumBugDB}
\newacronym{asa}{ASA}{Animal Sound Archive}
\newacronym{xc}{XC}{Xeno-Canto}
\newacronym{mac}{MAC}{Macaulay Library}
\newacronym{bs}{BS}{BirdSet}
\newacronym{ina}{INA}{iNaturalist}
\newacronym{ins}{INS}{iNatSounds}
\newacronym{mkt}{MKT}{MeerKAT}
\newacronym{is}{IS}{InsectSet459}
\newacronym{wmm}{WMM}{Watkins Marine Mammal - All Cut}
\newacronym{rs}{RS}{ReefSet}
\newacronym{lbl}{LBL}{LibriLight}
\newacronym{lsp}{LSP}{LibriSpeech}

\newacronym{per}{PER}{Amazon Basin}
\newacronym{nes}{NES}{Colombia Costa Rica}
\newacronym{uhh}{UHH}{Hawaiian Islands}
\newacronym{hsn}{HSN}{High Sierra Nevada}
\newacronym{nbp}{NBP}{NIPS4Bplus}
\newacronym{pow}{POW}{Powdermill Nature}
\newacronym{ssw}{SSW}{Sapsucker Woods}
\newacronym{sne}{SNE}{Sierra Nevada}
\newacronym{vox}{VOX}{BirdVox-DCASE-20k}

\newacronym{iam}{IAM}{IEMOCAP}
\newacronym{vc1}{VC1}{VoxCeleb V1}
\newacronym{sc1}{SC1}{SpeechCommandsV1}
\newacronym{sc2}{SC2}{SpeechCommandsV2}

\newacronym{bert}{BERT}{Bidirectional Encoder Representations from Transformers}
\newacronym{hubert}{HuBERT}{Hidden unit \gls{bert}}
\newacronym{ssast}{SSAST}{Self-Supervised Audio Spectrogram Transformer}
\newacronym{aves}{AVES}{Animal Vocalization Encoder Based on Self-Supervision}
\newacronym{beats}{BEATs}{Bidirectional Encoder representation from Audio Transformers}
\newacronym{eat}{EAT}{Efficient Audio Transformer}
\newacronym{mspm}{MSPM}{masked spectrogram patch modeling}
\newacronym{jepa}{JEPA}{joint embedding predictive architecture}
\newacronym{clap}{CLAP}{Contrastive Language-Audio Pretraining}
\newacronym{mae}{MAE}{Masked Autoencoder}
\newacronym{amae}{AudioMAE}{Audio Masked Autoencoder}
\newacronym{ltu}{LTU}{Listen, Think, and Understand}

\newcolumntype{L}[1]{>{\raggedright\arraybackslash}p{#1}}

\definecolor{lightgray}{gray}{0.9}
\definecolor{lightgreen}{RGB}{210,250,210}
\definecolor{dreamerblue}{RGB}{210,230,255}
\definecolor{muzeropurple}{RGB}{230,210,255}
\definecolor{harmonydreamorange}{RGB}{255,230,200}
\definecolor{sgigreen}{RGB}{220,255,220}
\definecolor{bbfyellow}{RGB}{255,255,200}
\definecolor{shift}{RGB}{200,200,255}
\definecolor{scale}{RGB}{255,200,255}

\definecolor{n_default}{RGB}{245,245,245}
\definecolor{n_gray}{RGB}{200,200,200}
\definecolor{n_brown}{RGB}{210,180,140}
\definecolor{n_orange}{RGB}{255,165,0}
\definecolor{n_yellow}{RGB}{255,255,0}
\definecolor{n_green}{RGB}{144,238,144}
\definecolor{n_blue}{RGB}{173,216,230}
\definecolor{n_purple}{RGB}{221,160,221}
\definecolor{n_pink}{RGB}{255,192,203}
\definecolor{n_red}{RGB}{255,99,99}

\newtcbox{\graybubble}{on line, colback=lightgray, colframe=lightgray,
boxrule=0pt, arc=3pt, left=1pt, right=1pt, top=0.5pt, bottom=0.5pt}

\newtcbox{\pinkbubble}{on line, colback=n_pink, colframe=n_pink,
boxrule=0pt, arc=3pt, left=1pt, right=1pt, top=0.5pt, bottom=0.5pt}

\newtcbox{\orangebubble}{on line, colback=n_orange, colframe=n_orange,
boxrule=0pt, arc=3pt, left=1pt, right=1pt, top=0.5pt, bottom=0.5pt}

\newtcbox{\yellowbubble}{on line, colback=n_yellow, colframe=n_yellow,
boxrule=0pt, arc=3pt, left=1pt, right=1pt, top=0.5pt, bottom=0.5pt}

\newtcbox{\greenbubble}{on line, colback=n_green, colframe=n_green,
boxrule=0pt, arc=3pt, left=1pt, right=1pt, top=0.5pt, bottom=0.5pt}

\newtcbox{\bluebubble}{on line, colback=n_blue, colframe=n_blue,
boxrule=0pt, arc=3pt, left=1pt, right=1pt, top=0.5pt, bottom=0.5pt}

\newtcbox{\purplebubble}{on line, colback=n_purple, colframe=n_purple,
boxrule=0pt, arc=3pt, left=1pt, right=1pt, top=0.5pt, bottom=0.5pt}

\newtcbox{\redbubble}{on line, colback=n_red, colframe=n_red,
boxrule=0pt, arc=3pt, left=1pt, right=1pt, top=0.5pt, bottom=0.5pt}

\newtcbox{\brownbubble}{on line, colback=n_brown, colframe=n_brown,
boxrule=0pt, arc=3pt, left=1pt, right=1pt, top=0.5pt, bottom=0.5pt}

\title{Foundation Models for Bioacoustics - a Comparative Review}

\usepackage{authblk}

\newbox{\orcid}\sbox{\orcid}{\includegraphics[scale=0.06]{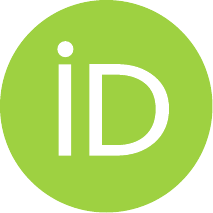}}
\author[1]{%
  \href{https://orcid.org/0009-0001-8519-3571}{\usebox{\orcid}~Raphael~Schwinger\thanks{\texttt{rsc@informatik.uni-kiel.de}}}%
}
\author[1]{%
  \href{https://orcid.org/0009-0007-8396-1585}{\usebox{\orcid}~Paria~Vali~Zadeh}%
}
\author[2]{%
  \href{https://orcid.org/0000-0002-6552-3270}{\usebox{\orcid}~Lukas~Rauch}%
}
\author[1]{%
  \href{https://orcid.org/0009-0007-2527-9078}{\usebox{\orcid}~Mats~Kurz}%
}
\author[1]{%
  \href{https://orcid.org/0009-0002-8928-0477}{\usebox{\orcid}~Tom~Hauschild}%
}
\author[3]{%
  \href{https://orcid.org/0000-0003-1637-6822}{\usebox{\orcid}~Sam~Lapp}%
}
\author[1]{%
  \href{https://orcid.org/0000-0002-5825-8915}{\usebox{\orcid}~Sven~Tomforde}%
}
\affil[1]{INS, Kiel University, Germany}
\affil[2]{IES, University of Kassel, Germany}
\affil[3]{University of Pittsburgh, Pittsburgh, PA, USA}

\renewcommand{\headeright}{Preprint}
\renewcommand{\undertitle}{Preprint}
\renewcommand{\shorttitle}{Foundation models for bioacoustics - a comparative review}

\hypersetup{
  pdftitle={Foundation Models for Bioacoustics - a Comparative Review},
  pdfsubject={cs.LG},
  pdfauthor={Raphael Schwinger, Paria Vali Zadeh, Lukas Rauch, Mats Kurz, Tom Hauschild, Sam Lapp, Sven Tomforde},
  pdfkeywords={foundation models, bioacoustics, transfer learning, self-supervised learning, classification},
  colorlinks=false,
}

\begin{document}
\maketitle

\begin{abstract}
  Automated bioacoustic analysis is essential for biodiversity monitoring and conservation, requiring advanced deep learning models that can adapt to diverse bioacoustic tasks. This article presents a comprehensive review of large-scale pretrained bioacoustic foundation models and systematically investigates their transferability across multiple bioacoustic classification tasks. We overview bioacoustic representation learning by analysing pretraining data sources and benchmarks. On this basis, we review bioacoustic foundation models, dissecting the models' training data, preprocessing, augmentations, architecture, and training paradigm. Additionally, we conduct an extensive empirical study of selected models on the BEANS and BirdSet benchmarks, evaluating generalisability under linear and attentive probing. Our experimental analysis reveals that Perch~2.0 achieves the highest BirdSet score (restricted evaluation) and the strongest linear probing result on BEANS, building on diverse multi-taxa supervised pretraining; that BirdMAE is the best model among probing-based strategies on BirdSet and second on BEANS after BEATs$_{NLM}$, the encoder of NatureLM-audio; that attentive probing is beneficial to extract the full performance of transformer-based models; and that general-purpose audio models trained with self-supervised learning on AudioSet outperform many specialised bird sound models on BEANS when evaluated with attentive probing. These findings provide valuable guidance for practitioners selecting appropriate models to adapt them to new bioacoustic classification tasks via probing.

\end{abstract}

\section{Introduction}
Monitoring biodiversity is essential for guiding conservation strategies and understanding ecological dynamics, providing crucial insights into the health and resilience of various ecosystems~\cite{schmeller_building_2017}. Such monitoring efforts enable researchers and policymakers to detect changes in species populations and ecological processes, thereby informing effective management and protection measures~\cite{margules_systematic_2000}. Monitoring bird population changes, for instance, indicates broader biodiversity shifts~\cite{sekercioglu2016}. \Gls{pam} provides a minimally invasive and scalable approach to monitoring sound-producing taxa, especially those that are rare or otherwise difficult to survey (e.g., nocturnal, visually cryptic, or in difficult-to-access environments)~\cite{Hoefer_PAM}. \Gls{pam} typically detects equivalent or more species than in-person surveys conducted by experts while achieving far greater temporal coverage~\cite{darras_comparing_2018, Hoefer_PAM}. Even though \Gls{pam} projects can efficiently collect thousands of hours of audio, expert annotations for every recording are infeasible. As a result, automated species detection methods have become a central objective for conducting biodiversity monitoring with \Gls{pam}.

\Gls{dl} models have demonstrated impressive performance in automatically detecting and classifying species by sound, making them invaluable tools in ecological monitoring and research~\cite{stowell_computational_2022}. \gls{dl} algorithms typically require large numbers of annotated examples of target vocalisations to achieve high accuracy and reliability. As a result, \gls{dl} methods may perform poorly for rare, endangered, cryptic, and understudied species---the same species for which \gls{pam} is particularly advantageous~\cite{sugai_pam,stowell_computational_2022}. Furthermore, ecological communities typically contain a few common species and many rare species that are more vulnerable to extinction. Thus, developing reliable automated species recognition systems with very limited training data remains a key challenge in leveraging \gls{pam} for conservation efforts. In addition, many bioacoustic recordings are captured with focal microphones, which have a higher signal-to-noise ratio than soundscape recording devices used in \gls{pam} applications~\cite{rauch2025birdset}.

Transfer learning is a set of techniques designed to address the issue of data scarcity in \gls{dl} by leveraging knowledge from related tasks~\cite{hosna2022transfer,ghaniImpactTransferLearning2025}. The key insight of transfer learning is that a \gls{dl} model trained on vast and diverse datasets can be adapted for specific tasks in a new domain using only a few training samples~\cite{bommasani_opportunities_2022,ghani_global_2023}. Under the paradigm of representation learning, the key to successful transfer learning is to train one \gls{dl} model, referred to as the \textit{foundation model}, that creates generalisable representations useful for a wide variety of downstream tasks~\cite{hu_pushing_2022}. Recent research demonstrates that foundation models trained on global repositories of birdsong recordings (e.g., \gls{xc}~\cite{vellinga_xeno-canto_2015}) can be adapted for accurate species classification of bird, frog, and mammal vocalisations with very few training samples \cite{ghani_global_2023, dumoulinSearchSquawkAgile2025, wedly_transferlearning_2025}.

Transfer learning strategies range in the degree to which the foundation model is preserved or modified. At one extreme, the entire \gls{dl} model can be "fine-tuned" by training all of its parameters (e.g., tens of millions of parameters) on the new data via standard backpropagation. When sufficient labelled training data is available, fine-tuning can yield strong in-domain performance~\cite{hosna2022transfer}, though it may distort pretrained features and underperform out-of-distribution~\cite{kumar2022finetuningdistortpretrainedfeatures}, and also incurs the highest computational cost. At the other extreme, probing methods utilise the foundation model as a frozen feature extractor. For instance, in linear probing, only the final layer of the network will be trained. One can also train a probe on the more flexible patch embeddings (i.e., before pooling embeddings across regions of the spectrogram). To keep the parameter count low, attentive probing~\cite{elnouby2024scalablepretraininglargeautoregressive} or prototypical probe~\cite{heinrich_audioProtoPNet,rauch2025maskedautoencoderslistenbirds} can be used. Intermediate strategies involve training some but not all of the foundation model's parameters with techniques like \gls{lora}. Building upon frozen representations has further computational benefits, as the representations can be cached during training, then used for search and retrieval tasks in large-scale datasets \cite{dumoulinSearchSquawkAgile2025} and edge deployment~\cite{rauch2024deepactivelearningavian}. The effectiveness of transfer learning therefore depends on the quality of the foundation model as a backbone. Training classification probes on fixed embeddings is an effective way to assess the generalisability of models~\cite{kumar2022finetuningdistortpretrainedfeatures, niizumo_byol_audio}.

This article aims to give practitioners and machine learning developers an overview of existing bioacoustic foundation models and the large-scale data sources these models are based on. Through our comparative analysis, we intend to give guidance on which model could be used in a probing-based classification scenario. By summarising the current \gls{sota} of bioacoustic foundation models, we aspire to foster future bioacoustic model development. Our contributions can be summarised as follows:
\begin{tcolorbox}[title=Contributions, boxrule=1.25pt, left=2pt, top=1pt, bottom=0.5pt]
  \begin{enumerate}[leftmargin=0.55cm, label=\customlabel{\arabic*}]
    \item We provide an overview of bioacoustic representation learning by analysing pretraining data sources and benchmarks, guiding researchers on what data resources to build on.
    \item We review bioacoustic foundation models, dissecting the models' training data, preprocessing, augmentations, architecture, and training paradigm.
    \item We conduct an extensive empirical study of selected models on the \texttt{BirdSet} and \texttt{BEANS} benchmarks, evaluating the models' generalisability under linear and attentive probing, revealing that:
      \begin{itemize}
        \item Perch~2.0 achieves the highest BirdSet score (restricted evaluation) and the strongest linear probing result on BEANS building on diverse multi-taxa supervised pretraining.
        \item BirdMAE is the best model among probing-based strategies on BirdSet and second on BEANS after BEATs$_{NLM}$, the encoder of NatureLM-audio.
        \item Attentive probing is beneficial to extract the full performance of transformer-based models.
        \item General-purpose audio models trained with self-supervised learning on AudioSet outperform many specialised bird sound models on BEANS when evaluated with attentive probing.
      \end{itemize}
    \item We provide a comprehensive \textbf{codebase}\footnote{\href{https://github.com/DBD-research-group/BioFoundation}{github.com/DBD-research-group/BioFoundation}} to support reproducibility and accessibility. We enhance transparency by providing detailed results and training logs via Weights and Biases\footnote{\href{https://wandb.ai/deepbirddetect/biofoundation}{wandb.ai/deepbirddetect/biofoundation}}~\cite{biewald_experiment_2020}.
  \end{enumerate}
\end{tcolorbox}

\section{Related work}
\label{sec:related-work}

The application of deep learning to bioacoustic analysis has rapidly evolved, driven by advances in representation learning and the development of standardised evaluation protocols. This section reviews relevant work in two key areas that directly inform our comparative review of foundation models for bioacoustic classification.

\paragraph{Bioacoustic representation learning}
Stowell~\cite{stowell_computational_2022} provides a comprehensive review of computational bioacoustics with deep learning, identifying key challenges including data scarcity, domain-specific requirements, and the need for robust evaluation practices. The field has responded with the development of standardised benchmarks (BEANS~\cite{hagiwara_beans_2023}, BIRB~\cite{hamer_birb_2023} and BirdSet~\cite{rauch2025birdset}) to enable systematic model comparison. In particular, BirdSet provides a comprehensive description of challenges in creating avian bioacoustic models that expand to other taxa. Van Merriënboer et al.~\cite{van_merrienboer_birds_2024} further emphasised the importance of robust evaluation protocols for assessing domain generalisation in bioacoustic foundation models, advocating for segment-based and event-based evaluation methodologies that better reflect real-world deployment scenarios.

\paragraph{Transfer learning and model comparison}
Several studies have systematically compared model performance and transfer learning strategies in bioacoustics. Ghani et al.~\cite{ghani_global_2023} demonstrated that embeddings from large-scale bird sound classifiers consistently outperform general audio models like AudioMAE~\cite{huang_masked_2022} and VGGish~\cite{audioset} across diverse bioacoustic tasks, establishing the value of domain-specific pretraining for few-shot transfer learning. While investigating AudioMAE more closely, they did not conduct experiments on the basis of the patch embeddings, which we found crucial for extracting the performance of general audio models. Their subsequent work~\cite{ghaniImpactTransferLearning2025} investigated various adaptation strategies including linear probing, fine-tuning, and knowledge distillation, finding that linear probing provides superior robustness for soundscape generalisation. Williams et al.~\cite{williams2024leveragingtropicalreefbird} extended cross-domain transfer learning to marine bioacoustics, comparing models trained on bird, reef, and general audio data, and demonstrating that multi-domain pretraining strategies can overcome domain-specific data limitations. Cauzinille et al.~\cite{cauzinille_investigating_2024} explored adapting self-supervised speech models (HuBERT~\cite{hsu_hubert_2021}, Wav2Vec2~\cite{baevski_wav2vec_2020}) for primate vocalisations, revealing that speech-based models exhibit superior robustness to background noise compared to traditional bioacoustic models through layer-wise performance analysis. Kath et al.~\cite{kath_active_2024} investigated the use of pretrained models as feature extractors in active learning settings, comparing BirdNET, VGGish, YAMNet~\cite{noauthor_google_nodate}, and \gls{cnn} architectures for efficient species identification with minimal labelling effort. Recent concurrent work by Kather et al.~\cite{kather_clustering_2025} evaluated feature extractors from 15 bioacoustic models using clustering approaches, identifying challenges in handling overlapping sounds and noisy environments across various model architectures and training paradigms.

\paragraph{Dataset construction and split protocols in bioacoustics} In bioacoustics, reported performance can be strongly affected by how train/validation/test splits are constructed~\cite{colonna_how_2016,stock_data_2023}. Random splits that ignore metadata (e.g., recorder identity, site, or temporal proximity) can inadvertently place highly correlated recordings into different splits, leading to optimistic estimates that do not reflect real-world deployment. In particular, since many animals produce a stereotyped sound in a repeated sequence, the inclusion of audio clips from the same recorder and similar time points (e.g. same audio file) across multiple data splits can be considered data leakage. This is particularly relevant for \gls{pam} applications, where protocol choices (including filtering and partitioning strategies that better mimic field conditions) can substantially change measured performance~\cite{aguiar_importance_2021}. Similar issues arise in bioacoustic classification settings when recordings are not independent across time or collection conditions; using stricter protocols (e.g., separating correlated segments or sessions) provides a more realistic assessment of generalisation~\cite{trapanotto_convolutional_2022}.
To improve comparability, recent benchmarks standardise dataset construction and evaluation protocols. BEANS aggregates multiple animal-sound datasets and provides pre-defined splits (train/validation/test) per dataset~\cite{hagiwara_beans_2023}. Most classification datasets in BEANS (Watkins, Dogs, HumBugDB) are randomly split into 60/20/20 train/validation/test portions with stratification to preserve class proportions. A notable exception is the \gls{cbi} dataset, which is split so that no recordist appears in more than one partition, thereby preventing data leakage from shared recording conditions. For the detection datasets, BEANS partitions long recordings temporally: continuous files are divided into 1-minute chunks and the first 60\% of chunks are assigned to training, the next 20\% to validation, and the final 20\% to testing (DCASE, ENAbirds), or files are assigned to splits directly (RFCX, HICEAS). The Gibbons dataset follows a day-level temporal split, using the first six recording days for training, the next three for validation, and the remainder for testing. Metadata influence can therefore not be entirely neglected when evaluating models on the BEANS benchmark.

BirdSet, on the other hand, enforces a more structured separation of training and evaluation data~\cite{rauch2025birdset}. Training is performed exclusively on weakly labelled focal recordings from \gls{xc}. Whereas for testing, BirdSet uses fully annotated soundscape recordings from seven geographically diverse \gls{pam} deployments. This clear separation of data sources removes any influence of metadata on the evaluation of the models.

In this work, we follow the official benchmark splits and evaluation protocols of BEANS and BirdSet to ensure that reported results remain comparable and representative.

\section{Data for Bioacoustic Representation Learning}
\label{sec:representationlearning}

This section provides an overview of available data for bioacoustic representation learning. We differentiate between \textit{pretraining} and \textit{evaluation} datasets. Whereas the size and diversity are the most important characteristics of pretraining data, for evaluation data, high annotation quality is important.

\begin{table}[t]
  \centering
  \caption{Large-scale audio datasets for audio representation learning, categorised into general, bioacoustic, and speech datasets. The table reports the number of samples, classes, and duration for each dataset. }
  \label{tab:pretraindatasets}
  \resizebox{0.9\textwidth}{!}{%
    \begin{tabular}{l}
      \begin{tabular}{lp{2cm} p{8.5cm} r r r c}
  \toprule
  \rule{0pt}{2em} %

  & \multicolumn{0}{c}{ \textbf{}}
  & \multicolumn{1}{c}{ \textbf{Name}}

  & \multicolumn{1}{c}{\textbf{\#Labels{\scriptsize\(\downarrow\)}}}

  & \multicolumn{1}{c}{\textbf{\#Classes} }
  & \multicolumn{1}{c}{ \textbf{Duration\scriptsize(h)} }

  \\

  \midrule
  & \multicolumn{5}{p{5cm}}{\textbf{\textit{General audio}}} \vspace{2mm}
  \\
  & AS                                                                    & \datalink{https://research.google.com/audioset/}{AudioSet}~\cite{audioset}                                                                & 2,100,000 & 527            & 5,800            \\
  \cmidrule(lr){2-7} %
  & \hspace{0.5mm}~\dirarrow~AC                                           & \datalink{https://github.com/cdjkim/audiocaps}{AudioCaps}~\cite{kim_audiocaps_2019}                                                     & 39,106    & -              & 108.6          & \\
  \cmidrule(lr){1-7} %

  & VGGS                                                                  & \datalink{https://www.robots.ox.ac.uk/~vgg/data/vggsound/}{VGGSound}~\cite{chen_vggsound_2020}                                                      & 200,000   & 310            & 550              \\
  \cmidrule(lr){1-7} %

  & FSD                                                                   & \datalink{https://zenodo.org/records/4060432}{FSD50k}~\cite{fsd50k}                                                                    & 51,197    & 200            & 108.3            \\

  \midrule
  & \multicolumn{5}{p{5cm}}{\textbf{\textit{Bioacoustic}}} \vspace{2mm}
  \\
  & MAC                                                                   & \datalink{https://www.macaulaylibrary.org/}{Macaulay Library\textsuperscript{*} \textsuperscript{+}}~\cite{noauthor_macaulay_nodate} & 2,699,789 & 10056          & >10,000          \\
  \cmidrule(lr){1-7} %

  & XC                                                                    & \datalink{https://xeno-canto.org/}{Xeno-Canto\textsuperscript{*}}~\cite{vellinga_xeno-canto_2015}                           & 1,668,986 & 12,514         & 17,221           \\
  \cmidrule(lr){2-7} %
  & \hspace{0.5mm}\dirarrow~BS                                            & \datalink{https://huggingface.co/datasets/DBD-research-group/BirdSet}{BirdSet}~\cite{rauch2025birdset}                                                         & 712,433   & 9,734          & $\approx$7,200   \\
  \cmidrule(lr){2-7} %
  & \hspace{0.5mm}\dirarrow~BIRB                                          & \datalink{https://github.com/google-research/perch}{Benchmark for Information Retrieval in Bioacoustics}~\cite{hamer_birb_2023}              & >750,000  & >10,000        & >10,000          \\
  \cmidrule(lr){1-7} %

  & INA                                                                   & \datalink{https://www.inaturalist.org/}{iNaturalist\textsuperscript{*}}~\cite{noauthor_inaturalist_nodate}                       & 1,142,635 & 12,838         & $\approx$5,962   \\
  \cmidrule(lr){2-7} %
  & \hspace{0.5mm}\dirarrow~INS                                           & \datalink{https://github.com/visipedia/inat_sounds}{iNatSounds}~\cite{chasmai_inaturalist_2024}                                              & 230,000   & $\approx$5,500 & 1,200            \\
  \cmidrule(lr){1-7} %

  & MKT                                                                   & \datalink{https://edmond.mpg.de/dataset.xhtml?persistentId=doi:10.17617/3.0J0DYB}{MeerKAT}~\cite{schaferzimmermann2024animal2vec}                                          & 184,000   & 12             & 184              \\
  \cmidrule(lr){1-7} %

  & ASA                                                                   & \datalink{https://www.museumfuernaturkunde.berlin/en/research/animal-sound-archive}{Animal Sound Archive}~\cite{noauthor_animal_nodate}                                      & ~25,438   & 991            & 1,284            \\
  \cmidrule(lr){1-7} %

  & IS                                                                    & \datalink{https://academictorrents.com/details/a8278b49a33cc05a23b14aebd2da75d693012678}{InsectSet459}~\cite{fais_insectset459_2025}                                              & 26,399    & 459            & 227.2            \\
  \cmidrule(lr){1-7} %
  & WMM                                                                   & \datalink{https://darchive.mblwhoilibrary.org/handle/1912/5877}{Watkins Marine Mammal - All Cut}~\cite{sayigh_watkins_2017}                              & 15,000    & 60             & 42               \\
  \cmidrule(lr){1-7} %
  & RS                                                                    & \datalink{https://arxiv.org/abs/2404.16436}{ReefSet}~\cite{williams2024leveragingtropicalreefbird}                                   & 13,000    & 38             & 156              \\

  \bottomrule
\end{tabular}%

                                                                                              \\
      \rule{0in}{1.2em}\textsuperscript{*} This entity is not a fixed dataset but a constantly growing collection of audio samples. \\
      \textsuperscript{+} This entity cannot be publicly accessed.                                                                 \\
    \end{tabular}

  }
\end{table}

\subsection{Pretraining datasets}
\label{sec:pretrainingdatasets}
Representations learned from large-scale datasets are crucial for training models capable of effectively generalising across diverse tasks. Table \ref{tab:pretraindatasets} provides an overview of the datasets most prominent and frequently employed in audio representation learning, categorised into general, bioacoustic, and speech datasets. We selected datasets used, either in the training of the bioacoustic foundation models analysed in Section~\ref{sec:models}, or for the baseline models selected for our experiments described in Section~\ref{sec:comparison}, or frequently referenced within bioacoustic research.

\paragraph{General datasets}
Here, we summarise key audio datasets for \gls{ml} model development that are not specific to bioacoustics.
\textit{\gls{as}}~\cite{audioset} is a dataset of over 2 million human-labelled 10-second sound clips sourced from YouTube videos, making it one of the largest and most diverse datasets available. It covers a wide range of sounds from 527 audio event classes, and is the most widely used dataset for training and benchmarking audio models. Since it is sourced from YouTube videos and officially only provides metadata, including the download links, some clips are no longer available. The dataset is divided into three distinct subsets: unbalanced, balanced, and testing. \textit{\gls{ac}}~\cite{kim_audiocaps_2019} is a small subset of AudioSet labelled with natural language captions.
\textit{\gls{vggs}}~\cite{chen_vggsound_2020} is a large-scale dataset containing 200,000 audiovisual clips from 310 classes, designed to facilitate the development of audiovisual models. Like \gls{as}, its 10-second clips are sourced from YouTube videos, and only metadata is provided.
The Freesound project\footnote{\href{https://www.freesound.org}{freesound.org} (last access: 2025-08-02)} collects and shares audio samples, including sound effects, field recordings, and music. The \textit{\gls{fsd}}~\cite{fsd50k} dataset is a subset of Freesound, containing 51,000 audio files annotated with 200 sound classes. It is designed to foster the development of general-purpose audio tagging systems, which are essential for tasks that require fine-grained audio understanding.

\paragraph{Bioacoustic datasets}

\begin{figure}[t]
  \centering
  \includegraphics[width=0.6\textwidth]{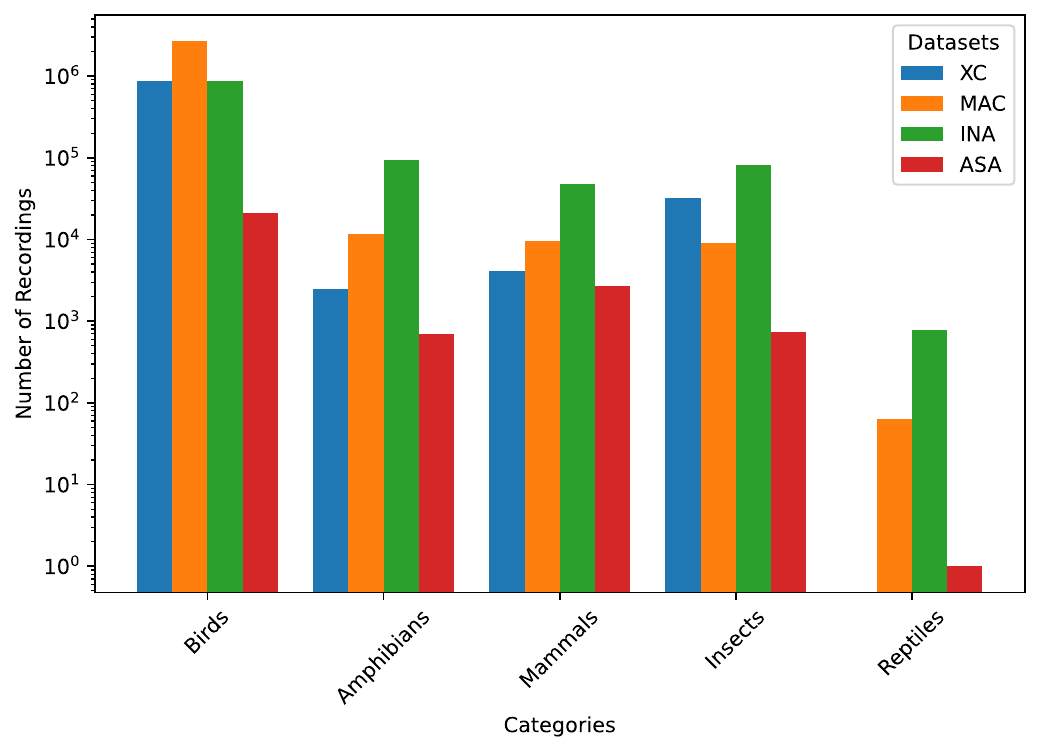}
  \caption{Taxonomy distribution (logarithmic scale) of the large bioacoustic data platforms—Xeno-Canto (XC), Macaulay Library (MAC), iNaturalist (INA), and Animal Sound Archive (ASA) —across five widely studied biological groups: Birds, Amphibians, Mammals, Insects, and Reptiles~\cite{noauthor_animal_nodate,noauthor_inaturalist_nodate,noauthor_macaulay_nodate,vellinga_xeno-canto_2015}.}
  \label{fig:taxonomy-distribution}
\end{figure}

There are several large-scale bioacoustic audio platforms, including the \textit{\gls{mac}}~\cite{noauthor_macaulay_nodate}, \textit{\gls{xc}}~\cite{vellinga_xeno-canto_2015} and \textit{\gls{ina}}~\cite{noauthor_inaturalist_nodate}, where professionals and citizen scientists can upload recordings. Of these, only \gls{xc} and \gls{ina} are fully accessible for public download and use. In total, these datasets contain millions of recordings covering more than 10,000 species. Bird sounds make up most of the recordings, but other animals are also represented; see Figure \ref{fig:taxonomy-distribution}. All recordings are weakly labelled, meaning that the primary vocalising species are assigned to the entire variable-length clip, but specific annotations are not provided. Sometimes, additional background species are also labelled. The sheer size and diversity of these collections make them ideal for pretraining bioacoustic models, while the weak labels limit their utility for model evaluation. Some other online repositories contain weakly labelled recordings of specific taxonomic groups, such as fonozoo\footnote{\href{http://www.fonozoo.com}{fonozoo.com} (last access: 2025-08-02)} for amphibians, and ChiroVox\footnote{\href{https://obm.ecolres.hu/projects/chirovox/index.php/}{obm.ecolres.hu/projects/chirovox} (last access: 2025-08-02)} for bats. The BioAcoustic-Ai project\footnote{\href{http://bioacoustic-ai.github.io/bioacoustics-datasets/}{bioacoustic-ai.github.io/bioacoustics-datasets} (last access: 2025-08-02)} collects and classifies datasets by taxonomic class and duration.

Specific datasets have been created to provide standardised evaluation benchmarks to foster the development and comparability of bioacoustic classification and detection models. The \textit{\gls{bs}}~\cite{rauch2025birdset} dataset is a large-scale dataset for bird sound classification, curating over 0.5 million samples from \gls{xc} for training and seven fully annotated, strongly labelled soundscape test datasets. \textit{\gls{ins}}~\cite{chasmai_inaturalist_2024} is a large-scale weakly-labelled dataset for animal sound classification, containing over 200 thousand samples from \gls{ina} covering more than 5,000 species. We provide more detail on BirdSet and \gls{ins} in Section \ref{sec:evaldatasets}. \textit{\gls{is}}~\cite{fais_insectset459_2025} curates a collection of insect sounds from \gls{xc} (Orthoptera), \gls{ina} (Orthoptera \& Cicadidae) and BioAcoustica (Cicadidae)\footnote{\href{http://bio.acousti.ca}{bio.acousti.ca} (last access: 2025-08-02)}. \textit{\Gls{mkt}}~\cite{schaferzimmermann2024animal2vec} provides recordings of meerkat vocalisations with millisecond-resolution annotations. The \textit{\gls{asa}}~\cite{noauthor_animal_nodate} collects and shares animal sounds, covering a wide range of species and sound types. Not all recordings are annotated or publicly available, but annotations that do exist are provided by experts and are of high quality. \textit{\gls{wmm}}~\cite{sayigh_watkins_2017} is a collection of various marine mammal recordings, covering 60 species and 15 thousand recordings. \textit{\Gls{rs}} is a collection of reef sound recordings collected around the globe.

\subsection{Bioacoustic Benchmarks and Evaluation Datasets} \label{sec:evaldatasets}

\begin{table}[t]
  \centering
  \caption{Datasets and benchmarks for model evaluation, grouped by general, bioacoustic and speech content. The number of labels in each split, the number of classes in the test set and the duration of the test set in hours are listed.}
  \label{tab:evaldatasets}
  \setlength{\tabcolsep}{0.35em}
  \resizebox{0.8\textwidth}{!}{%
    \begin{tabular}{lp{2.2cm} p{5.5cm} r r r c}
    \toprule
    \rule{0pt}{2em} %

     & \multicolumn{1}{c}{ \textbf{Name}}
     & \multicolumn{1}{c}{ \textbf{Notes}}

     & \multicolumn{1}{c}{\textbf{\#Train/Valid/Test{\scriptsize\(\downarrow\)}}}

     & \multicolumn{1}{c}{\textbf{\#Classes} }
     & \multicolumn{1}{c}{ \textbf{Duration\scriptsize(h)} }

    \\

    \midrule
     & \multicolumn{5}{p{5cm}}{\textbf{\textit{General audio}}} \vspace{2mm}
    \\
     & AS-2M~\cite{audioset}                                                      & \datalink{https://research.google.com/audioset/}{unbalanced train set}                                         & 2,041,786 / - / 20,383   & 527   & 56.6  & \\
    \cmidrule(lr){1-7} %
     & AS-20k~\cite{audioset}                                                     & \datalink{https://research.google.com/audioset/}{balanced train set}                                           & 22,160 / - / 20,383      & 527   & 56.6  & \\
    \cmidrule(lr){1-7} %

     & ESC-50~\cite{piczak_esc_2015}                                              & \datalink{https://github.com/karolpiczak/ESC-50}{evaluated with 5-fold cross validation}                       & 2,000                    & 50    & 2.8     \\

    \midrule
     & \multicolumn{5}{p{5cm}}{\textbf{\textit{Bioacoustic}}} \vspace{2mm}
    \\

     & \datalink{https://huggingface.co/datasets/DBD-research-group/BirdSet}{BirdSet}~\cite{rauch2025birdset}  \vspace{1mm}                                                                                                                                          \\
     & \hspace{0.5mm}\dirarrow~PER                                                & \datalink{https://zenodo.org/records/7079124}{Amazon Basin}~\cite{hopping_collection_2022}                  & 16,802 / - / 14,798      & 132   & 21      \\
    \cmidrule(lr){2-7} %
     & \hspace{0.5mm}\dirarrow~NES                                                & \datalink{https://zenodo.org/records/7525349}{Colombia Costa Rica}~\cite{vega-hidalgo_collection_2023}      & 16,117 / - / 6,952       & 89    & 34      \\
    \cmidrule(lr){2-7} %
     & \hspace{0.5mm}\dirarrow~UHH                                                & \datalink{https://zenodo.org/records/7078499}{Hawaiian Islands}~\cite{navine_collection_2022}               & 3,626 / - / 59,583       & 25    & 50.9    \\
    \cmidrule(lr){2-7} %
     & \hspace{0.5mm}\dirarrow~HSN                                                & \datalink{https://zenodo.org/records/7525805}{High Sierra Nevada}~\cite{clapp_collection_2023}              & 5,460 / - / 10,296       & 21    & 16.7    \\
    \cmidrule(lr){2-7} %
     & \hspace{0.5mm}\dirarrow~NBP                                                & \datalink{https://github.com/fbravosanchez/NIPS4Bplus}{NIPS4BPlus}~\cite{fbravosanchez_fbravosancheznips4bplus_2024} & 24,327 / - / 14,798      & 51    & 0.8     \\
    \cmidrule(lr){2-7} %
     & \hspace{0.5mm}\dirarrow~POW                                                & \datalink{https://doi.org/10.5061/dryad.d2547d81z}{Powdermill Nature Reserve}~\cite{chronister_annotated_2021}   & 14,911 / - / 16,052      & 48    & 6.3     \\
    \cmidrule(lr){2-7} %
     & \hspace{0.5mm}\dirarrow~SSW                                                & \datalink{https://zenodo.org/records/7018484}{Sapsucker Woods}~\cite{kahl_collection_2022}                  & 28,403 / - / 50,760      & 81    & 285     \\
    \cmidrule(lr){2-7} %
     & \hspace{0.5mm}\dirarrow~SNE                                                & \datalink{https://zenodo.org/records/7050014}{Sierra Nevada}~\cite{kahl_collection_2022-1}                  & 19,390 / - / 20,147      & 56    & 33      \\

    \cmidrule(lr){1-7} %

     & \datalink{https://github.com/google-research/perch}{BIRB}~\cite{hamer_birb_2023} \vspace{1mm}                                                                                                                                               \\

     & \hspace{0.5mm}\dirarrow~POW                                                & \datalink{https://doi.org/10.5061/dryad.d2547d81z}{Powdermill Nature Reserve}~\cite{chronister_annotated_2021}   & – / 16,052 / –           & 48    & 6.3     \\
    \cmidrule(lr){2-7}

     & \hspace{0.5mm}\dirarrow~SSW                                                & \datalink{https://zenodo.org/records/7018484}{Sapsucker Woods}~\cite{kahl_collection_2022}                  & – / – / 50,760           & 96    & 285     \\
    \cmidrule(lr){2-7}

     & \hspace{0.5mm}\dirarrow~UHH                                                & \datalink{https://zenodo.org/records/7078499}{Hawaiian Islands}~\cite{navine_collection_2022}               & – / – / 59,583           & 27    & 50.9    \\
    \cmidrule(lr){2-7}

     & \hspace{0.5mm}\dirarrow~NES                                                & \datalink{https://zenodo.org/records/7525349}{Colombia Costa Rica}~\cite{vega-hidalgo_collection_2023}      & – / – / 6,952            & 89    & 34      \\
    \cmidrule(lr){2-7}

     & \hspace{0.5mm}\dirarrow~HSN                                                & \datalink{https://zenodo.org/records/7525805}{High Sierra Nevada}~\cite{clapp_collection_2023}              & – / – / 10,296           & 19    & 16.7    \\
    \cmidrule(lr){2-7}

     & \hspace{0.5mm}\dirarrow~SNE                                                & \datalink{https://zenodo.org/records/7050014}{Sierra Nevada}~\cite{kahl_collection_2022-1}                  & – / – / 20,147           & 56    & 33      \\
    \cmidrule(lr){2-7}

     & \hspace{0.5mm}\dirarrow~PER                                                & \datalink{https://zenodo.org/records/7079124}{Amazon Basin}~\cite{hopping_collection_2022}                  & – / – / 14,798           & 132   & 21      \\

    \midrule
     & \datalink{https://github.com/earthspecies/beans}{BEANS}~\cite{hagiwara_beans_2023}  \vspace{1mm}                                                                                                                                         \\
     & \hspace{0.5mm}\dirarrow~WTK                                                & \datalink{https://darchive.mblwhoilibrary.org/handle/1912/5877}{Watkins - best cut}~\cite{sayigh_watkins_2017}                & 1017 / 339 / 339         & 31    & 1.1     \\
    \cmidrule(lr){2-7} %
     & \hspace{0.5mm}\dirarrow~BAT                                                & \datalink{https://www.nature.com/articles/sdata2017143}{Bats}~\cite{prat_annotated_2017}                              & 6000 / 2000 / 2000       & 10    & 1.0     \\
    \cmidrule(lr){2-7} %
     & \hspace{0.5mm}\dirarrow~CBI                                                & \datalink{https://www.kaggle.com/competitions/birdsong-recognition}{Cornell Bird Identification}~\cite{birdsong-recognition}      & 14207 / 3548 / 3620      & 264   & 9.6     \\
    \cmidrule(lr){2-7} %
     & \hspace{0.5mm}\dirarrow~DOG                                                & \datalink{https://github.com/earthspecies/beans}{Dogs}~\cite{yin_barking_2004}                                 & 415 / 139 / 139          & 10    & 0.5     \\
    \cmidrule(lr){2-7} %
     & \hspace{0.5mm}\dirarrow~HUM                                                & \datalink{https://zenodo.org/record/4904800}{HumBugDB}~\cite{kiskin_humbugdb_2021}                         & 9293 / 1859 / 1859       & 14    & 6.7     \\

    \cmidrule(lr){1-7} %
     & \datalink{https://github.com/visipedia/inat_sounds}{INS}~\cite{chasmai_inaturalist_2024}                                        & iNat Sounds test and val subset                              & 137,012/ 45,698 / 49,527 & 1,212 & 137.6   \\
    \cmidrule(lr){1-7} %
     & \datalink{https://zenodo.org/records/8342596}{AnuraSet}~\cite{canas_dataset_2023}                                         &                                                              & 65,365/ - / 28,013       & 42    & 27      \\

    \bottomrule
\end{tabular}%

  }

\end{table}
Benchmarks play an important role in the development and evaluation of \gls{ml} models by providing standardised datasets and protocols for a fair and reproducible comparison. They enable researchers to systematically evaluate model performance, identify strengths and weaknesses, and drive progress through transparent reporting of results. In this study, we detail the bioacoustic benchmarks used in the surveyed models in Section~\ref{sec:models}. Table~\ref{tab:evaldatasets} summarises the key properties of the evaluation datasets.

\paragraph{The BEANS benchmark~\cite{hagiwara_beans_2023}} aims to facilitate accurate evaluation and comparison of \gls{ml} models using a diverse collection of bioacoustic datasets spanning a wide range of species. It focuses on two core tasks in bioacoustics, classification and detection, and includes twelve datasets covering birds, land and marine mammals, amphibians, and insects. Specifically, five datasets are designated for classification: \textit{\gls{wtk}}~\cite{sayigh_watkins_2017}, derived from \gls{wmm}; \textit{\gls{bat}}~\cite{prat_annotated_2017}; \textit{\gls{cbi}}~\cite{birdsong-recognition}, part of \gls{xc}; \textit{\gls{dog}}~\cite{yin_barking_2004}; and \textit{\gls{hum}}~\cite{kiskin_humbugdb_2021}. Additionally, five datasets are designated for detection: \textit{dcase}~\cite{morfi_few-shot_2021}, \textit{enabirds}~\cite{chronister_annotated_2021}, \textit{hiceas}~\cite{noaa2022hiceas}, \textit{rfcx}~\cite{lebien_pipeline_2020}, and \textit{gibbons}~\cite{dufourq_automated_2021}. Furthermore, two auxiliary datasets, \textit{ESC-50}~\cite{piczak_esc_2015} and \textit{SC1}~\cite{warden_speech_2018}, are provided for tasks such as training, augmentation, or validation. The classification task is framed as a multi-class problem to either classify the species (\gls{wtk}, \gls{hum}, \gls{cbi}) or individual animals (\gls{bat}, \gls{dog}).

\paragraph{BirdSet benchmark~\cite{rauch2025birdset}} comprises approximately 520,000 global bird recordings for training and over 400 hours of \gls{pam} recordings for testing. The dataset is organised into three components: training, auxiliary, and test sets. The training set contains weakly labelled focal recordings sourced from \gls{xc}. The auxiliary set supports model development through data augmentation and validation, incorporating non-bird soundscape recordings from the \textit{\gls{vox}}~\cite{lostanlen_birdvox-dcase-20k_2018} dataset, as well as \textit{\gls{pow}}~\cite{chronister_annotated_2021}, a small, fully annotated bird soundscape dataset. The test set consists of fully annotated soundscapes and is framed as a multi-label classification task, spanning diverse acoustic environments including \textit{\gls{per}}~\cite{hopping_collection_2022}, \textit{\gls{nes}}~\cite{vega-hidalgo_collection_2023}, \textit{\gls{uhh}}~\cite{navine_collection_2022}, \textit{\gls{hsn}}~\cite{clapp_collection_2023}, \textit{\gls{nbp}}~\cite{fbravosanchez_fbravosancheznips4bplus_2024}, \textit{\gls{ssw}}~\cite{kahl_collection_2022}, and \textit{\gls{sne}}~\cite{kahl_collection_2022-1}. These subsets represent a wide range of geographic regions and recording conditions. BirdSet's training protocol is tailored for a multi-label classification problem, (pre)training a model on the \gls{xc} training set with 528,434 or 89,798 samples for the XC-large (XCL) or XC-medium (XCM) set, respectively. In addition, for each evaluation set, a dedicated training set (DT) covering the species present in the evaluation set is provided from the XCL set. BirdSet's evaluation protocol states a multi-label classification task by analysing the entire soundscape dataset using 5-second segments. The training on focal recordings (\gls{xc} contains mostly focal recordings) and testing on soundscape data reflects a common and challenging scenario for practical \gls{pam} applications.

\paragraph{BIRB benchmark~\cite{hamer_birb_2023}} presents a generalisation benchmark for information retrieval in bioacoustics, designed to evaluate model performance under real-world conditions. The benchmark is structured as a retrieval task: models trained on weakly labelled focal recordings from the XC corpus must retrieve relevant vocalisations from downstream corpora using a small number of exemplar recordings per species. BIRB systematically evaluates three key generalisation challenges: out-of-distribution retrieval from passive soundscapes, few-shot learning of novel species, and robustness to class imbalance and label shift. The upstream training data is drawn from \gls{xc}. \gls{pow} is used exclusively for validation and is not part of the evaluation set. The evaluation datasets include soundscape corpora such as \gls{ssw}, \gls{uhh}, \gls{nes}, \gls{hsn}, \gls{sne}, and \gls{per}. In addition, the evaluation set also includes carefully curated subsets of \gls{xc} recordings held out from training, such as artificially rare species from New York and species from held-out regions like Hawai’i and Colombia. BIRB integrates these heterogeneous datasets by aligning species taxonomies, resolving label format inconsistencies, extracting fixed-length audio slices via peak-finding, and converting time-boxed annotations into slice-level labels. The same soundscape corpus can thus appear with different class counts across benchmarks (e.g., \gls{ssw} has 81 classes in BirdSet and 96 in BIRB).

\paragraph{iNatSounds Benchmark~\cite{chasmai_inaturalist_2024}} introduces a large-scale, taxonomically diverse collection of animal sound recordings, encompassing approximately 5,500 species from a wide range of geographic regions. The dataset includes vocalisations from birds, mammals, insects, reptiles, and amphibians, with audio samples and species labels derived from observations submitted to \gls{ina}~\cite{noauthor_inaturalist_nodate}. Each recording is annotated with a single species, regardless of potential background sounds or overlapping vocalisations, resulting in a weakly labelled dataset. Nevertheless, Chasmai et al.~\cite{chasmai_inaturalist_2024} demonstrated that its scale and diversity make it a valuable resource for pretraining bioacoustic models—especially when used in combination with downstream datasets containing strong, time-stamped annotations. Despite its promise, the dataset presents several limitations: geographic representation is biased towards accessible regions such as North America and Europe, and the absence of precise temporal labels complicates certain modelling tasks.

\paragraph{AnuraSet~\cite{canas_dataset_2023}} presents a large-scale, multi-species dataset of anuran amphibian calls, comprising 27 hours of expert, human-generated annotations for 42 different species from 12 genera and 5 families, across two Neotropical Brazilian biomes. Given the complexity of tropical acoustic environments and the scarcity of manually annotated datasets, AnuraSet can accelerate the development of robust \gls{ml} models for wildlife monitoring in biodiversity hotspots. The dataset frames the species identification problem as a multi-label classification task, considering the common occurrence of call overlap in \gls{pam}.

\paragraph{General datasets} In addition to domain-specific datasets, several general-purpose audio datasets have been widely used to evaluate audio classification models across diverse tasks. While \gls{as}~\cite{audioset} was introduced in Section~\ref{sec:representationlearning}, it is worth noting that it is commonly used in two distinct forms: the full dataset (\textit{AS-2M}), which includes over 2 million clips with an imbalanced class distribution, and a smaller balanced subset (\textit{AS-20K}) comprising around 22,000 samples. The latter is often employed in settings that require uniform class representation. Both training subsets provide the same test set with around 20,000 samples. \textit{\gls{esc}}~\cite{piczak_esc_2015} is another widely used dataset in this domain. It contains 2,000 short audio clips evenly distributed across 50 sound event categories, including animal vocalisations, natural sounds, human activities, and domestic environments. Despite its limited scale, ESC-50 serves as a standard test bed for small-scale audio classification due to its well-structured design and high-quality annotations.

\section{Review of Bioacoustic Models}
\label{sec:models}

In this section, we review large-scale bioacoustic species classification models. We conducted a keyword-based literature search on the OpenAlex database~\cite{priem2022openalexfullyopenindexscholarly} using the following search query:

\begin{verbatim}
    ((bioacoustic* OR "animal vocal*"  OR "xeno-canto" OR "xeno canto" OR inaturalist
            OR "macaulay library"  OR watkins OR "animal sound archive")
        AND
        ("foundation model" OR "deep learning" OR "self-supervised learning"
            OR pretraining OR "deep neural network*"))
    OR ("birdset" OR "inatsounds" OR "InsectSet459")
\end{verbatim}

We selected models that were trained on large-scale bioacoustic datasets (as described in Section \ref{sec:pretrainingdatasets}) and therefore could serve as foundation models for transfer learning applications. In addition, references, including citations, from the selected papers were included. The models covered in this review are: Animal2Vec~\cite{schaferzimmermann2024animal2vec},
AudioMAE~\cite{huang_masked_2022},
AVES~\cite{hagiwara_aves_2023},
BEATs~\cite{chen_beats_2022},
BioLingual~\cite{robinson_transferable_2024}, BirdAVES~\cite{birdaves}, BirdMAE~\cite{rauch2025maskedautoencoderslistenbirds}, BirdNET~\cite{kahl_birdnet_2021}, ConvNext$_{BS}$~\cite{rauch2025birdset},
EAT~\cite{chen_eat_2024},
NatureLM-audio~\cite{robinson2025naturelmaudio}, Perch~\cite{hamer_birb_2023}, Perch~2.0~\cite{vanmerrienboer_perch_2025}, ProtoCLR~\cite{moummad2024dirlbs}, SurfPerch~\cite{williams2024leveragingtropicalreefbird}, and ViT$_{INS}$~\cite{chasmai_inaturalist_2024}. \\
In the following, we summarise the key design decisions of these models, categorised into training data, preprocessing steps, augmentations, architectures, and training paradigms. See Table~\ref{tab:modeldetails} for a quick overview.

\subsection{Training data}

\begin{table}[t]
  \caption{Analysis of pretraining and evaluation datasets per model. The pretrain row indicates if a model has been pretrained on the specific dataset, the eval row indicates if the model has been evaluated on the dataset. The performance is reported from the original works using the standard metric for each dataset: $\text{cmAP}$ for multi-label BS, and $Acc$ for multi-class ESC, INS, MKT, BEANS, and SC2. \textsuperscript{1} Trained on eval datasets. \textsuperscript{2} Results reported in~\cite{rauch2025birdset}. \textsuperscript{3} Zero Shot evaluation.}
  \centering

  \setlength{\tabcolsep}{0.25em} %
  \resizebox{\textwidth}{!}{ %
    \begin{tabular}
  {
    p{3.2cm}
    p{1.0cm}
    p{0.8cm}
    p{0.8cm}
    p{0.8cm}
    p{0.8cm}
    p{0.8cm}
    p{0.8cm}
    p{0.8cm}
    p{0.8cm}
    p{0.8cm}
    p{0.8cm}
    p{0.8cm}
    p{0.8cm}
    p{0.8cm}
    p{0.8cm}
    p{0.8cm}
  }
  \toprule

  \multirow{2}{*}{\textbf{Model}}                                                   & \multirow{1}{*}{\textbf{Usage}} & \multicolumn{5}{c}{\textbf{General}} & \multicolumn{10}{c}{\textbf{Bioacoustic}}                                                                                                                                                                                                                                                                          \\
  \cmidrule(lr){3-7} \cmidrule(lr){8-17}
  &                                 & {\scriptsize \textbf{AS-2M}}         & \textbf{AC}                               & {\scriptsize \textbf{VGGS}} & \textbf{FSD} & \textbf{ESC} & \textbf{MAC} & \textbf{XC} & \textbf{BS}                             & \textbf{INA} & \textbf{INS}        & \textbf{MKT}        & \textbf{ASA} & \textbf{WMM} & \textbf{RS} & {\scriptsize \textbf{BEANS}} \\
  \midrule
  \multicolumn{4}{p{5cm}}{\textit{Pure bioacoustic models}} \vspace{0.5mm}                                                                                                                                                                                                                                                                                                                                                                                                        \\
  \multirow{2}{*}{\textbf{Animal2Vec} \cite{schaferzimmermann2024animal2vec}}       & pretrain                        & \no                                  & \no                                       & \no                         & \no          & \no          & \no          & \no         & \no                                     & \no          & \no                 & \yes                & \no          & \no          & \no         & \no                          \\ \cline{2-17}
  & eval                            & \no                                  & \no                                       & \no                         & \no          & \no          & \no          & \no         & \no                                     & \no          & \no                 & \yes{\textbf{0.91}} & \no          & \no          & \no         & \no                          \\ \hline
  \multirow{2}{*}{\textbf{BirdMAE} \cite{rauch2025maskedautoencoderslistenbirds}}   & pretrain                        & \no                                  & \no                                       & \no                         & \no          & \no          & \no          & \yes        & \yes                                    & \no          & \no                 & \no                 & \no          & \no          & \no         & \no                          \\ \cline{2-17}
  & eval                            & \no                                  & \no                                       & \no                         & \no          & \no   & \no          & \no         & \yes{44.0}                     & \no          & \no                 & \no                 & \no          & \no          & \no         & \no                          \\ \hline
  \multirow{2}{*}{\textbf{BirdNET v2.4} \cite{kahl_birdnet_2021}}                   & pretrain                        & \no                                  & \no                                       & \no                         & \no          & \no          & \yes         & \yes        & \yes{\textsuperscript{1}}               & \yes         & \no                 & \no                 & \no          & \yes         & \no         & \no                          \\ \cline{2-17}
  & eval                            & \no                                  & \no                                       & \no                         & \no          & \no          & \no          & \no         & \no                                     & \no          & \no                 & \no                 & \no          & \no          & \no         & \no                          \\ \hline
  \multirow{2}{*}{\textbf{ConvNext$_{BS}$} \cite{rauch2025birdset}}                 & pretrain                        & \no                                  & \no                                       & \no                         & \no          & \no          & \no          & \no         & \yes                                    & \no          & \no                 & \no                 & \no          & \no          & \no         & \no                          \\ \cline{2-17}
  & eval                            & \no                                  & \no                                       & \no                         & \no          & \no          & \no          & \no         & \yes{\underline{36}}                    & \no          & \no                 & \no                 & \no          & \no          & \no         & \no                          \\ \hline
  \multirow{2}{*}{\textbf{Perch} \cite{hamer_birb_2023}}                            & pretrain                        & \no                                  & \no                                       & \no                         & \no          & \no          & \no          & \yes        & \no                                     & \no          & \no                 & \no                 & \no          & \no          & \no         & \no                          \\ \cline{2-17}
  & eval                            & \no                                  & \no                                       & \no                         & \no          & \no          & \no          & \no         & \yes{36\textsuperscript{2}} & \no          & \no                 & \no                 & \no          & \no          & \no         & \no                          \\ \hline
  \multirow{2}{*}{\textbf{ProtoCLR} \cite{moummad2024dirlbs}}                       & pretrain                        & \no                                  & \no                                       & \no                         & \no          & \no          & \no          & \yes        & \no                                     & \no          & \no                 & \no                 & \no          & \no          & \no         & \no                          \\ \cline{2-17}
  & eval                            & \no                                  & \no                                       & \no                         & \no          & \no          & \no          & \no         & \no                                     & \no          & \no                 & \no                 & \no          & \no          & \no         & \no                          \\ \hline
  \multirow{2}{*}{\textbf{ViT$_{INS}$} \cite{chasmai_inaturalist_2024}}             & pretrain                        & \no                                  & \no                                       & \no                         & \no          & \no          & \no          & \no         & \no                                     & \yes         & \yes                & \no                 & \no          & \no          & \no         & \no                          \\ \cline{2-17}
  & eval                            & \no                                  & \no                                       & \no                         & \no          & \no          & \no          & \no         & \no                                     & \no          & \yes{\textbf{60.3}} & \no                 & \no          & \no          & \no         & \no                          \\
  \midrule
  \multicolumn{4}{p{5cm}}{\textit{Mixed-source models}} \vspace{0.5mm}                                                                                                                                                                                                                                                                                                                                                                                                            \\
  \multirow{2}{*}{\textbf{AVES} \cite{hagiwara_aves_2023}}                          & pretrain                        & \yes                                 & \no                                       & \yes                        & \yes         & \no          & \no          & \no         & \no                                     & \no          & \no                 & \no                 & \no          & \no          & \no         & \no                          \\ \cline{2-17}
  & eval                            & \no                                  & \no                                       & \no                         & \no          & \yes{77.3}   & \no          & \no         & \no                                     & \no          & \no                 & \no                 & \no          & \no          & \no         & \yes{52.8}                   \\ \hline
  \multirow{2}{*}{\textbf{BirdAVES} \cite{birdaves}}                                & pretrain                        & \yes                                 & \no                                       & \yes                        & \yes         & \no          & \no          & \yes        & \no                                     & \yes         & \no                 & \no                 & \no          & \no          & \no         & \no                          \\ \cline{2-17}
  & eval                            & \no                                  & \no                                       & \no                         & \no          & \no          & \no          & \no         & \no                                     & \no          & \no                 & \no                 & \no          & \no          & \no         & \yes{55.1}                   \\ \hline
  \multirow{2}{*}{\textbf{BioLingual} \cite{robinson_transferable_2024}}            & pretrain                        & \yes                                 & \yes                                      & \no                         & \no          & \no          & \no          & \yes        & \no                                     & \yes         & \no                 & \no                 & \yes         & \yes         & \no         & \no                          \\ \cline{2-17}
  & eval                            & \no                                  & \no                                       & \no                         & \no          & \no          & \no          & \no         & \no                                     & \no          & \no                 & \no                 & \no          & \no          & \no         & \yes{\underline{83.8}}          \\ \hline
  \multirow{2}{*}{\textbf{NatureLM-audio} \cite{robinson2025naturelmaudio}}         & pretrain                        & \yes                                 & \yes                                      & \no                         & \no          & \no          & \no          & \yes        & \yes{\textsuperscript{1}}               & \yes         & \no                 & \yes                & \yes         & \yes         & \no         & \no                          \\ \cline{2-17}
  & eval                            & \no                                  & \no                                       & \no                         & \no          & \no          & \no          & \no         & \yes{\textsuperscript{3}}               & \no          & \no                 & \no                 & \no          & \no          & \no         & \yes{\textsuperscript{3}}    \\ \hline
  \multirow{2}{*}{\textbf{Perch 2.0} \cite{vanmerrienboer_perch_2025}}              & pretrain                        & \no                                  & \no                                       & \no                         & \yes         & \no          & \no          & \yes        & \no                                     & \yes         & \no                 & \no                 & \yes         & \no          & \no         & \no                          \\ \cline{2-17}
  & eval                            & \no                                  & \no                                       & \no                         & \no          & \no          & \no          & \no         & \yes{\underline{43.1}}                    & \no          & \no                 & \no                 & \no          & \no          & \no         & \yes{\textbf{84.0}}                   \\ \hline
  \multirow{2}{*}{\textbf{SurfPerch} \cite{williams2024leveragingtropicalreefbird}} & pretrain                        & \no                                  & \no                                       & \no                         & \yes         & \no          & \no          & \yes        & \no                                     & \no          & \no                 & \no                 & \no          & \no          & \yes        & \no                          \\ \cline{2-17}
  & eval                            & \no                                  & \no                                       & \no                         & \no          & \no          & \no          & \no         & \no                                     & \no          & \no                 & \no                 & \no          & \no          & \yes        & \no                          \\ \hline
  \bottomrule
\end{tabular}

  }
  \label{tab:model-datasets}
\end{table}

The choice of training data is a key factor in model development. We summarise the key data sources of each model in Table~\ref{tab:model-datasets}.
The models can be broadly categorised into two groups: pure bioacoustic models and mixed-source models.

\paragraph{Pure bioacoustic models} are trained exclusively on bioacoustic datasets, each on a single data source. \gls{xc} is the most commonly used data source, exclusively used for training BirdMAE, ConvNext$_{BS}$, Perch, and ProtoCLR. The dedicated BirdSet training split is used to train BirdMAE and ConvNext$_{BS}$, while ProtoCLR uses the BIRB train subset and Perch uses a custom one. Animal2Vec is trained exclusively on the \gls{mkt} dataset, whereas ViT$_{INS}$ is trained on the INS dataset. BirdNET v2.4 utilises a custom \gls{xc} training split as well as \gls{mac}, the soundscape evaluation subsets from BirdSet, and project-internal data.

\paragraph{Mixed-source models} exploit a wider range of datasets to improve model generalisation. AVES is trained on bioacoustic portions of the general audio datasets \gls{as}, \gls{vggs} and \gls{fsd}, while BirdAVES also includes avian sounds from \gls{xc}. SurfPerch extends the \gls{xc} training data of Perch with data from \gls{fsd} and, most importantly, aquatic soundscapes from \gls{rs}. Perch~2.0 builds on the bioacoustic datasets \gls{xc}, \gls{ina} and \gls{asa} and also incorporates general audio from \gls{fsd}. BioLingual curates the custom text-audio pair dataset AnimalSpeak for training. This collection includes data from \gls{as}, \gls{ac}, \gls{xc}, \gls{ina}, \gls{asa}, and \gls{wmm}. The text labels are derived from the metadata of the audio files, providing a rich source of information for training. NatureLM-audio uses a further diversified set of datasets, including \gls{ac}, \gls{xc}, \gls{ina}, \gls{wmm}, \gls{asa}, as well as music and speech datasets. The corresponding metadata includes \gls{llm}-generated text labels, derived from existing audio metadata and used to construct additional training data via mixing.

\subsection{Preprocessing}

\begin{table}[t]
  \centering
  \caption{Overview of bioacoustic and baseline general audio models and their characteristics. For each model, we indicate the year of release, the number of classes the model is trained to classify, training method—supervised learning (SL) or self‐supervised learning (SSL)—as well as the architecture, number of parameters, embedding size, input duration (in seconds), input type, sample rate (in kHz) and used augmentations during pretraining.}
  \label{tab:modeldetails}
  \setlength{\tabcolsep}{0.25em}
  \begin{adjustbox}{max width=\textwidth}
    \setlength{\extrarowheight}{0.5em}
\begin{tabular}{lp{3.5cm} l c c c c c c c c c }
    \toprule
    \rule{0pt}{2em} %

     & \multicolumn{1}{c}{ \textbf{Model}}
     & \multicolumn{1}{c}{ \textbf{Year}}
     & \multicolumn{1}{c}{\textbf{Classes}}
     & \multicolumn{1}{c}{\begin{tabular}[c]{@{}c@{}} \textbf{Training} \\ \textbf{Method} \end{tabular}}
     & \multicolumn{1}{c}{\textbf{Architecture}}
     & \multicolumn{1}{c}{\begin{tabular}[c]{@{}c@{}} \textbf{Parameters} \\ \textbf{\scriptsize(M)} \end{tabular}}
     & \multicolumn{1}{c}{\begin{tabular}[c]{@{}c@{}} \textbf{Embedding} \\ \textbf{Size} \end{tabular}}
     & \multicolumn{1}{c}{\begin{tabular}[c]{@{}c@{}} \textbf{Input} \\ \textbf{Duration \scriptsize(s)} \end{tabular}}
     & \multicolumn{1}{c}{\begin{tabular}[c]{@{}c@{}} \textbf{Input} \\ \textbf{Type} \end{tabular}}
     & \multicolumn{1}{c}{\begin{tabular}[c]{@{}c@{}} \textbf{Sample} \\ \textbf{Rate \scriptsize(kHz)} \end{tabular}}
     & \multicolumn{1}{c}{\textbf{Augmentations}}
    \\
    \midrule
     & \multicolumn{4}{p{5cm}}{\textit{General audio models}} \vspace{0.5mm}                                                                                                                                                                                                                                                                         \\
     & \textbf{AudioMAE}~\cite{huang_masked_2022}                                                                       & \scriptsize2022 & -           & SSL & ViT-B                    & 86    & 768  & 10       & Spectrogram & 16    & \pinkbubble{Masking}                                                                                      \\ \cmidrule(lr){2-12}
     & \textbf{BEATs}~\cite{chen_beats_2022}                                                                            & \scriptsize2022 & -           & SSL & ViT-B                    & 90    & 768  & 10       & Spectrogram & 16    & \pinkbubble{Masking}, \graybubble{Mixup}, \bluebubble{SpecAug}, \purplebubble{Roll}                       \\ \cmidrule(lr){2-12}
     & \textbf{EAT}~\cite{chen_eat_2024}                                                                                & \scriptsize2024 & -           & SSL & CNN + Transformer        & 88    & 768  & 10       & Spectrogram & 16    & \pinkbubble{Masking}, \graybubble{Mixup}, \bluebubble{SpecAug}, \purplebubble{Roll}, \redbubble{Droppath} \\ \midrule
     & \multicolumn{5}{p{5cm}}{\textit{Bioacoustics foundation models}} \vspace{0.5mm}                                                                                                                                                                                                                                                               \\
     & \textbf{Animal2Vec}~\cite{schaferzimmermann2024animal2vec}                                                       & \scriptsize2024 & -           & SSL & SincNet + Transformer    & 315   & 768  & 10       & Waveform    & 8     & \graybubble{Mixup}                                                                                        \\ \cmidrule(lr){2-12}
     & \textbf{AVES}~\cite{hagiwara_aves_2023}                                                                          & \scriptsize2023 & -           & SSL & CNN + Transformer        & 95    & 768  & variable & Waveform    & 16    & -                                                                                                         \\ \cmidrule(lr){2-12}
     & \textbf{BioLingual}~\cite{robinson_transferable_2024}                                                            & \scriptsize2024 & -           & SSL & HTS-AT + RoBERTa         & 190.8 & 1024 & 10       & Spectrogram & 48    & -                                                                                                         \\ \cmidrule(lr){2-12}
     & \textbf{BirdAVES}~\cite{birdaves}                                                                                & \scriptsize2024 & -           & SSL & CNN + Transformer        & 316   & 768  & variable & Waveform    & 16    & -                                                                                                         \\ \cmidrule(lr){2-12}
     & \textbf{BirdMAE}~\cite{rauch2025maskedautoencoderslistenbirds}                                                   & \scriptsize2025 & -           & SSL & ViT-L                    & 300   & 1024 & 5        & Spectrogram & 32    & \pinkbubble{Masking}                                                                                      \\ \cmidrule(lr){2-12}
     & \textbf{BirdNET} v2.4~\cite{kahl_birdnet_2021}                                                                   & \scriptsize2023 & 6,522       & SL  & EfficientNetB0-like      & 5.3   & 1024 & 3        & Spectrogram & 48    & -                                                                                                         \\ \cmidrule(lr){2-12}
     & \textbf{ConvNext$_{BS}$}~\cite{rauch2025birdset}                                                                 & \scriptsize2025 & 9,734       & SL  & ConvNext                 & 88    & 768  & 5        & Spectrogram & 32    & \pinkbubble{Masking}, \graybubble{Mixup}, \bluebubble{SpecAug}, \brownbubble{Gain}                        \\ \cmidrule(lr){2-12}
     & \textbf{NatureLM-audio}~\cite{robinson2025naturelmaudio}                                                         & \scriptsize2024 & -           & SSL & BEATs + U-Former + LLaMA & 665   & 768  & 10       & Spectrogram & 16    & \graybubble{Mixup}, \yellowbubble{Scale}                                                                  \\ \cmidrule(lr){2-12}
     & \textbf{Perch}~\cite{hamer_birb_2023}                                                                     & \scriptsize2023 & 10,932      & SL  & EfficientNetB1           & 8     & 1280 & 5        & Spectrogram & 32    & \graybubble{Mixup}, \brownbubble{Gain}, \greenbubble{Shift}, \orangebubble{Low-pass}                                                                                    \\ \cmidrule(lr){2-12}
     & \textbf{Perch 2.0}~\cite{vanmerrienboer_perch_2025}                                                              & \scriptsize2025 & 14,795      & SL  & EfficientNetB3          & 12    & 1536 & 5        & Spectrogram & 32    & \graybubble{Mixup}                                                                                        \\ \cmidrule(lr){2-12}
     & \textbf{ProtoCLR}~\cite{moummad2024dirlbs}                                                                       & \scriptsize2024 & -           & SSL & CvT-13                   & 20    & 384  & 6        & Spectrogram & 16    & \greenbubble{Shift}, \bluebubble{SpecAug}, \graybubble{Mixup}                                             \\ \cmidrule(lr){2-12}
     & \textbf{SurfPerch}~\cite{williams2024leveragingtropicalreefbird}                                                 & \scriptsize2024 & 10,932 + 38 & SL  & EfficientNetB1           & 8     & 1280 & 5        & Spectrogram & 32    & \graybubble{Mixup}, \brownbubble{Gain}                                                                    \\ \cmidrule(lr){2-12}
     & \textbf{ViT$_{INS}$}~\cite{chasmai_inaturalist_2024}                                                             & \scriptsize2024 & 5,569       & SL  & ViT-B                    & 87    & 768  & 3        & Spectrogram & 22.05 & \pinkbubble{Masking}, \graybubble{Mixup}, \bluebubble{SpecAug}                                            \\
    \bottomrule
\end{tabular}%

  \end{adjustbox}
\end{table}

Preprocessing pipelines vary significantly across bioacoustic foundation models, reflecting diverse architectural requirements, input modalities, and domain-specific adaptations to handle the unique challenges of animal vocalisations.

\paragraph{Resampling and input standardisation} The models exhibit substantial variation in sampling rate requirements, ranging from 8 kHz to 48 kHz. The sample rate is selected based on the frequency range of relevant biological signals, according to the Nyquist theorem, which states that the highest frequencies retained in an audio signal are half of the sample rate. Animal2Vec operates at the lowest sampling rate of 8 kHz optimised for meerkat vocalisation events, while BirdNET and BioLingual use the highest rate of 48 kHz to preserve high frequency components of bird vocalisations. The other models either standardise at 16 kHz (AVES, BirdAVES, NatureLM-audio, ProtoCLR) or 32 kHz (BirdMAE, ConvNext$_{BS}$, Perch, Perch~2.0, SurfPerch) with ViT$_{INS}$ using 22.05 kHz.

\paragraph{Fixed-length segmentation and temporal windowing} All models implement fixed-length input processing, which facilitates batch training. Varying temporal windows are used: BirdNET and ViT$_{INS}$ use 3-second segments, ProtoCLR doubles this to 6-second segments. Animal2Vec, BioLingual, and NatureLM-audio process 10-second chunks originating from AudioSet's clip length, while other models standardise on 5-second windows. When using bioacoustic data sources (e.g., \gls{xc}, \gls{ina}) with variable length audio recordings, it is important to select the segments with meaningful vocalisations. BirdNET uses a signal strength detector, and Perch uses a peak-finding algorithm for this purpose. In contrast, Perch~2.0 reverts to random window selection, which performs on par. BirdMAE and ConvNext$_{BS}$ use the BirdSet \gls{xc} training data selection that provides a list of detected events per file, originating from the bambird detector~\cite{michaud2023}. NatureLM-audio and ViT$_{INS}$ stride with half of their window length over the recordings.

\begin{figure}[t]
  \centering
  \includegraphics[width=0.9\textwidth]{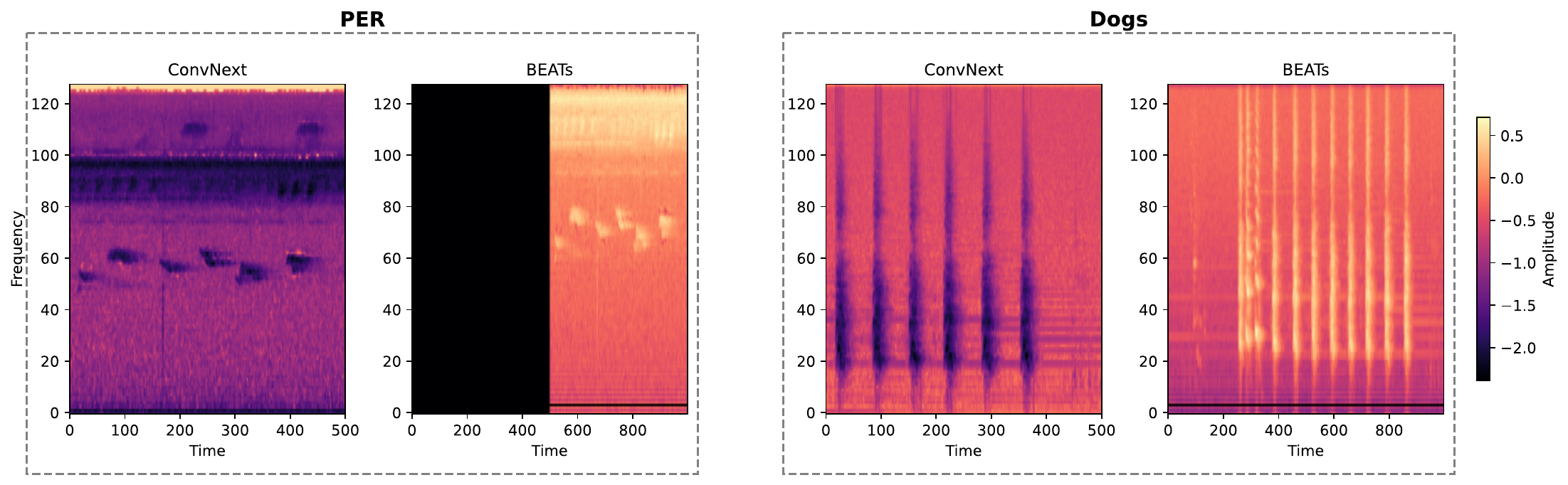}
  \caption{A sample from the PER dataset of BirdSet~\cite{rauch2025birdset} and from the Dogs of BEANS~\cite{hagiwara_beans_2023} preprocessed according to the preprocessing pipelines of ConvNext$_{BS}$~\cite{rauch2025birdset} and BEATs~\cite{chen_beats_2022}. The sample is displayed as a mel-spectrogram with the time dimension at the x-axis and the frequency dimension at the y-axis.}
  \label{fig:spectrogram}
\end{figure}

\paragraph{Spectrogram-based preprocessing} Most models (BirdMAE, BirdNET, BioLingual, ConvNext$_{BS}$, NatureLM-audio, ProtoCLR, Perch models, ViT$_{INS}$) convert raw audio to time-frequency representations using mel-scale spectrograms. These models employ \gls{stft} with diverse technical configurations tailored to bioacoustic signal characteristics. BirdNET v2.4 implements a dual mel-spectrogram approach optimised for bird vocalisations: the first spectrogram covers low frequencies (0-3 kHz) using n\_fft=2048, hop\_length=278, and 96 mel bins to capture fundamental frequencies and harmonic structure, while the second spectrogram targets higher frequencies (0.5-15 kHz) using n\_fft=1024, hop\_length=280, and 96 mel bins to preserve fine temporal details in bird calls. BirdMAE and ConvNext$_{BS}$ utilise n\_fft=1024 with hop\_length=320 samples, generating 128 mel bands covering 0-16 kHz at 32 kHz sampling rate for enhanced temporal resolution. Perch employs n\_fft=2048, hop\_length=512 configuration with 96 mel bands spanning 0-11.025 kHz, incorporating Per-Channel Energy Normalisation (PCEN) for robust feature extraction across varying recording conditions. Perch~2.0 uses n\_fft=1024 with hop\_length=320 and a 640 sample Hann window, producing 128 mel bands with log-magnitude scaling. ProtoCLR uses n\_fft=1024, hop\_length=320 with 128 mel. Most models apply logarithmic scaling to the output of the \gls{stft} to compress the dynamic range and MEL transformation to the frequency axis to emphasise perceptually relevant spectral features. Figure~\ref{fig:spectrogram} illustrates the spectrogram preprocessing of ConvNext$_{BS}$ and BEATs.

Animal2Vec, AVES, and BirdAVES process raw audio directly rather than converting the signal to a spectrogram. We will go into more detail when discussing the architecture required for processing the large quantity of raw data.

\paragraph{Normalisation} The majority of models (BioLingual, BirdMAE, ConvNext$_{BS}$, NatureLM-audio, ProtoCLR) employ standardisation, normalising spectrograms to zero mean and unit variance. BirdNET uses min-max normalisation to scale spectrograms to a fixed range ($[-1,1]$). Perch and SurfPerch implement Per-Channel Energy Normalisation (PCEN), a robust normalisation technique specifically designed for audio processing that provides adaptive gain control and noise suppression, making it particularly effective for handling varying recording conditions in bioacoustic data. Perch~2.0 omits PCEN and applies only log-magnitude scaling (log with floor $10^{-5}$, then scale by 0.1) to the mel-spectrogram. ViT$_{INS}$ applies rescaling to map spectrogram values to the range [0,255], following computer vision conventions. For raw waveform processing models, Animal2Vec employs instance-wise standardisation, while AVES and BirdAVES do not specify explicit normalisation steps, relying on the inherent normalisation properties of their transformer-based architectures.

\subsection{Augmentations}

Data augmentation is critical for improving model robustness and generalisation across bioacoustic foundation models. We detail augmentations in the pretraining stage and categorise into waveform-level and spectrogram-level techniques.

\paragraph{Waveform-level augmentations} Several models apply augmentations directly to raw audio signals before spectrogram conversion. \textit{Mixup}~\cite{zhang2017mixup} is extensively used across models (Animal2Vec, BirdMAE, ConvNext$_{BS}$, NatureLM-audio, Perch, Perch~2.0, SurfPerch, ViT$_{INS}$) to combine multiple audio samples including the corresponding label information, creating synthetic training examples that improve generalisation. Perch~2.0 generalises mixup to more than two components: the number of components is sampled from a beta-binomial distribution and weights from a symmetric Dirichlet distribution, with multi-hot targets reflecting all vocalisations in the mixed window. NatureLM-audio and Perch~\cite{hamer_birb_2023} employ \textit{noise mixing} at random \gls{snr} levels (Perch uses diverse noise sources including DCASE, BBC Nature Sound Effects, and Common Voice at \gls{snr} 0--40\,dB); NatureLM-audio additionally uses \textit{time scaling} to capture temporal variations and \textit{silence insertion} to model natural gaps in vocalisations. Animal2Vec introduces \textit{between-classes-learning (BCL)} augmentation with A-weighted stochastic mixing, which combines samples from different classes to improve inter-class discrimination. Furthermore, ConvNext$_{BS}$, Perch, and ProtoCLR adjust the \textit{gain} of the audio signal to simulate varying recording conditions, which is particularly important in bioacoustic applications where environmental noise can significantly impact model performance. Perch additionally applies \textit{random time shift} by sampling a random 5-second window from each 6-second training example.

\paragraph{Spectrogram-level augmentations} Most models apply augmentations to time-frequency representations. \textit{SpecAugment}~\cite{park2019specaugment} is adopted across models (ProtoCLR, ViT$_{INS}$, ConvNext$_{BS}$), applying \textit{frequency masking} and \textit{time masking} to simulate missing spectral content and temporal gaps, effectively simulating real-world recording artefacts and missing data. Perch applies a \textit{random low-pass filter} by scaling frequency bands in the mel-spectrogram to simulate the low-pass effect of increased distance from the sound source. Some models, such as ProtoCLR, incorporate additional \textit{temporal shift} augmentations and domain-specific transformations to simulate natural acoustic variability in different recording environments.

\subsection{Model architecture}
The surveyed models employ a diverse range of neural architectures. An overview of the selected models is provided in Table~\ref{tab:modeldetails}. We broadly differentiate between five categories:

\paragraph{Convolutional Neural Networks (CNNs)}
BirdNET, ConvNext$_{BS}$, Perch, Perch~2.0 and SurfPerch utilise CNN-based architectures, leveraging convolutional layers to extract local features from spectrogram inputs. BirdNET v2.4 employs an EfficientNetB0-like~\cite{efficientnet} backbone architecture, which has approximately 5.3 million parameters and a final embedding size of 1,024. Perch implements an EfficientNet-B1 architecture with a backbone of approximately 8 million parameters, while the complete model is much bigger ($\approx$ 80 million) due to the multiple classification heads for taxonomic classification. Perch~2.0 uses an EfficientNet-B3~\cite{efficientnet} backbone with approximately 12 million parameters and a 1,536-dimensional embedding, reflecting the increased scale of its multi-taxa training data. SurfPerch adopts the same EfficientNet-B1 foundation as Perch. ConvNext$_{BS}$ utilises the ConvNext-Base~\cite{convnext} architecture with approximately 88 million parameters, featuring hierarchical feature extraction through downsampling residual blocks, depthwise convolutions, and global log-mean-exponential pooling for robust multi-label classification capabilities.

\paragraph{Transformer-based models}
In bioacoustics, transformer-based architectures have gained prominence for their ability to model long-range dependencies and capture complex temporal patterns in audio data~\cite{birdaves, rauch2025maskedautoencoderslistenbirds}.

\textit{\textbf{Vision Transformers (ViT):}}
BirdMAE and ViT$_{INS}$ utilise \gls{vit}~\cite{vit} architectures. BirdMAE uses an encoder-decoder architecture also based on the ViT architecture in the variants Base, Large and Huge. The Large variant, with 300 million parameters, achieves the best performance. ViT$_{INS}$ adapts a smaller ViT-Base architecture, which has 86 million parameters. The models use an embedding size of 1024 and 768, respectively.

\textit{\textbf{Feature Extractor + Transformer Encoder:}}
Animal2Vec, AVES, BirdAVES and ProtoCLR employ hybrid architectures that combine a parameterised feature extractor layer with a transformer encoder. Animal2Vec utilises SincNet-style~\cite{ravanelli2019speakerrecognitionrawwaveform} filterbanks to process raw waveforms, followed by a transformer encoder, totalling 315 million parameters. AVES and BirdAVES adapt the HuBERT~\cite{hsu_hubert_2021} architecture for bioacoustics, featuring a CNN token extractor followed by a transformer with a total of 95 and 316 million parameters, respectively. Both the AVES models and Animal2Vec use an embedding size of 768.
ProtoCLR employs a Convolutional Vision Transformer (CvT)~\cite{ctv} architecture with 20 million parameters, where the CvT-13 backbone integrates convolutional operations within transformer blocks to extract both local and global features from spectrograms. It comprises 13 transformer blocks that incorporate convolutional projections and convolutional feed-forward networks, enabling efficient processing of visual features with a final embedding size of 384.

\paragraph{Audio-Language Models}
BioLingual and NatureLM-audio represent the first generation of audio-language foundation models designed explicitly for bioacoustics, employing multimodal architectures that combine audio encoders with language models to enable cross-modal understanding and generation.
BioLingual combines an HTS-AT~\cite{hts-at} audio encoder with a RoBERTa~\cite{roberta} text encoder. The HTS-AT component processes mel-spectrograms through hierarchical token-semantic audio transformers, while RoBERTa handles text captions. Both encoders are connected through a \gls{mlp} layer that projects embeddings into a shared 1,024-dimensional space, totalling 190 million parameters.
NatureLM-audio adopts a generative audio-language architecture that combines a BEATs~\cite{chen_beats_2022} audio encoder with a Llama 3.1-8B~\cite{llama3} \gls{llm}. The BEATs encoder (90 million parameters) processes audio inputs and produces window-level embeddings, which are then processed by a Q-Former~\cite{blip2} adapter to convert audio representations into text-compatible tokens. The Q-Former applies learnable queries to audio embeddings, enabling flexible audio-to-text alignment. The Llama 3.1-8B model is fine-tuned using \gls{lora}~\cite{lora} on all attention layers while keeping the base model parameters frozen. This architecture enables the model to process audio inputs alongside text instructions and generate natural language responses for tasks such as species classification, detection, and audio captioning. In total, this model features 665 million trained parameters, keeping the original 8 billion parameters of the \gls{llm} frozen.

\subsection{Training paradigm}

The pretraining paradigms employed by bioacoustic foundation models can be broadly categorised into \gls{sl} and \gls{ssl} approaches:

\paragraph{Supervised learning}
\gls{sl} models rely on labelled datasets where each audio sample is associated with explicit annotations such as species identity, call type, or behavioural context. BirdNET, ConvNext$_{BS}$, and ViT$_{INS}$ predict class labels directly from spectrograms, employing binary cross-entropy loss for multi-label classification tasks, covering 6{,}522, 9{,}734, and 5{,}569 classes, respectively.
Perch and SurfPerch extend this approach to hierarchical classification, predicting not only species but also family and order labels using a hierarchical binary cross-entropy loss function. This multi-level taxonomy structure captures the hierarchical relationships between species, families, and orders, enhancing classification accuracy in complex bioacoustic datasets.
Perch~2.0 trains on supervised species and sound-class classification (14{,}795 classes) and augments this with self-distillation: a prototype-learning classifier~\cite{chen_protopnet_2019} acts as teacher, with stop-gradient so its predictions provide soft targets for the main linear classifier. An auxiliary source-prediction loss (DIET~\cite{balestriero_diet_2023}) asks the model to predict the source recording from the embedding, encouraging representations that capture recording-level structure.

\paragraph{Self-supervised learning}
\gls{ssl} approaches leverage unlabelled audio data by designing pretext tasks that enable models to learn meaningful representations without explicit annotations. These methods address the significant challenge of annotation scarcity in bioacoustics while potentially capturing richer acoustic patterns.

\textit{\textbf{Masked Language Modeling (MLM):}} (Bird)AVES pioneered the application of HuBERT~\cite{hsu_hubert_2021}, a \gls{ssl} framework, to animal vocalisations. The model employs a masked language modelling objective where discrete acoustic units are first discovered through k-means clustering of mel-spectrogram features. During training, random portions of the input spectrogram are masked, and the model learns to predict the corresponding acoustic unit labels, effectively learning to model the distributional properties of animal vocalisations.

\textit{\textbf{Masked Autoencoding:}} BirdMAE adapts the \gls{mae}~\cite{huang_masked_2022} paradigm specifically for bird sound classification. Mel-spectrograms are divided into patches, a subset of which is masked during training. The encoder processes only visible patches, while the decoder reconstructs the complete spectrogram from the encoder's outputs and mask tokens. The approach is adapted to bird vocalisations by increasing the number of pretraining epochs and batch size, and adjusting the masking ratio to 75\% to account for the sparsity of bird calls. Furthermore, increasing the mixup ratio improves the model's robustness to background noise.

\textit{\textbf{Mean Teacher Self-Distillation:}} Animal2Vec introduces a self-supervised approach specifically designed for sparse bioacoustic data characteristics. The method employs mean teacher self-distillation~\cite{data2vec} combined with masked prediction objectives, where a teacher network generates soft targets for a student network learning to predict masked portions of input spectrograms. This approach is particularly suited for handling the temporal sparsity and irregular occurrence patterns typical of animal vocalisations in field recordings.

\textit{\textbf{Contrastive Learning:}} BioLingual demonstrates the application of contrastive language-audio pretraining to bioacoustics. The model learns joint representations of audio and text by maximising agreement between paired audio-caption embeddings while minimising agreement between unpaired combinations. Similarly, ProtoCLR employs contrastive learning within a prototypical framework, learning discriminative representations by contrasting positive and negative prototype-sample pairs.

\textit{\textbf{Audio-Language Models:}} NatureLM-audio combines audio and language modelling for bioacoustics. The model employs a next-token prediction loss to train a \gls{llm}, Q-Former, and audio encoder end-to-end. Given a prompt and an audio clip, the model's task is to predict fitting text tokens that match the text pairs in the training data. The \gls{llm} is trained exclusively using \gls{lora}, and crucially adapting the audio encoder is essential for performance. Curriculum learning~\cite{soviany_curriculum} is used to first learn perception by classifying species from focal recordings. This is followed by generalisation fine-tuning on multiple bioacoustic tasks such as detection, captioning, life-stage prediction or call-type prediction. The multimodal approach enables sophisticated zero-shot capabilities and natural language interaction with bioacoustic data.

\paragraph{Downstream task adaptation}

We briefly summarise the downstream task adaptation techniques used in the original works.
BirdNET, Perch and ViT$_{INS}$ do not follow a pretraining-fine-tuning scheme and are trained and evaluated directly for the task they are trained on.
ConvNext$_{BS}$ and Perch (2.0) are trained for bird species classification, where they can distinguish thousands of species. Both use \textit{logit restriction} to improve the performance on specific evaluation datasets with a small set of different classes. In this restriction, only the logits representing classes in the evaluation dataset are taken into account.

Following the pretraining, (Bird)AVES, BirdMAE and BioLingual are fine-tuned using supervision on the evaluation benchmarks. Whereas for (Bird)AVES and BioLingual, a simple linear classification head is added, BirdMAE uses a more sophisticated prototypical pooling layer on top of the patch embeddings, followed by a linear layer~\cite{heinrich_audioProtoPNet}. Uniquely, only BirdMAE employs domain-specific augmentations (time shift, mixup, gain adjustments, time / frequency masking) following BirdSet~\cite{rauch2025birdset} for downstream adaptation.

The audio-text inputs of BioLingual and NatureLM-audio enable prompt-based zero-shot evaluation. For BioLingual, texts with the corresponding labels are embedded alongside the audio recording, then the similarity between the texts and audio embeddings is calculated. As NatureLM-audio generates text output, any arbitrary prompt can be used for evaluation, e.g., outputting the scientific name of the species present in an audio recording.

ProtoCLR and SurfPerch are adapted in a few-shot setting, both keep the encoder frozen. ProtoCLR uses the SimpleShot approach~\cite{simpleshot}, whereas SurfPerch fits a linear layer.

\section{Comparative Analysis}

\begin{figure}
  \begin{floatrow}
    \ffigbox{
      \includegraphics[width=0.20\textwidth]{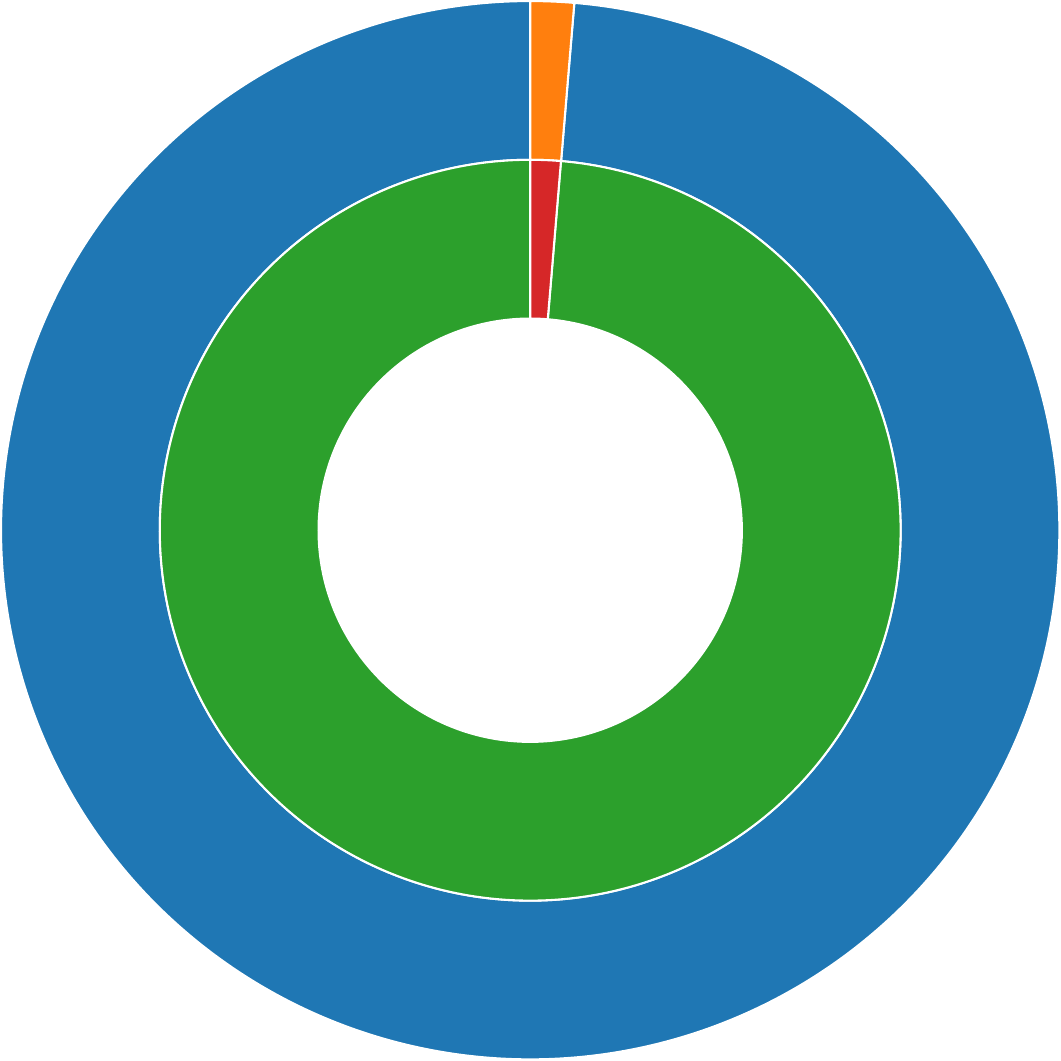}
    }{
      \caption{Comparison of the number of network parameters used for probing the BEATs model on HSN dataset (21 classes) \textcolor[HTML]{1f77b4}{\rule{1ex}{1ex}} Encoder (90.3M),
        \textcolor[HTML]{ff7f0e}{\rule{1ex}{1ex}} Trainable parameters (1.22M),
        \textcolor[HTML]{2ca02c}{\rule{1ex}{1ex}} Attentive pooling (1.2M),
      \textcolor[HTML]{d62728}{\rule{1ex}{1ex}} Linear classifier (16.1k).\label{fig:parameter-distribution}}
    }
    \capbtabbox{%
      \setlength{\tabcolsep}{0.25em} %
      \resizebox{0.45\textwidth}{!}{ %
        \begin{tabular}
    {
        p{2.5cm}
        p{1.0cm}
        p{0.8cm}
        p{0.8cm}
        p{0.8cm}
    }
    \toprule

    \multirow{2}{*}{\textbf{Model}}                             & \multirow{1}{*}{\textbf{Usage}} & \multicolumn{3}{c}{\textbf{General}}                                                          \\
    \cmidrule(lr){3-5}
                                                                &                                 & {\scriptsize \textbf{AS-2M}}         & {\scriptsize \textbf{AS-20k}} & \textbf{ESC}           \\ \hline
    \multirow{1}{*}{\textbf{AudioMAE} \cite{huang_masked_2022}} & eval                            & \yes{\underline{47.3}}               & \yes{37.0}                    & \yes{94.1}             \\ \hline
    \multirow{1}{*}{\textbf{BEATs} \cite{chen_beats_2022}}      & eval                            & \textbf{\yes{48.6}}                  & \yes{\underline{38.9}}        & \yes{\textbf{98.1}}    \\ \hline
    \multirow{1}{*}{\textbf{EAT} \cite{chen_eat_2024}}          & eval                            & \textbf{\yes{48.6}}                  & \textbf{\yes{40.2}}           & \yes{\underline{95.9}} \\
    \bottomrule
\end{tabular}

      }
    }
    {%
      \caption{Overview of reported results of general audio models trained on AudioSet. The metric $mAP$ for AS-20k, and $Acc$ for ESC is used. \label{tab:audioset-model-datasets}}%
    }
  \end{floatrow}
\end{figure}

\label{sec:comparison}

To assess how well the reviewed models transfer to downstream tasks, we present a comprehensive empirical evaluation of foundation models for bioacoustic classification, investigating which model yields the best generalisation performance when adapting to a bioacoustic classification task. We detail our experimental design and then present the results of our experiments. Finally, we discuss the implications of our findings for future research and applications in bioacoustic classification.

\subsection{Experimental design}

We will first detail the benchmark tasks we selected for our evaluation, then describe the model selection process, and finally outline the training protocol used for adaptation.

\paragraph{Selected classification benchmarks}

We evaluated models on two established bioacoustic benchmarks that cover complementary aspects of the field. \textbf{BEANS}~\cite{hagiwara_beans_2023} offers a diverse collection of classification tasks spanning multiple taxonomic groups (birds, mammals, amphibians, and insects), representing various bioacoustic challenges, including species and individual classification. \textbf{BirdSet}~\cite{rauch2025birdset} focuses specifically on bird species classification and provides weakly-labelled training data from \gls{xc} and strongly-labelled multi-label soundscape test sets from \gls{pam} scenarios. BirdSet uses a windowed evaluation protocol, analysing coherent audio recordings, which could also be framed as a detection task. Therefore, we omit the detection tasks from BEANS to keep the number of experiments tractable.

\paragraph{Model selection}

We cover all bioacoustic models described in Section~\ref{sec:models} except for BirdNET, as it is already trained on the BirdSet evaluation datasets. The generative audio-language model NatureLM-audio is not directly comparable with the other models because it is designed to be used with textual queries. We therefore extracted the audio encoder from NatureLM-audio, and used it as a feature extractor. We denote this model BEATs$_{NLM}$. Note that this model was exposed to the \gls{ssw} evaluation dataset of BirdSet during training; its strong performance on SSW and the resulting BirdSet aggregate should be interpreted with this potential data leakage in mind. Animal2Vec's \gls{xc}-pretrained model, which is more interesting for our experimental framework than the \gls{mkt} version, is not publicly available; we therefore excluded it from our evaluation. When more than one model variant is available, we chose the one with the best reported performance.
As a baseline, we also include three general audio models trained on AudioSet: AudioMAE~\cite{huang_masked_2022}, BEATs~\cite{chen_beats_2022} and EAT~\cite{chen_eat_2024}.
Table \ref{tab:audioset-model-datasets} reports their performance on \gls{as}, AS-20k and \gls{esc}. We used the checkpoints fine-tuned on AudioSet with supervised learning.

\paragraph{Training protocol}

\begin{table}[htbp]
  \centering
  \caption{Hyperparameter settings for different benchmark and probing strategies. LP denotes Linear Probing and AP denotes Attentive Probing.}
  \label{tab:hyperparameters}
  \setlength{\tabcolsep}{0.8em}
  \begin{tabular}{lcccc}
    \toprule
    \textbf{Hyperparameter} & \textbf{BEANS LP} & \textbf{BEANS AP} & \textbf{BirdSet LP} & \textbf{BirdSet AP} \\
    \midrule
    Learning Rate           & 0.01              & 0.0013            & 0.005               & 0.0013              \\
    Weight Decay            & 0.0005            & 0.007             & 0.0005              & 0.0005              \\
    Max Epochs              & 50                & 50                & 15                  & 20                  \\
    Batch Size              & 128               & 128               & 128                 & 128                 \\
    Monitor                 & Val/Acc           & Val/Acc           & Val/Loss            & Val/Loss            \\
    Patience                & 5                 & 5                 & 3                   & 5                   \\
    Min Delta               & 0.001             & 0.001             & 0.0001              & 0.001               \\
    \bottomrule
\end{tabular}

\end{table}

The goal is to adapt the pretrained models to the bioacoustic classification tasks defined by the BEANS and BirdSet benchmarks. For each task, a dedicated training set is provided. Table~\ref{tab:evaldatasets} lists the number of labels in each split. As BirdSet does not provide a fixed validation split for each individual task, we use 20\% of the train split as validation data. The table also provides information on the number of classes and the total duration of the test set. To assess the generalisability of the models, we keep the feature extractor frozen~\cite{kumar2022finetuningdistortpretrainedfeatures}. We apply two different adaptation strategies:

\textit{\textbf{Linear probing}} trains a single linear layer on top of 1D embeddings extracted from the pretrained models. The CLS-token of the transformer architectures or the global average pooling output of the CNNs is used for each audio sample. This is the most parameter-efficient adaptation technique, as for a classifier mapping embeddings of size $d$ to $C$ classes, only $C d$ parameters have to be trained.

\textit{\textbf{Attentive probing}} extends linear probing by building on the layer before the 1D embeddings. For the transformer architecture we use the patch tokens as input to a trainable multi-head attention layer. The output is then fed into a single linear layer. This enables the model to learn more complex relationships between different parts of the input, while maintaining a low number of trainable parameters. In this setting $2d^2+(C+1)d+C$ parameters have to be trained. In comparison with the frozen feature extractor this is a tiny fraction as Figure~\ref{fig:parameter-distribution} visualises. Additionally, we conducted attentive probing experiments with the \gls{cnn} model ConvNext$_{BS}$, where we use the output of the last convolutional layer as an input to the attention layer. (Surf)Perch does not offer access to the patch embeddings and we therefore could not conduct attentive probing experiments.

\textit{\textbf{Restricted.}} ConvNext$_{BS}$ and (Surf)Perch are trained to classify thousands of bird species, including those present in BirdSet evaluation tasks. We therefore add experiments of evaluating these models as-is by restricting the output logits to the classes present in the test set. For classes that are not represented by a logit, a large negative value ($-10$) is set. This represents a baseline performance of existing models without additional training. This is only the case for the (Surf)Perch models for two species in the \gls{nbp} set\footnote{eBird codes of missing species in (Surf)Perch: crelar1, easwar1}.

\textit{\textbf{Preprocessing.}} We follow the protocols outlined by BEANS~\cite{hagiwara_beans_2023} and BirdSet~\cite{rauch2025birdset} to prepare training and evaluation audio samples. Initially, the audio samples are adjusted to match the input length required by the model, either by padding or truncating. Next, the audio data is resampled to the specific sampling rate used during the model's training, ensuring compatibility with the model's parameters. Subsequently and if necessary, features are derived by transforming the raw waveform into a spectrogram, adhering to the unique preprocessing requirements of each model.

\textit{\textbf{Augmentations.}} During training of every experiment, several augmentations are applied to the audio data to enhance model robustness and generalisation. Following BirdSet's training protocol~\cite{rauch2025birdset}, we apply augmentation on the waveforms. \textit{Mixup} includes additional sounds and, in the case of BirdSet's multi-label evaluation, their corresponding labels to create augmented samples. This technique encourages the model to learn more generalised representations by exposing it to mixed audio signals and their associated multi-label annotations. \textit{Background noise} and \textit{coloured noise} augmentations simulate real-world acoustic environments, thereby improving the model's ability to handle noisy conditions. \textit{Gain} augmentation adjusts the amplitude of the audio signal, enabling the model to become invariant to variations in recording volume. For the BirdSet tasks we additionally mix in samples without any calls from the \gls{vox} dataset.

\textit{\textbf{Metric.}} We use the Area Under the Receiver Operating Characteristic (AUROC) curve as our primary evaluation metric across all experiments, as it is a threshold-free metric that is less sensitive to class count than accuracy-based metrics and does not require threshold selection~\cite{hamer_birb_2023}. AUROC measures the trade-off between true positive rate (sensitivity) and false positive rate (1 - specificity) across all classification thresholds, providing a threshold-independent assessment of model performance. For both multi-class and multi-label classification tasks, we compute the macro-averaged AUROC as:
$$\text{AUROC}_{\text{macro}} = \frac{1}{C} \sum_{c=1}^{C} \int_0^1 \text{TPR}_c(\text{FPR}_c^{-1}(t)) \, dt$$
where $C$ is the number of classes, $\text{TPR}_c$ is the true positive rate for class $c$, and $\text{FPR}_c$ is the false positive rate for class $c$, both computed in a one-vs-rest manner. This approach treats each class as an independent binary classification problem, making it suitable for both single-label tasks (BEANS) and multi-label tasks (BirdSet). AUROC values range from 0.5 (random performance) to 1.0 (perfect classification), making it particularly suitable for comparing model performance across diverse bioacoustic classification tasks with varying class distributions. Furthermore, we report the standard evaluation metrics for each benchmark in Appendix Table~\ref{tab:results-standard-metrics}: Top-1 Accuracy for BEANS and cmAP5 for BirdSet.

\textit{\textbf{Training.}} We opt for fixed hyperparameters for each model to improve comparability and tractability. Training is conducted until convergence, employing early stopping when the validation metric does not improve for a specified minimum of epochs. We use AdamW~\cite{adamw} as the optimiser. All training is performed on a single GPU, with a fixed batch size. The hyperparameter settings for each configuration are summarised in Table~\ref{tab:hyperparameters}. Full details are available in the experiments tracked in Weights and Biases\footnote{https://wandb.ai/deepbirddetect/biofoundation}.

\begin{table}[t]
  \centering
  \caption{The AUROC results of our models on the BirdSet and BEANS benchmark. The best results per pooling strategy are highlighted in \textbf{bold}, and the second best are \underline{underlined}. We also calculate an averaged score for each model and benchmark; for BirdSet, POW is excluded.}
  \resizebox{0.9\textwidth}{!}{%
    \renewcommand{\arraystretch}{1.5} %
\setlength{\tabcolsep}{2pt}

\begin{tabular}{>{\centering\arraybackslash}p{0.7cm} p{1.5cm} | ccccc | >{\centering\arraybackslash}p{0.8cm} !{\vrule width 1.3pt} cccccccc | >{\centering\arraybackslash}p{0.8cm}}
  \toprule
  \multicolumn{2}{c}{}                 & \multicolumn{6}{c}{\textbf{BEANS}}       & \multicolumn{9}{c}{\makecell[c]{\textbf{BirdSet}                                                                                                                                                                                                                                                                                                                                                                                                                                                                                                                                                                                                                         \\[-12pt] \hspace{-7.5cm} {\color{gray}\scriptsize VAL}}}                                                                                                                                                                                                                                                                                                                                                                                  \\
  \addlinespace[2pt]
  \cline{3-17} %
  \addlinespace[2pt]

  \multicolumn{2}{c}{\textbf{Setting}} & \cellcolor{gray!25}\textbf{\textsc{WTK}} & \cellcolor{gray!25}\textbf{\textsc{BAT}}         & \cellcolor{gray!25}\textbf{\textsc{CBI}} & \cellcolor{gray!25}\textbf{\textsc{DOG}} & \cellcolor{gray!25}\textbf{\textsc{HUM}} & \cellcolor{gray!25}\textbf{\underline{Score}} & \cellcolor{gray!25}\textbf{\textsc{POW}} & \cellcolor{gray!25}\textbf{\textsc{PER}} & \cellcolor{gray!25}\textbf{\textsc{NES}} & \cellcolor{gray!25}\textbf{\textsc{UHH}} & \cellcolor{gray!25}\textbf{\textsc{HSN}} & \cellcolor{gray!25}\textbf{\textsc{NBP}} & \cellcolor{gray!25}\textbf{\textsc{SSW}} & \cellcolor{gray!25}\textbf{\textsc{SNE}} & \cellcolor{gray!25}\textbf{\underline{Score}}                                \\
  \addlinespace[2pt]
  \cline{3-17} %
  \addlinespace[2pt]
  \midrule
  \multicolumn{17}{p{5cm}}{\textit{Baseline general audio models}} \vspace{0.5mm}                                                                                                                                                                                                                                                                                                                                                                                                                                                                                                                                                                                                                                                                            \\
  \multirow{2}{*}{\textbf{\rotatebox[origin=c]{90}{\renewcommand{\arraystretch}{1}
        \begin{tabular}{@{}c@{}}Audio \\ MAE
      \end{tabular}}
    }
  }                                   & {Linear}                                 & 88.52 & 87.97 & 92.95 & 59.80 & 93.09 & \heatgreen{84.47} &68.51 & 62.46 & 80.19 & \underline{77.03} & 76.87 & 75.56 & 82.02 & 72.88 & \heatgreen{75.29} \\
  & {Attentive}                              & 99.02 & 95.12 & 95.83 & 99.05 & 96.91 & \heatblue{97.19} &76.36 & 69.47 & 86.21 & \textbf{82.32} & 83.42 & 83.67 & 83.84 & 78.43 & \heatblue{81.05} \\
  \hline
  \multirow{2}{*}{\textbf{\rotatebox[origin=c]{90}{BEATs}
    }
  }                                   & {Linear}                                 & 98.75 & 91.16 & 88.89 & 95.24 & \underline{96.46} & \heatgreen{94.10} &66.63 & 62.13 & 77.17 & 69.03 & 74.29 & 75.34 & 77.83 & 73.09 & \heatgreen{72.70} \\
  & {Attentive}                              & 99.32 & 96.60 & 97.01 & 99.43 & 97.53 & \heatblue{97.98} &76.93 & 70.93 & 87.13 & \underline{81.69} & 83.60 & 84.98 & 89.10 & 78.53 & \heatblue{82.28} \\
  \hline
  \multirow{2}{*}{\textbf{\rotatebox[origin=c]{90}{EAT}
    }
  }                                   & {Linear}                                 & 98.69 & 90.82 & 95.43 & 98.37 & 96.31 & \heatgreen{95.93} &69.82 & 64.32 & 79.75 & 65.87 & 68.29 & 80.52 & 77.96 & 70.86 & \heatgreen{72.51} \\
  & {Attentive}                              & 98.78 & 95.33 & 96.89 & 98.55 & \textbf{98.00} & \heatblue{97.51} &74.62 & 70.78 & 87.38 & 80.29 & 80.10 & 85.16 & 85.31 & 78.40 & \heatblue{81.06} \\
  \midrule
  \multicolumn{17}{p{5cm}}{\textit{Bioacoustic foundation models}} \vspace{0.5mm}                                                                                                                                                                                                                                                                                                                                                                                                                                                                                                                                                                                                                                                                            \\
  \multirow{2}{*}{\textbf{\rotatebox[origin=c]{90}{AVES}
    }
  }                                   & {Linear}                                 & 96.33 & 87.79 & 84.04 & 86.94 & 94.72 & \heatgreen{89.96} &60.14 & 55.03 & 69.91 & 66.31 & 60.38 & 62.96 & 71.14 & 60.87 & \heatgreen{63.80} \\
  & {Attentive}                              & 98.78 & 95.26 & 95.44 & 99.23 & 97.11 & \heatblue{97.16} &71.48 & 59.47 & 83.67 & 76.55 & 76.67 & 75.93 & 80.59 & 68.48 & \heatblue{74.48} \\
  \hline
  \multirow{2}{*}{\textbf{\rotatebox[origin=c]{90}{\renewcommand{\arraystretch}{1}
        \begin{tabular}{@{}c@{}}BEATs \\ NLM
      \end{tabular}}
    }
  }                                   & {Linear}                                 & \underline{98.95} & 92.50 & 93.06 & 93.76 & 95.95 & \heatgreen{94.84} &77.99 & 66.05 & 84.06 & 73.10 & 84.42 & 85.25 & 89.55 & 78.30 & \heatgreen{80.10} \\
  & {Attentive}                              & \underline{99.48} & \textbf{96.85} & \textbf{98.89} & \textbf{99.82} & \underline{97.80} & \heatblue[bold]{98.57} &\underline{83.10} & \underline{72.95} & \textbf{89.24} & 80.73 & \underline{84.46} & \underline{90.14} & \textbf{93.22} & \underline{81.12} & \heatblue[underline]{84.55} \\
  \hline
  \multirow{2}{*}{\textbf{\rotatebox[origin=c]{90}{\renewcommand{\arraystretch}{1}
        \begin{tabular}{@{}c@{}}Biolin \\ gual
      \end{tabular}}
    }
  }                                   & {Linear}                                 & 98.32 & 89.14 & 93.61 & 92.13 & 92.27 & \heatgreen{93.09} &70.13 & 58.56 & 75.89 & 61.83 & 77.70 & 74.65 & 77.71 & 70.18 & \heatgreen{70.93} \\
  & {Attentive}                              & 99.10 & 94.81 & \underline{98.52} & 99.35 & 97.03 & \heatblue{97.76} &78.06 & 68.47 & 87.13 & 79.60 & 82.51 & 88.60 & 90.87 & 80.44 & \heatblue{82.51} \\
  \hline
  \multirow{2}{*}{\textbf{\rotatebox[origin=c]{90}{\renewcommand{\arraystretch}{1}
        \begin{tabular}{@{}c@{}}Bird \\ AVES
      \end{tabular}}
    }
  }                                   & {Linear}                                 & 95.61 & 89.37 & 83.20 & 78.06 & 93.20 & \heatgreen{87.89} &62.90 & 57.28 & 72.97 & 57.85 & 67.05 & 65.73 & 73.37 & 64.81 & \heatgreen{65.58} \\
  & {Attentive}                              & 97.67 & 95.44 & 96.01 & \underline{99.45} & 96.98 & \heatblue{97.11} &76.03 & 63.61 & 88.45 & 74.49 & 82.55 & 81.82 & 85.21 & 75.99 & \heatblue{78.87} \\
  \hline
  \multirow{2}{*}{\textbf{\rotatebox[origin=c]{90}{\renewcommand{\arraystretch}{1}
        \begin{tabular}{@{}c@{}}Bird \\ MAE
      \end{tabular}}
    }
  }                                   & {Linear}                                 & 97.29 & 91.99 & 96.51 & 89.73 & 96.34 & \heatgreen{94.37} &77.84 & 68.59 & 86.65 & 75.47 & 73.36 & 81.99 & 83.21 & 74.36 & \heatgreen{77.66} \\
  & {Attentive}                              & \textbf{99.51} & \underline{96.76} & 97.99 & 99.33 & 97.30 & \heatblue[underline]{98.18} &\textbf{83.85} & \textbf{78.20} & \underline{88.56} & 81.54 & \textbf{89.11} & \textbf{92.17} & \underline{92.35} & \textbf{83.83} & \heatblue[bold]{86.54} \\
  \hline
  \multirow{2}{*}{\textbf{\rotatebox[origin=c]{90}{\renewcommand{\arraystretch}{1}
        \begin{tabular}{@{}c@{}}Conv \\ Next$_{BS}$
      \end{tabular}}
    }
  }                                   & {Linear}                                 & 98.90 & 93.73 & 98.92 & 99.35 & 96.21 & \heatgreen[underline]{97.42} &83.87 & \textbf{72.28} & 88.66 & \textbf{78.49} & \underline{90.76} & \underline{92.27} & 92.49 & \textbf{85.29} & \heatgreen[bold]{85.75} \\
  & {Restricted}                             & - & - & 99.17 & - & - & - &81.73 & 72.54 & 87.75 & 77.71 & 89.62 & 91.58 & 93.44 & 82.70 & \heatblue{85.05} \\
  \hline
  \multirow{2}{*}{\textbf{\rotatebox[origin=c]{90}{Perch}
    }
  }                                   & {Linear}                                 & 98.40 & 88.98 & \underline{99.00} & \underline{99.49} & 95.64 & \heatgreen{96.30} &\underline{85.14} & 72.06 & \textbf{91.68} & 75.26 & \textbf{91.40} & \textbf{92.46} & \underline{92.75} & \underline{83.81} & \heatgreen[underline]{85.63} \\
  & {Restricted}                             & - & - & 99.33 & - & - & - &83.60 & 70.49 & 90.78 & 76.15 & 86.25 & 90.42 & 90.91 & 82.59 & \heatblue{83.94} \\
  \hline
  \multirow{2}{*}{\textbf{\rotatebox[origin=c]{90}{\renewcommand{\arraystretch}{1}
        \begin{tabular}{@{}c@{}}Perch \\ 2.0
      \end{tabular}}
    }
  }                                   & {Linear}                                 & \textbf{99.15} & \textbf{94.70} & \textbf{99.12} & 99.43 & \textbf{96.52} & \heatgreen[bold]{97.78} &\textbf{87.18} & \underline{72.26} & \underline{91.43} & 75.50 & 83.85 & 91.88 & \textbf{93.35} & 83.19 & \heatgreen{84.49} \\
  & {Restricted}                             & - & - & 98.75 & - & - & - &92.09 & 78.56 & 95.26 & 91.17 & 91.49 & 93.32 & 97.33 & 88.29 & \heatblue{90.78} \\
  \hline
  \multirow{2}{*}{\textbf{\rotatebox[origin=c]{90}{\renewcommand{\arraystretch}{1}
        \begin{tabular}{@{}c@{}}Proto \\ CLR
      \end{tabular}}
    }
  }                                   & {Linear}                                 & 98.31 & \underline{93.92} & 97.87 & \textbf{99.55} & 96.41 & \heatgreen{97.21} &76.03 & 68.08 & 81.40 & 71.23 & 76.42 & 80.95 & 80.93 & 72.52 & \heatgreen{75.93} \\
  & {Attentive}                              & 97.87 & 94.18 & 97.62 & 99.43 & 96.73 & \heatblue{97.17} &76.39 & 67.85 & 86.05 & 73.59 & 80.69 & 84.84 & 84.65 & 74.97 & \heatblue{78.95} \\
  \hline
  \multirow{2}{*}{\textbf{\rotatebox[origin=c]{90}{\renewcommand{\arraystretch}{1}
        \begin{tabular}{@{}c@{}}Surf \\ Perch
      \end{tabular}}
    }
  }                                   & {Linear}                                 & 98.75 & 89.42 & 97.58 & 96.27 & 96.15 & \heatgreen{95.63} &77.12 & 65.74 & 87.01 & 73.62 & 82.08 & 79.26 & 83.35 & 74.76 & \heatgreen{77.97} \\
  & {Restricted}                             & - & - & 98.30 & - & - & - &74.83 & 64.16 & 88.31 & 78.40 & 85.68 & 74.64 & 86.38 & 79.28 & \heatblue{79.55} \\
  \hline
  \multirow{2}{*}{\textbf{\rotatebox[origin=c]{90}{\renewcommand{\arraystretch}{1}
        \begin{tabular}{@{}c@{}}ViT \\ INS
      \end{tabular}}
    }
  }                                   & {Linear}                                 & 97.27 & 87.12 & 82.78 & 89.38 & 93.52 & \heatgreen{90.01} &64.96 & 59.11 & 72.46 & 65.07 & 67.56 & 69.59 & 72.84 & 66.34 & \heatgreen{67.57} \\
  & {Attentive}                              & 97.37 & 93.51 & 88.09 & 89.01 & 95.42 & \heatblue{92.68} &68.25 & 60.65 & 77.03 & 68.76 & 66.77 & 75.98 & 77.40 & 70.42 & \heatblue{71.00} \\
  \bottomrule
\end{tabular}

  }
  \label{tab:results}
\end{table}

\subsection{Results and Discussion}

Table~\ref{tab:results} reports results for adapting the selected models to the BEANS and BirdSet benchmarks using linear and attentive probing. The results are presented as AUROC scores, averaged across all tasks and two seeds in each benchmark, excluding the \gls{pow} validation task for BirdSet. We present results including standard deviation in Appendix Table~\ref{tab:results-std} and results using the cmAP5 metric for BirdSet and Acc for BEANS in Table~\ref{tab:results-standard-metrics}. BEANS scores are considerably higher than BirdSet scores, reflecting the more complex nature of the BirdSet tasks, which involve multi-label classification in soundscapes compared to the multi-class classification in BEANS. The best performing models on BEANS are BEATs$_{NLM}$ (98.57 AUROC) and BirdMAE (98.18 AUROC) with attentive probing; among linear probing only, Perch~2.0 achieves the highest BEANS score (97.78 AUROC). The worst performing models on BEANS are AudioMAE (84.47 AUROC with linear probing) and BirdAVES (87.89 AUROC with linear probing). On BirdSet, Perch~2.0 with restricted evaluation achieves the highest overall score (90.78 AUROC). Among probing-based strategies, BirdMAE (86.54 AUROC with attentive probing) is best, followed by ConvNext$_{BS}$ (85.75 AUROC with linear probing) and Perch (85.63 AUROC with linear probing), while the bottom performers are AVES (63.80 AUROC with linear probing) and BirdAVES (65.58 AUROC with linear probing).

\paragraph{Probing strategy}
Transformer-based models (except ProtoCLR) benefit significantly from the added parameters in the attentive probing strategy across all benchmarks, outperforming linear probing by substantial margins. For example, AudioMAE improves from 84.47 to 97.19 AUROC on BEANS and from 75.29 to 81.05 AUROC on BirdSet when using attentive probing. Similarly, BEATs shows dramatic improvements from 94.10 to 97.98 AUROC on BEANS and from 72.70 to 82.28 AUROC on BirdSet. This improvement is likely due to the 1D embeddings of the CLS-token not being well-aligned for the bioacoustic classification task, as indicated by the lower improvement of ViT$_{INS}$ (90.01 to 92.68 AUROC on BEANS), which is trained for classification in a supervised manner. \gls{cnn}-based models (ConvNext$_{BS}$) do not benefit from the attentive probing strategy, which we could only test experimentally with ConvNext$_{BS}$ (see Table~\ref{app:tab:cnn_attentive}) as we could not access the patch embeddings of the Perch models. Simply mean pooling the last convolutional layer's output outperforms a parameterised attentive pooling layer. Notably, linear probing outperforms direct evaluation with logits restriction for ConvNext$_{BS}$ (85.75 vs. 85.05 AUROC on BirdSet) and Perch (85.63 vs. 83.94 AUROC on BirdSet). Perch~2.0 reverses this pattern: its restricted evaluation (90.78 AUROC on BirdSet) substantially outperforms linear probing (84.49 AUROC), suggesting that its pretrained classification head generalises exceptionally well to the BirdSet soundscape tasks.

\begin{table}[htbp]
  \centering
  \caption{AUROC results of the ConvNext$_{BS}$ with linear and attentive probing strategy. The best results are highlighted in \textbf{bold}.}
  \resizebox{0.9\textwidth}{!}{%
    \renewcommand{\arraystretch}{1.5} %
\setlength{\tabcolsep}{2pt}

\begin{tabular}{>{\centering\arraybackslash}p{0.7cm} p{1.5cm} | ccccc | >{\centering\arraybackslash}p{0.8cm} !{\vrule width 1.3pt} cccccccc | >{\centering\arraybackslash}p{0.8cm}}
    \toprule
    \multicolumn{2}{c}{} & \multicolumn{6}{c}{\textbf{BEANS}}       & \multicolumn{9}{c}{\makecell[c]{\textbf{BirdSet}                                                                                                                                                                                                                                                                                                                                                                                                                                                                                                                                                                                                                   \\[-12pt] \hspace{-7.5cm} {\color{gray}\scriptsize VAL}}}                                                                                                                                                                                                                                                                                                                                                                                  \\
    \addlinespace[2pt]
    \cline{3-17} %
    \addlinespace[2pt]

    \multicolumn{2}{c}{} & \cellcolor{gray!25}\textbf{\textsc{WTK}} & \cellcolor{gray!25}\textbf{\textsc{BAT}}         & \cellcolor{gray!25}\textbf{\textsc{CBI}} & \cellcolor{gray!25}\textbf{\textsc{DOG}} & \cellcolor{gray!25}\textbf{\textsc{HUM}} & \cellcolor{gray!25}\textbf{\underline{Score}} & \cellcolor{gray!25}\textbf{\textsc{POW}} & \cellcolor{gray!25}\textbf{\textsc{PER}} & \cellcolor{gray!25}\textbf{\textsc{NES}} & \cellcolor{gray!25}\textbf{\textsc{UHH}} & \cellcolor{gray!25}\textbf{\textsc{HSN}} & \cellcolor{gray!25}\textbf{\textsc{NBP}} & \cellcolor{gray!25}\textbf{\textsc{SSW}} & \cellcolor{gray!25}\textbf{\textsc{SNE}} & \cellcolor{gray!25}\textbf{\underline{Score}}                          \\
    \addlinespace[2pt]
    \cline{3-17} %
    \addlinespace[2pt]

    \multirow{2}{*}{\textbf{\rotatebox[origin=c]{90}{\renewcommand{\arraystretch}{1}\begin{tabular}{@{}c@{}}Conv \\ Next$_{BS}$\end{tabular}}
    }}                   & {Linear}                                 & \textbf{98.9}                                    & \textbf{93.2}                            & \textbf{98.9}                            & \textbf{99.2}                            & \textbf{96.2}                                 & \heatgreen[bold]{97.3}                   & \textbf{83.7  }                          & \textbf{72.4}                            & \textbf{88.6}                            & \textbf{78.1}                            & \textbf{90.6}                            & \textbf{92.3}                            & \textbf{92.5}                            & \textbf{85.3}                                 & \heatgreen[bold]{85.7} \\
                         & {Attentive}                              & 97.7                                             & 87.0                                     & 96.8                                     & 98.5                                     & 94.5                                          & \heatblue{94.9}                          & 78.9                                     & 63.5                                     & 85.3                                     & 73.6                                     & 70.3                                     & 79.8                                     & 80.7                                     & 69.1                                          & \heatblue{74.6}
    \\
    \bottomrule
\end{tabular}

  }
  \label{app:tab:cnn_attentive}
\end{table}

\paragraph{Training data}
On BEANS, the baseline models trained on AudioSet are competitive and outperform many bioacoustic models. BEATs achieves the third-best performance (97.98 AUROC with attentive probing), only outperformed by BEATs$_{NLM}$ (98.57 AUROC) and BirdMAE (98.18 AUROC). BEATs$_{NLM}$ is further aligned using a large amount of bioacoustic data. Among linear probing only, the mixed-source model Perch~2.0 (trained on \gls{xc}, \gls{ina}, \gls{asa}, and \gls{fsd}) achieves the strongest BEANS score (97.78 AUROC), competitive with attentive probing of the top models. Contrary to the results of Ghani et al.~\cite{ghani_global_2023}, bird-trained models do not outperform general audio models when those are evaluated with attentive probing. The representations learned from AudioSet are therefore applicable to bioacoustics when extracted in a more sophisticated manner. Bioacoustic pretraining data does not guarantee better performance, as shown by ViT$_{INS}$, which is trained on the INS dataset but achieves only 92.68 AUROC with attentive probing, underperforming general audio models like BEATs.

On BirdSet, Perch~2.0 with restricted evaluation far surpasses all other models (90.78 AUROC), indicating that its diverse multi-taxa supervised training (\gls{xc}, \gls{ina}, \gls{asa}, \gls{fsd}) produces a classification head that generalises strongly to bird soundscape tasks. Among probing-based strategies, the specialised bird sound classification models (BirdMAE, ConvNext$_{BS}$, Perch) excel, setting the high scores of 86.54, 85.75, and 85.63 AUROC respectively. BEATs$_{NLM}$ with its diverse training set shows improvements (84.55 AUROC) compared to BEATs, which is trained solely on AudioSet (82.28 AUROC). The general audio models (AudioMAE, BEATs, EAT) perform well, but do not reach the performance of the bird-specific models. BioLingual also shows good performance (82.51 AUROC), while the bioacoustic pretraining data of AVES (74.48 AUROC), SurfPerch (79.55 AUROC), and ViT$_{INS}$ (71.00 AUROC) does not lead to better performance. The addition of more diverse data, including \gls{rs}, is associated with lower performance of SurfPerch compared to Perch. Surprisingly, BirdAVES (78.87 AUROC) and ProtoCLR (78.95 AUROC), both using large amounts of bird sound data, do not perform particularly well, showing that training data alone is not a guarantee for success.

\paragraph{Preprocessing}
Models using higher sampling rates generally demonstrate superior performance on bird-focused tasks. The top-performing models on BirdSet include BirdMAE (86.54 AUROC), ConvNext$_{BS}$ (85.75 AUROC), Perch (85.63 AUROC), and Perch~2.0 (90.78 AUROC restricted), which all utilise 32 kHz sampling rates, enabling capture of high-frequency bird vocalisations up to 16 kHz. This potentially explains the lower performance of models using lower sampling rates and may be a contributing factor to BEATs$_{NLM}$ being outperformed by the top-performing models on bird classification tasks. A further difference of these models is the use of 5-second windows, which is optimised for the average duration of bird calls~\cite{rauch2025birdset}. Models processing raw waveforms (AVES: 74.48 AUROC, BirdAVES: 78.87 AUROC) consistently underperform compared to their spectrogram-based counterparts on both benchmarks.

\paragraph{Model architecture}
\Gls{vit} architectures prove to be appropriate for bioacoustic tasks, with most transformer-based models achieving strong performance when combined with attentive probing. Likewise, \glspl{cnn} can succeed in bioacoustic classification tasks. However, a bigger model size does not lead to a clear advantage, as demonstrated by Perch (8M parameters, 85.63 AUROC) versus ConvNext$_{BS}$ (88M parameters, 85.75 AUROC) or BirdMAE (300M parameters, 86.54 AUROC) achieving comparable performance on BirdSet despite the significant difference in parameter count. Perch~2.0 (12M parameters) reaches 90.78 AUROC on BirdSet with restricted evaluation, showing that a moderately sized \gls{cnn} can excel when its pretrained head is well aligned to the evaluation setting.

\paragraph{Training paradigm}
On BEANS, SSL-learned representations such as those from BirdMAE (98.18 AUROC) or BEATs (97.98 AUROC) achieve excellent performance. Additional alignment of BEATs for bioacoustics in the BEATs$_{NLM}$ model leads to the best performance (98.57 AUROC). The \gls{sl} model Perch~2.0, which augments supervised species classification with self-distillation, achieves the best linear probing result on BEANS (97.78 AUROC) and the highest BirdSet score overall (90.78 AUROC with restricted evaluation), substantially exceeding the SSL-based BirdMAE (86.54 AUROC with attentive probing). BirdMAE (86.54 AUROC) outperforms the models pretrained solely on \gls{xc} with supervision (ConvNext$_{BS}$ with 85.75 AUROC, Perch with 85.63 AUROC) when evaluated with attentive probing. This suggests that \gls{ssl} methods can learn more generalisable acoustic representations by capturing intrinsic patterns in spectrograms without being constrained by specific classification objectives. The masked autoencoding approach of BirdMAE or the masked audio modeling objective of the BEATs model may enable it to learn robust features that better transfer to the challenging multi-label soundscape classification task, where understanding temporal and spectral relationships is crucial for detecting overlapping vocalisations. At the same time, Perch~2.0 demonstrates that supervised learning combined with self-distillation and diverse multi-taxa data can yield representations whose pretrained classification heads generalise exceptionally well to bird soundscape tasks.

\paragraph{Implications for bioacoustic model development}

While our comparison of pretrained bioacoustic foundation models does not offer a direct ablation study of the different design decisions, we can draw several suggestions for future model development. First, advances in general audio understanding, particularly on the \gls{as} benchmark, translate effectively to bioacoustic tasks, as demonstrated by the superior performance of BEATs over AudioMAE. This suggests that sophisticated \gls{ssl} methods developed for broader audio domains can be successfully leveraged for biological sound analysis, and that pretraining models with such methods on bioacoustic data promises further performance increases. Second, selecting a high sample rate (32 kHz) and a suitable window length (5 seconds) seems beneficial, while advantages in using raw waveforms over spectrograms are not evident in our experiments.

Future research could investigate whether combining general audio data with bioacoustic data during pretraining enables models to develop more robust auditory representations that generalise across diverse acoustic environments, as demonstrated by the performance of Perch~2.0 compared to Perch.
However, dataset scale alone does not guarantee superior performance—curation quality proves equally critical~\cite{oquab2024dinov2learningrobustvisual,rauch2025maskedautoencoderslistenbirds}. For practitioners developing new bioacoustic models, we recommend leveraging established datasets such as \gls{as}, \gls{bs}, and \gls{ins}, which provide diverse acoustic coverage, include quality curation, and offer accessibility for research. A notable gap remains in large-scale \gls{pam} datasets that would better reflect real-world deployment scenarios. The absence of models trained on such a dataset in our evaluation highlights an important direction for future data collection and model development efforts, as such models would likely achieve better ecological monitoring performance and close the gap between controlled laboratory settings (BEANS) and real-world applications (BirdSet).

\paragraph{Advice for model selection}

For practitioners selecting foundation models for bioacoustic applications, several considerations emerge from our evaluation. Perch~2.0 achieves the highest BirdSet score (90.78 AUROC with restricted evaluation) and the strongest linear probing result on BEANS (97.78 AUROC), while having a small computational footprint due to the small model size with 12M parameters, making it a clear recommendation. When additional task alignment with attentive probing is possible, BirdMAE provides meaningful bioacoustic representations to build on. It is a close second after BEATs$_{NLM}$ on the BEANS benchmark and the second highest score on BirdSet. Both models require attentive probing to extract the full potential of their representations. ConvNext$_{BS}$ provides a complete open-source training pipeline that offers advantages in transparency and customisation for research applications. Both Perch (2.0) and ConvNext$_{BS}$ can be effectively adapted using linear probing, a computationally efficient approach that requires minimal additional training data. Storage and computational efficiency considerations also influence model selection. Models that perform well on smaller averaged 1D embeddings offer significant advantages over those requiring 3D patch embeddings for attentive probing, as compact representations are easier to store and process, a critical factor for applications that involve vector databases or edge device deployment~\cite{dumoulinSearchSquawkAgile2025}. Finally, BEATs$_{NLM}$, serving as the encoder for NatureLM-audio, demonstrates impressive performance, while the full model enables text-based interaction through its integration into an audio-\gls{llm} framework. This accessibility feature represents a substantial advantage for citizen science platforms and educational applications, where natural language interfaces can lower barriers to acoustic analysis.

\paragraph{Limitations}
We evaluate our models solely with frozen encoders. Unfreezing them and fully fine-tuning the models could further improve the performance~\cite{kumar2022finetuningdistortpretrainedfeatures}. We omitted such experiments not only because of the substantial increase in computational requirements but also because of the sensitivity to hyperparameter adjustments. Furthermore, our probing-based experimental results are influenced by the choice of hyperparameters, and keeping them fixed across models and benchmarks could favour some models. A computationally intensive model- and dataset-based hyperparameter optimisation could therefore improve the results.

\section{Conclusion}
\label{sec:conclusion}

This work presents a comprehensive review and comparative analysis of thirteen foundation models for bioacoustic classification. We detailed major pretraining data sources and evaluation benchmarks, reviewed large-scale bioacoustic models analysing their key design decisions, and compared selected models on the BEANS and BirdSet benchmarks using linear and attentive probing techniques. \\
Our systematic experimental analysis reveals four key findings about bioacoustic foundation models. First, Perch~2.0, a mixed-source supervised model trained on \gls{xc}, \gls{ina}, \gls{asa}, and \gls{fsd} with self-distillation, achieves the highest BirdSet score (90.78 AUROC with restricted evaluation) and the strongest linear probing result on BEANS (97.78 AUROC), demonstrating that diverse multi-taxa supervised pretraining can yield representations whose classification heads generalise exceptionally well. Second, BirdMAE trained on large-scale bird song data with self-supervision is the best model among probing-based strategies on BirdSet and second on BEANS after BEATs$_{NLM}$, the encoder of NatureLM-audio. Third, attentive probing is beneficial to extract the full performance of transformer-based models. Fourth, general-purpose audio models trained with self-supervised learning on AudioSet outperform many specialised bird sound models on the diverse BEANS benchmark when evaluated with attentive probing. \\
These findings have critical implications for practitioners selecting models for bioacoustic classification tasks. Perch~2.0 is a strong default choice: it leads on BirdSet with restricted evaluation and on BEANS under linear probing while having a small footprint (12M parameters), making it well-suited for resource-constrained and deployment-oriented applications. For taxa-wide classification with attentive probing, BEATs$_{NLM}$ is preferable. For bird sound classification when additional task alignment via attentive probing is feasible, BirdMAE provides the best probing-based performance on BirdSet; ConvNext$_{BS}$ provides a complete open-source training pipeline. The dramatic improvements from attentive probing highlight the importance of adaptive attention mechanisms in transferring audio representations. \\
Looking forward, key research directions include developing large-scale foundation models trained on passive acoustic monitoring data, investigating optimal combinations of general audio and bioacoustic data during pretraining (as exemplified by Perch~2.0 versus Perch), and exploring more sophisticated adaptation strategies beyond attentive probing such as prototypical probing and \gls{lora} for bioacoustic classification tasks.

\section*{Data availability}

All data and code for this study are publicly available. Source code, evaluation scripts, and documentation are hosted at \url{https://github.com/DBD-research-group/BioFoundation}; training logs and detailed results are at \url{https://wandb.ai/deepbirddetect/biofoundation}.

\section*{Acknowledgments}

This research has been funded by the German Ministry for the Environment, Nature Conservation, Nuclear Safety, and Consumer Protection through the project "DeepBirdDetect - Automatic Bird Detection of Endangered Species Using Deep Neural Networks" (67KI31040C).

\printbibliography

\newpage
\appendix

\section*{Appendix}

\begin{table}[h!]
  \centering
  \caption{The results of our models on the BirdSet and BEANS benchmark where for BEANS we report the AUROC. The best results are highlighted in \textbf{bold}, and the second-best results are \underline{underlined}. We also calculate a score for each model and benchmark but for BirdSet POW is excluded.}
  \resizebox{\textwidth}{!}{%
    \renewcommand{\arraystretch}{1.5} %
\setlength{\tabcolsep}{2pt}

\begin{tabular}{>{\centering\arraybackslash}p{0.7cm} p{1.5cm} | ccccc | >{\centering\arraybackslash}p{0.8cm} !{\vrule width 1.3pt} cccccccc | >{\centering\arraybackslash}p{0.8cm}}
  \toprule
  \multicolumn{2}{c}{}                 & \multicolumn{6}{c}{\textbf{BEANS}}       & \multicolumn{9}{c}{\makecell[c]{\textbf{BirdSet}                                                                                                                                                                                                                                                                                                                                                                                                                                                                                                                                                                                                                         \\[-12pt] \hspace{-7.5cm} {\color{gray}\scriptsize VAL}}}                                                                                                                                                                                                                                                                                                                                                                                  \\
  \addlinespace[2pt]
  \cline{3-17} %
  \addlinespace[2pt]

  \multicolumn{2}{c}{\textbf{Setting}} & \cellcolor{gray!25}\textbf{\textsc{WTK}} & \cellcolor{gray!25}\textbf{\textsc{BAT}}         & \cellcolor{gray!25}\textbf{\textsc{CBI}} & \cellcolor{gray!25}\textbf{\textsc{DOG}} & \cellcolor{gray!25}\textbf{\textsc{HUM}} & \cellcolor{gray!25}\textbf{\underline{Score}} & \cellcolor{gray!25}\textbf{\textsc{POW}} & \cellcolor{gray!25}\textbf{\textsc{PER}} & \cellcolor{gray!25}\textbf{\textsc{NES}} & \cellcolor{gray!25}\textbf{\textsc{UHH}} & \cellcolor{gray!25}\textbf{\textsc{HSN}} & \cellcolor{gray!25}\textbf{\textsc{NBP}} & \cellcolor{gray!25}\textbf{\textsc{SSW}} & \cellcolor{gray!25}\textbf{\textsc{SNE}} & \cellcolor{gray!25}\textbf{\underline{Score}}                                \\
  \addlinespace[2pt]
  \cline{3-17} %
  \addlinespace[2pt]
  \midrule
  \multicolumn{17}{p{5cm}}{\textit{Baseline general audio models}} \vspace{0.5mm}                                                                                                                                                                                                                                                                                                                                                                                                                                                                                                                                                                                                                                                                            \\
  \multirow{2}{*}{\textbf{\rotatebox[origin=c]{90}{\renewcommand{\arraystretch}{1}
        \begin{tabular}{@{}c@{}}Audio \\ MAE
      \end{tabular}}
    }
  }                                   & {Linear}                                 & 88.52\scriptsize{$\pm$ 8.37} & 87.97\scriptsize{$\pm$ 0.62} & 92.95\scriptsize{$\pm$ 0.14} & 59.80\scriptsize{$\pm$ 5.33} & 93.09\scriptsize{$\pm$ 1.55} & \heatgreen{84.47} &68.51\scriptsize{$\pm$ 0.41} & 62.46\scriptsize{$\pm$ 0.03} & 80.19\scriptsize{$\pm$ 0.04} & \underline{77.03\scriptsize{$\pm$ 0.09}} & 76.87\scriptsize{$\pm$ 0.21} & 75.56\scriptsize{$\pm$ 0.15} & 82.02\scriptsize{$\pm$ 0.03} & 72.88\scriptsize{$\pm$ 0.04} & \heatgreen{75.29} \\
  & {Attentive}                              & 99.02\scriptsize{$\pm$ 0.01} & 95.12\scriptsize{$\pm$ 0.12} & 95.83\scriptsize{$\pm$ 0.04} & 99.05\scriptsize{$\pm$ 0.11} & 96.91\scriptsize{$\pm$ 0.02} & \heatblue{97.19} &76.36\scriptsize{$\pm$ 0.32} & 69.47\scriptsize{$\pm$ 0.39} & 86.21\scriptsize{$\pm$ 0.16} & \textbf{82.32\scriptsize{$\pm$ 0.36}} & 83.42\scriptsize{$\pm$ 0.48} & 83.67\scriptsize{$\pm$ 0.33} & 83.84\scriptsize{$\pm$ 0.10} & 78.43\scriptsize{$\pm$ 0.25} & \heatblue{81.05} \\
  \hline
  \multirow{2}{*}{\textbf{\rotatebox[origin=c]{90}{BEATs}
    }
  }                                   & {Linear}                                 & 98.75\scriptsize{$\pm$ 0.03} & 91.16\scriptsize{$\pm$ 0.28} & 88.89\scriptsize{$\pm$ 0.11} & 95.24\scriptsize{$\pm$ 1.99} & \underline{96.46\scriptsize{$\pm$ 0.48}} & \heatgreen{94.10} &66.63\scriptsize{$\pm$ 0.06} & 62.13\scriptsize{$\pm$ 0.13} & 77.17\scriptsize{$\pm$ 0.03} & 69.03\scriptsize{$\pm$ 0.07} & 74.29\scriptsize{$\pm$ 1.23} & 75.34\scriptsize{$\pm$ 0.05} & 77.83\scriptsize{$\pm$ 0.02} & 73.09\scriptsize{$\pm$ 0.49} & \heatgreen{72.70} \\
  & {Attentive}                              & 99.32\scriptsize{$\pm$ 0.10} & 96.60\scriptsize{$\pm$ 0.61} & 97.01\scriptsize{$\pm$ 0.56} & 99.43\scriptsize{$\pm$ 0.36} & 97.53\scriptsize{$\pm$ 0.62} & \heatblue{97.98} &76.93\scriptsize{$\pm$ 0.09} & 70.93\scriptsize{$\pm$ 0.03} & 87.13\scriptsize{$\pm$ 0.18} & \underline{81.69\scriptsize{$\pm$ 0.01}} & 83.60\scriptsize{$\pm$ 0.22} & 84.98\scriptsize{$\pm$ 0.14} & 89.10\scriptsize{$\pm$ 0.15} & 78.53\scriptsize{$\pm$ 0.06} & \heatblue{82.28} \\
  \hline
  \multirow{2}{*}{\textbf{\rotatebox[origin=c]{90}{EAT}
    }
  }                                   & {Linear}                                 & 98.69\scriptsize{$\pm$ 0.01} & 90.82\scriptsize{$\pm$ 0.01} & 95.43\scriptsize{$\pm$ 0.02} & 98.37\scriptsize{$\pm$ 0.08} & 96.31\scriptsize{$\pm$ 0.01} & \heatgreen{95.93} &69.82\scriptsize{$\pm$ 0.52} & 64.32\scriptsize{$\pm$ 0.02} & 79.75\scriptsize{$\pm$ 0.43} & 65.87\scriptsize{$\pm$ 0.72} & 68.29\scriptsize{$\pm$ 0.84} & 80.52\scriptsize{$\pm$ 0.03} & 77.96\scriptsize{$\pm$ 0.49} & 70.86\scriptsize{$\pm$ 0.28} & \heatgreen{72.51} \\
  & {Attentive}                              & 98.78\scriptsize{$\pm$ 0.07} & 95.33\scriptsize{$\pm$ 0.96} & 96.89\scriptsize{$\pm$ 0.03} & 98.55\scriptsize{$\pm$ 0.39} & \textbf{98.00\scriptsize{$\pm$ 0.05}} & \heatblue{97.51} &74.62\scriptsize{$\pm$ 1.34} & 70.78\scriptsize{$\pm$ 0.00} & 87.38\scriptsize{$\pm$ 0.70} & 80.29\scriptsize{$\pm$ 0.19} & 80.10\scriptsize{$\pm$ 1.46} & 85.16\scriptsize{$\pm$ 0.32} & 85.31\scriptsize{$\pm$ 0.50} & 78.40\scriptsize{$\pm$ 0.21} & \heatblue{81.06} \\
  \midrule
  \multicolumn{17}{p{5cm}}{\textit{Bioacoustic foundation models}} \vspace{0.5mm}                                                                                                                                                                                                                                                                                                                                                                                                                                                                                                                                                                                                                                                                            \\
  \multirow{2}{*}{\textbf{\rotatebox[origin=c]{90}{AVES}
    }
  }                                   & {Linear}                                 & 96.33\scriptsize{$\pm$ 0.44} & 87.79\scriptsize{$\pm$ 0.88} & 84.04\scriptsize{$\pm$ 0.07} & 86.94\scriptsize{$\pm$ 0.42} & 94.72\scriptsize{$\pm$ 0.06} & \heatgreen{89.96} &60.14\scriptsize{$\pm$ 0.31} & 55.03\scriptsize{$\pm$ 0.41} & 69.91\scriptsize{$\pm$ 0.20} & 66.31\scriptsize{$\pm$ 1.12} & 60.38\scriptsize{$\pm$ 0.01} & 62.96\scriptsize{$\pm$ 0.82} & 71.14\scriptsize{$\pm$ 0.29} & 60.87\scriptsize{$\pm$ 0.26} & \heatgreen{63.80} \\
  & {Attentive}                              & 98.78\scriptsize{$\pm$ 0.06} & 95.26\scriptsize{$\pm$ 0.36} & 95.44\scriptsize{$\pm$ 0.17} & 99.23\scriptsize{$\pm$ 0.30} & 97.11\scriptsize{$\pm$ 0.07} & \heatblue{97.16} &71.48\scriptsize{$\pm$ 0.42} & 59.47\scriptsize{$\pm$ 0.54} & 83.67\scriptsize{$\pm$ 0.25} & 76.55\scriptsize{$\pm$ 1.14} & 76.67\scriptsize{$\pm$ 0.27} & 75.93\scriptsize{$\pm$ 0.03} & 80.59\scriptsize{$\pm$ 0.14} & 68.48\scriptsize{$\pm$ 1.75} & \heatblue{74.48} \\
  \hline
  \multirow{2}{*}{\textbf{\rotatebox[origin=c]{90}{\renewcommand{\arraystretch}{1}
        \begin{tabular}{@{}c@{}}BEATs \\ NLM
      \end{tabular}}
    }
  }                                   & {Linear}                                 & \underline{98.95\scriptsize{$\pm$ 0.01}} & 92.50\scriptsize{$\pm$ 0.02} & 93.06\scriptsize{$\pm$ 0.01} & 93.76\scriptsize{$\pm$ 0.59} & 95.95\scriptsize{$\pm$ 0.02} & \heatgreen{94.84} &77.99\scriptsize{$\pm$ 0.01} & 66.05\scriptsize{$\pm$ 0.29} & 84.06\scriptsize{$\pm$ 0.13} & 73.10\scriptsize{$\pm$ 0.51} & 84.42\scriptsize{$\pm$ 0.00} & 85.25\scriptsize{$\pm$ 0.33} & 89.55\scriptsize{$\pm$ 0.13} & 78.30\scriptsize{$\pm$ 0.17} & \heatgreen{80.10} \\
  & {Attentive}                              & \underline{99.48\scriptsize{$\pm$ 0.17}} & \textbf{96.85\scriptsize{$\pm$ 0.01}} & \textbf{98.89\scriptsize{$\pm$ 0.00}} & \textbf{99.82\scriptsize{$\pm$ 0.06}} & \underline{97.80\scriptsize{$\pm$ 0.12}} & \heatblue[bold]{98.57} &\underline{83.10\scriptsize{$\pm$ 0.20}} & \underline{72.95\scriptsize{$\pm$ 1.04}} & \textbf{89.24\scriptsize{$\pm$ 0.26}} & 80.73\scriptsize{$\pm$ 0.02} & \underline{84.46\scriptsize{$\pm$ 0.68}} & \underline{90.14\scriptsize{$\pm$ 0.14}} & \textbf{93.22\scriptsize{$\pm$ 0.06}} & \underline{81.12\scriptsize{$\pm$ 0.89}} & \heatblue[underline]{84.55} \\
  \hline
  \multirow{2}{*}{\textbf{\rotatebox[origin=c]{90}{\renewcommand{\arraystretch}{1}
        \begin{tabular}{@{}c@{}}Biolin \\ gual
      \end{tabular}}
    }
  }                                   & {Linear}                                 & 98.32\scriptsize{$\pm$ 0.15} & 89.14\scriptsize{$\pm$ 0.45} & 93.61\scriptsize{$\pm$ 0.86} & 92.13\scriptsize{$\pm$ 2.60} & 92.27\scriptsize{$\pm$ 0.06} & \heatgreen{93.09} &70.13\scriptsize{$\pm$ 0.04} & 58.56\scriptsize{$\pm$ 0.07} & 75.89\scriptsize{$\pm$ 0.32} & 61.83\scriptsize{$\pm$ 0.26} & 77.70\scriptsize{$\pm$ 0.23} & 74.65\scriptsize{$\pm$ 0.53} & 77.71\scriptsize{$\pm$ 0.75} & 70.18\scriptsize{$\pm$ 0.02} & \heatgreen{70.93} \\
  & {Attentive}                              & 99.10\scriptsize{$\pm$ 0.03} & 94.81\scriptsize{$\pm$ 0.18} & \underline{98.52\scriptsize{$\pm$ 0.02}} & 99.35\scriptsize{$\pm$ 0.41} & 97.03\scriptsize{$\pm$ 0.03} & \heatblue{97.76} &78.06\scriptsize{$\pm$ 1.32} & 68.47\scriptsize{$\pm$ 0.12} & 87.13\scriptsize{$\pm$ 0.60} & 79.60\scriptsize{$\pm$ 0.42} & 82.51\scriptsize{$\pm$ 0.51} & 88.60\scriptsize{$\pm$ 0.07} & 90.87\scriptsize{$\pm$ 0.05} & 80.44\scriptsize{$\pm$ 0.05} & \heatblue{82.51} \\
  \hline
  \multirow{2}{*}{\textbf{\rotatebox[origin=c]{90}{\renewcommand{\arraystretch}{1}
        \begin{tabular}{@{}c@{}}Bird \\ AVES
      \end{tabular}}
    }
  }                                   & {Linear}                                 & 95.61\scriptsize{$\pm$ 0.02} & 89.37\scriptsize{$\pm$ 0.15} & 83.20\scriptsize{$\pm$ 0.01} & 78.06\scriptsize{$\pm$ 1.32} & 93.20\scriptsize{$\pm$ 0.34} & \heatgreen{87.89} &62.90\scriptsize{$\pm$ 0.31} & 57.28\scriptsize{$\pm$ 0.10} & 72.97\scriptsize{$\pm$ 0.13} & 57.85\scriptsize{$\pm$ 0.59} & 67.05\scriptsize{$\pm$ 0.22} & 65.73\scriptsize{$\pm$ 0.03} & 73.37\scriptsize{$\pm$ 0.16} & 64.81\scriptsize{$\pm$ 0.19} & \heatgreen{65.58} \\
  & {Attentive}                              & 97.67\scriptsize{$\pm$ 0.59} & 95.44\scriptsize{$\pm$ 0.47} & 96.01\scriptsize{$\pm$ 0.05} & \underline{99.45\scriptsize{$\pm$ 0.07}} & 96.98\scriptsize{$\pm$ 0.09} & \heatblue{97.11} &76.03\scriptsize{$\pm$ 0.09} & 63.61\scriptsize{$\pm$ 0.04} & 88.45\scriptsize{$\pm$ 0.15} & 74.49\scriptsize{$\pm$ 0.20} & 82.55\scriptsize{$\pm$ 1.15} & 81.82\scriptsize{$\pm$ 0.29} & 85.21\scriptsize{$\pm$ 0.33} & 75.99\scriptsize{$\pm$ 0.05} & \heatblue{78.87} \\
  \hline
  \multirow{2}{*}{\textbf{\rotatebox[origin=c]{90}{\renewcommand{\arraystretch}{1}
        \begin{tabular}{@{}c@{}}Bird \\ MAE
      \end{tabular}}
    }
  }                                   & {Linear}                                 & 97.29\scriptsize{$\pm$ 0.33} & 91.99\scriptsize{$\pm$ 0.35} & 96.51\scriptsize{$\pm$ 0.01} & 89.73\scriptsize{$\pm$ 3.63} & 96.34\scriptsize{$\pm$ 0.54} & \heatgreen{94.37} &77.84\scriptsize{$\pm$ 0.15} & 68.59\scriptsize{$\pm$ 0.03} & 86.65\scriptsize{$\pm$ 0.07} & 75.47\scriptsize{$\pm$ 0.40} & 73.36\scriptsize{$\pm$ 0.24} & 81.99\scriptsize{$\pm$ 0.22} & 83.21\scriptsize{$\pm$ 0.03} & 74.36\scriptsize{$\pm$ 0.13} & \heatgreen{77.66} \\
  & {Attentive}                              & \textbf{99.51\scriptsize{$\pm$ 0.02}} & \underline{96.76\scriptsize{$\pm$ 0.11}} & 97.99\scriptsize{$\pm$ 0.03} & 99.33\scriptsize{$\pm$ 0.06} & 97.30\scriptsize{$\pm$ 0.36} & \heatblue[underline]{98.18} &\textbf{83.85\scriptsize{$\pm$ 0.06}} & \textbf{78.20\scriptsize{$\pm$ 0.95}} & \underline{88.56\scriptsize{$\pm$ 0.29}} & 81.54\scriptsize{$\pm$ 0.58} & \textbf{89.11\scriptsize{$\pm$ 0.14}} & \textbf{92.17\scriptsize{$\pm$ 0.52}} & \underline{92.35\scriptsize{$\pm$ 0.05}} & \textbf{83.83\scriptsize{$\pm$ 0.40}} & \heatblue[bold]{86.54} \\
  \hline
  \multirow{2}{*}{\textbf{\rotatebox[origin=c]{90}{\renewcommand{\arraystretch}{1}
        \begin{tabular}{@{}c@{}}Conv \\ Next$_{BS}$
      \end{tabular}}
    }
  }                                   & {Linear}                                 & 98.90\scriptsize{$\pm$ 0.01} & 93.73\scriptsize{$\pm$ 0.76} & 98.92\scriptsize{$\pm$ 0.02} & 99.35\scriptsize{$\pm$ 0.16} & 96.21\scriptsize{$\pm$ 0.03} & \heatgreen[underline]{97.42} &83.87\scriptsize{$\pm$ 0.27} & \textbf{72.28\scriptsize{$\pm$ 0.21}} & 88.66\scriptsize{$\pm$ 0.02} & \textbf{78.49\scriptsize{$\pm$ 0.61}} & \underline{90.76\scriptsize{$\pm$ 0.21}} & \underline{92.27\scriptsize{$\pm$ 0.00}} & 92.49\scriptsize{$\pm$ 0.02} & \textbf{85.29\scriptsize{$\pm$ 0.07}} & \heatgreen[bold]{85.75} \\
  & {Restricted}                             & - & - & 99.17\scriptsize{$\pm$ 0.00} & - & - & - &81.73\scriptsize{$\pm$ 0.00} & 72.54\scriptsize{$\pm$ 0.00} & 87.75\scriptsize{$\pm$ 0.00} & 77.71\scriptsize{$\pm$ 0.00} & 89.62\scriptsize{$\pm$ 0.00} & 91.58\scriptsize{$\pm$ 0.00} & 93.44\scriptsize{$\pm$ 0.00} & 82.70\scriptsize{$\pm$ 0.00} & \heatblue{85.05} \\
  \hline
  \multirow{2}{*}{\textbf{\rotatebox[origin=c]{90}{Perch}
    }
  }                                   & {Linear}                                 & 98.40\scriptsize{$\pm$ 0.10} & 88.98\scriptsize{$\pm$ 0.32} & \underline{99.00\scriptsize{$\pm$ 0.00}} & \underline{99.49\scriptsize{$\pm$ 0.20}} & 95.64\scriptsize{$\pm$ 0.12} & \heatgreen{96.30} &\underline{85.14\scriptsize{$\pm$ 0.04}} & 72.06\scriptsize{$\pm$ 0.16} & \textbf{91.68\scriptsize{$\pm$ 0.02}} & 75.26\scriptsize{$\pm$ 0.45} & \textbf{91.40\scriptsize{$\pm$ 0.10}} & \textbf{92.46\scriptsize{$\pm$ 0.19}} & \underline{92.75\scriptsize{$\pm$ 0.14}} & \underline{83.81\scriptsize{$\pm$ 0.28}} & \heatgreen[underline]{85.63} \\
  & {Restricted}                             & - & - & 99.33\scriptsize{$\pm$ 0.00} & - & - & - &83.60\scriptsize{$\pm$ 0.00} & 70.49\scriptsize{$\pm$ 0.00} & 90.78\scriptsize{$\pm$ 0.00} & 76.15\scriptsize{$\pm$ 0.00} & 86.25\scriptsize{$\pm$ 0.00} & 90.42\scriptsize{$\pm$ 0.00} & 90.91\scriptsize{$\pm$ 0.00} & 82.59\scriptsize{$\pm$ 0.00} & \heatblue{83.94} \\
  \hline
  \multirow{2}{*}{\textbf{\rotatebox[origin=c]{90}{\renewcommand{\arraystretch}{1}
        \begin{tabular}{@{}c@{}}Perch \\ 2.0
      \end{tabular}}
    }
  }                                   & {Linear}                                 & \textbf{99.15\scriptsize{$\pm$ 0.10}} & \textbf{94.70\scriptsize{$\pm$ 0.44}} & \textbf{99.12\scriptsize{$\pm$ 0.02}} & 99.43\scriptsize{$\pm$ 0.17} & \textbf{96.52\scriptsize{$\pm$ 0.00}} & \heatgreen[bold]{97.78} &\textbf{87.18\scriptsize{$\pm$ 0.08}} & \underline{72.26\scriptsize{$\pm$ 0.05}} & \underline{91.43\scriptsize{$\pm$ 0.02}} & 75.50\scriptsize{$\pm$ 0.21} & 83.85\scriptsize{$\pm$ 0.44} & 91.88\scriptsize{$\pm$ 0.30} & \textbf{93.35\scriptsize{$\pm$ 0.01}} & 83.19\scriptsize{$\pm$ 0.74} & \heatgreen{84.49} \\
  & {Restricted}                             & - & - & 98.75\scriptsize{$\pm$ 0.00} & - & - & - &92.09\scriptsize{$\pm$ 0.00} & 78.56\scriptsize{$\pm$ 0.00} & 95.26\scriptsize{$\pm$ 0.00} & 91.17\scriptsize{$\pm$ 0.00} & 91.49\scriptsize{$\pm$ 0.00} & 93.32\scriptsize{$\pm$ 0.00} & 97.33\scriptsize{$\pm$ 0.00} & 88.29\scriptsize{$\pm$ 0.00} & \heatblue{90.78} \\
  \hline
  \multirow{2}{*}{\textbf{\rotatebox[origin=c]{90}{\renewcommand{\arraystretch}{1}
        \begin{tabular}{@{}c@{}}Proto \\ CLR
      \end{tabular}}
    }
  }                                   & {Linear}                                 & 98.31\scriptsize{$\pm$ 0.01} & \underline{93.92\scriptsize{$\pm$ 0.01}} & 97.87\scriptsize{$\pm$ 0.07} & \textbf{99.55\scriptsize{$\pm$ 0.02}} & 96.41\scriptsize{$\pm$ 0.14} & \heatgreen{97.21} &76.03\scriptsize{$\pm$ 0.02} & 68.08\scriptsize{$\pm$ 0.13} & 81.40\scriptsize{$\pm$ 0.03} & 71.23\scriptsize{$\pm$ 0.02} & 76.42\scriptsize{$\pm$ 0.04} & 80.95\scriptsize{$\pm$ 0.02} & 80.93\scriptsize{$\pm$ 0.01} & 72.52\scriptsize{$\pm$ 0.02} & \heatgreen{75.93} \\
  & {Attentive}                              & 97.87\scriptsize{$\pm$ 0.03} & 94.18\scriptsize{$\pm$ 0.20} & 97.62\scriptsize{$\pm$ 0.02} & 99.43\scriptsize{$\pm$ 0.12} & 96.73\scriptsize{$\pm$ 0.14} & \heatblue{97.17} &76.39\scriptsize{$\pm$ 0.57} & 67.85\scriptsize{$\pm$ 0.11} & 86.05\scriptsize{$\pm$ 0.46} & 73.59\scriptsize{$\pm$ 0.45} & 80.69\scriptsize{$\pm$ 0.18} & 84.84\scriptsize{$\pm$ 0.26} & 84.65\scriptsize{$\pm$ 0.09} & 74.97\scriptsize{$\pm$ 0.33} & \heatblue{78.95} \\
  \hline
  \multirow{2}{*}{\textbf{\rotatebox[origin=c]{90}{\renewcommand{\arraystretch}{1}
        \begin{tabular}{@{}c@{}}Surf \\ Perch
      \end{tabular}}
    }
  }                                   & {Linear}                                 & 98.75\scriptsize{$\pm$ 0.00} & 89.42\scriptsize{$\pm$ 0.31} & 97.58\scriptsize{$\pm$ 0.03} & 96.27\scriptsize{$\pm$ 1.62} & 96.15\scriptsize{$\pm$ 0.01} & \heatgreen{95.63} &77.12\scriptsize{$\pm$ 1.92} & 65.74\scriptsize{$\pm$ 0.00} & 87.01\scriptsize{$\pm$ 0.22} & 73.62\scriptsize{$\pm$ 1.40} & 82.08\scriptsize{$\pm$ 0.23} & 79.26\scriptsize{$\pm$ 0.03} & 83.35\scriptsize{$\pm$ 0.64} & 74.76\scriptsize{$\pm$ 1.61} & \heatgreen{77.97} \\
  & {Restricted}                             & - & - & 98.30\scriptsize{$\pm$ 0.00} & - & - & - &74.83\scriptsize{$\pm$ 0.00} & 64.16\scriptsize{$\pm$ 0.00} & 88.31\scriptsize{$\pm$ 0.00} & 78.40\scriptsize{$\pm$ 0.00} & 85.68\scriptsize{$\pm$ 0.00} & 74.64\scriptsize{$\pm$ 0.00} & 86.38\scriptsize{$\pm$ 0.00} & 79.28\scriptsize{$\pm$ 0.00} & \heatblue{79.55} \\
  \hline
  \multirow{2}{*}{\textbf{\rotatebox[origin=c]{90}{\renewcommand{\arraystretch}{1}
        \begin{tabular}{@{}c@{}}ViT \\ INS
      \end{tabular}}
    }
  }                                   & {Linear}                                 & 97.27\scriptsize{$\pm$ 0.31} & 87.12\scriptsize{$\pm$ 0.73} & 82.78\scriptsize{$\pm$ 0.04} & 89.38\scriptsize{$\pm$ 0.13} & 93.52\scriptsize{$\pm$ 0.01} & \heatgreen{90.01} &64.96\scriptsize{$\pm$ 0.09} & 59.11\scriptsize{$\pm$ 0.50} & 72.46\scriptsize{$\pm$ 0.09} & 65.07\scriptsize{$\pm$ 0.17} & 67.56\scriptsize{$\pm$ 0.33} & 69.59\scriptsize{$\pm$ 0.31} & 72.84\scriptsize{$\pm$ 0.07} & 66.34\scriptsize{$\pm$ 0.26} & \heatgreen{67.57} \\
  & {Attentive}                              & 97.37\scriptsize{$\pm$ 0.87} & 93.51\scriptsize{$\pm$ 0.10} & 88.09\scriptsize{$\pm$ 0.55} & 89.01\scriptsize{$\pm$ 4.77} & 95.42\scriptsize{$\pm$ 0.40} & \heatblue{92.68} &68.25\scriptsize{$\pm$ 0.04} & 60.65\scriptsize{$\pm$ 0.04} & 77.03\scriptsize{$\pm$ 0.14} & 68.76\scriptsize{$\pm$ 0.03} & 66.77\scriptsize{$\pm$ 0.09} & 75.98\scriptsize{$\pm$ 0.02} & 77.40\scriptsize{$\pm$ 0.03} & 70.42\scriptsize{$\pm$ 0.01} & \heatblue{71.00} \\
  \bottomrule
\end{tabular}

  }
  \label{tab:results-std}
\end{table}

\begin{table}[h!]
  \centering
  \caption{The results of our models on the BirdSet and BEANS benchmark where for BEANS we report the Top1-Accuracy and for BirdSet the cmAP5. The best results are highlighted in \textbf{bold}, and the second-best results are \underline{underlined}. We also calculate a score for each model and benchmark but for BirdSet POW is excluded.}
  \resizebox{\textwidth}{!}{%
    \renewcommand{\arraystretch}{1.5} %
\setlength{\tabcolsep}{2pt}

\begin{tabular}{>{\centering\arraybackslash}p{0.7cm} p{1.5cm} | ccccc | >{\centering\arraybackslash}p{0.8cm} !{\vrule width 1.3pt} cccccccc | >{\centering\arraybackslash}p{0.8cm}}
  \toprule
  \multicolumn{2}{c}{}                 & \multicolumn{6}{c}{\textbf{BEANS}}       & \multicolumn{9}{c}{\makecell[c]{\textbf{BirdSet}                                                                                                                                                                                                                                                                                                                                                                                                                                                                                                                                                                                                                         \\[-12pt] \hspace{-7.5cm} {\color{gray}\scriptsize VAL}}}                                                                                                                                                                                                                                                                                                                                                                                  \\
  \addlinespace[2pt]
  \cline{3-17} %
  \addlinespace[2pt]

  \multicolumn{2}{c}{\textbf{Setting}} & \cellcolor{gray!25}\textbf{\textsc{WTK}} & \cellcolor{gray!25}\textbf{\textsc{BAT}}         & \cellcolor{gray!25}\textbf{\textsc{CBI}} & \cellcolor{gray!25}\textbf{\textsc{DOG}} & \cellcolor{gray!25}\textbf{\textsc{HUM}} & \cellcolor{gray!25}\textbf{\underline{Score}} & \cellcolor{gray!25}\textbf{\textsc{POW}} & \cellcolor{gray!25}\textbf{\textsc{PER}} & \cellcolor{gray!25}\textbf{\textsc{NES}} & \cellcolor{gray!25}\textbf{\textsc{UHH}} & \cellcolor{gray!25}\textbf{\textsc{HSN}} & \cellcolor{gray!25}\textbf{\textsc{NBP}} & \cellcolor{gray!25}\textbf{\textsc{SSW}} & \cellcolor{gray!25}\textbf{\textsc{SNE}} & \cellcolor{gray!25}\textbf{\underline{Score}}                                \\
  \addlinespace[2pt]
  \cline{3-17} %
  \addlinespace[2pt]
  \midrule
  \multicolumn{17}{p{5cm}}{\textit{Baseline general audio models}} \vspace{0.5mm}                                                                                                                                                                                                                                                                                                                                                                                                                                                                                                                                                                                                                                                                            \\
  \multirow{2}{*}{\textbf{\rotatebox[origin=c]{90}{\renewcommand{\arraystretch}{1}
        \begin{tabular}{@{}c@{}}Audio \\ MAE
      \end{tabular}}
    }
  }                                   & {Linear}                                 & 44.69\scriptsize{$\pm$ 19.82} & 45.60\scriptsize{$\pm$ 1.13} & 35.80\scriptsize{$\pm$ 0.63} & 24.46\scriptsize{$\pm$ 4.07} & 74.69\scriptsize{$\pm$ 0.49} & \heatgreen{45.05} &14.98\scriptsize{$\pm$ 0.02} & 5.75\scriptsize{$\pm$ 0.03} & 11.34\scriptsize{$\pm$ 0.40} & 14.66\scriptsize{$\pm$ 0.03} & 14.61\scriptsize{$\pm$ 0.09} & 25.96\scriptsize{$\pm$ 0.05} & 8.73\scriptsize{$\pm$ 0.02} & 11.83\scriptsize{$\pm$ 0.02} & \heatgreen{13.27} \\
  & {Attentive}                              & 83.33\scriptsize{$\pm$ 0.21} & 67.02\scriptsize{$\pm$ 0.25} & 47.89\scriptsize{$\pm$ 2.79} & 85.97\scriptsize{$\pm$ 1.53} & 79.53\scriptsize{$\pm$ 0.57} & \heatblue{72.75} &27.16\scriptsize{$\pm$ 0.13} & 13.83\scriptsize{$\pm$ 0.06} & 25.48\scriptsize{$\pm$ 0.35} & 20.35\scriptsize{$\pm$ 0.14} & 29.10\scriptsize{$\pm$ 0.79} & 44.66\scriptsize{$\pm$ 0.98} & 22.09\scriptsize{$\pm$ 0.49} & 20.22\scriptsize{$\pm$ 0.21} & \heatblue{25.10} \\
  \hline
  \multirow{2}{*}{\textbf{\rotatebox[origin=c]{90}{BEATs}
    }
  }                                   & {Linear}                                 & \underline{83.19\scriptsize{$\pm$ 0.00}} & 57.57\scriptsize{$\pm$ 0.60} & 19.25\scriptsize{$\pm$ 0.94} & 75.90\scriptsize{$\pm$ 6.61} & \textbf{78.46\scriptsize{$\pm$ 0.49}} & \heatgreen{62.87} &14.54\scriptsize{$\pm$ 0.08} & 5.76\scriptsize{$\pm$ 0.05} & 8.16\scriptsize{$\pm$ 0.04} & 9.83\scriptsize{$\pm$ 0.08} & 13.06\scriptsize{$\pm$ 0.48} & 23.61\scriptsize{$\pm$ 0.03} & 6.67\scriptsize{$\pm$ 0.00} & 11.38\scriptsize{$\pm$ 0.02} & \heatgreen{11.21} \\
  & {Attentive}                              & 85.69\scriptsize{$\pm$ 1.04} & 73.27\scriptsize{$\pm$ 3.08} & 54.67\scriptsize{$\pm$ 3.05} & 90.65\scriptsize{$\pm$ 7.12} & 81.93\scriptsize{$\pm$ 0.99} & \heatblue{77.24} &25.67\scriptsize{$\pm$ 0.15} & 14.91\scriptsize{$\pm$ 0.03} & 24.51\scriptsize{$\pm$ 0.71} & 20.85\scriptsize{$\pm$ 0.00} & 30.29\scriptsize{$\pm$ 0.00} & 42.49\scriptsize{$\pm$ 0.34} & 20.35\scriptsize{$\pm$ 0.09} & 22.17\scriptsize{$\pm$ 0.08} & \heatblue{25.08} \\
  \hline
  \multirow{2}{*}{\textbf{\rotatebox[origin=c]{90}{EAT}
    }
  }                                   & {Linear}                                 & 82.74\scriptsize{$\pm$ 0.21} & 56.78\scriptsize{$\pm$ 0.18} & 40.68\scriptsize{$\pm$ 0.25} & 85.25\scriptsize{$\pm$ 0.51} & 77.57\scriptsize{$\pm$ 0.08} & \heatgreen{68.60} &18.43\scriptsize{$\pm$ 0.07} & 7.58\scriptsize{$\pm$ 0.05} & 12.96\scriptsize{$\pm$ 0.09} & 10.62\scriptsize{$\pm$ 0.47} & 16.72\scriptsize{$\pm$ 0.30} & 31.60\scriptsize{$\pm$ 0.28} & 8.61\scriptsize{$\pm$ 0.02} & 12.19\scriptsize{$\pm$ 0.04} & \heatgreen{14.33} \\
  & {Attentive}                              & 83.78\scriptsize{$\pm$ 0.42} & 67.82\scriptsize{$\pm$ 3.08} & 49.96\scriptsize{$\pm$ 1.66} & 85.61\scriptsize{$\pm$ 2.03} & \textbf{82.81\scriptsize{$\pm$ 0.34}} & \heatblue{74.00} &25.45\scriptsize{$\pm$ 1.46} & 13.38\scriptsize{$\pm$ 0.09} & 24.62\scriptsize{$\pm$ 1.29} & 19.44\scriptsize{$\pm$ 0.65} & 27.34\scriptsize{$\pm$ 3.84} & 43.91\scriptsize{$\pm$ 0.10} & 17.26\scriptsize{$\pm$ 0.08} & 18.99\scriptsize{$\pm$ 0.78} & \heatblue{23.56} \\
  \midrule
  \multicolumn{17}{p{5cm}}{\textit{Bioacoustic foundation models}} \vspace{0.5mm}                                                                                                                                                                                                                                                                                                                                                                                                                                                                                                                                                                                                                                                                            \\
  \multirow{2}{*}{\textbf{\rotatebox[origin=c]{90}{AVES}
    }
  }                                   & {Linear}                                 & 64.60\scriptsize{$\pm$ 6.67} & 46.77\scriptsize{$\pm$ 0.88} & 7.54\scriptsize{$\pm$ 0.12} & 60.79\scriptsize{$\pm$ 0.51} & 73.67\scriptsize{$\pm$ 0.11} & \heatgreen{50.68} &11.40\scriptsize{$\pm$ 0.10} & 3.03\scriptsize{$\pm$ 0.04} & 4.03\scriptsize{$\pm$ 0.01} & 9.54\scriptsize{$\pm$ 0.51} & 4.86\scriptsize{$\pm$ 0.04} & 13.75\scriptsize{$\pm$ 0.24} & 2.87\scriptsize{$\pm$ 0.02} & 6.01\scriptsize{$\pm$ 0.01} & \heatgreen{6.30} \\
  & {Attentive}                              & 78.91\scriptsize{$\pm$ 1.46} & 67.77\scriptsize{$\pm$ 1.66} & 47.93\scriptsize{$\pm$ 0.20} & 86.33\scriptsize{$\pm$ 3.05} & 79.59\scriptsize{$\pm$ 0.42} & \heatblue{72.11} &19.14\scriptsize{$\pm$ 0.32} & 5.83\scriptsize{$\pm$ 0.19} & 16.84\scriptsize{$\pm$ 0.18} & 14.59\scriptsize{$\pm$ 0.25} & 21.27\scriptsize{$\pm$ 0.50} & 31.30\scriptsize{$\pm$ 0.49} & 11.38\scriptsize{$\pm$ 0.26} & 13.98\scriptsize{$\pm$ 0.08} & \heatblue{16.46} \\
  \hline
  \multirow{2}{*}{\textbf{\rotatebox[origin=c]{90}{\renewcommand{\arraystretch}{1}
        \begin{tabular}{@{}c@{}}BEATs \\ NLM
      \end{tabular}}
    }
  }                                   & {Linear}                                 & \textbf{83.48\scriptsize{$\pm$ 2.09}} & 57.50\scriptsize{$\pm$ 0.28} & 32.89\scriptsize{$\pm$ 0.14} & 71.58\scriptsize{$\pm$ 4.58} & 76.90\scriptsize{$\pm$ 0.11} & \heatgreen{64.47} &23.46\scriptsize{$\pm$ 0.04} & 6.90\scriptsize{$\pm$ 0.04} & 14.01\scriptsize{$\pm$ 0.11} & 12.02\scriptsize{$\pm$ 0.14} & 23.23\scriptsize{$\pm$ 1.03} & 41.57\scriptsize{$\pm$ 0.01} & 21.13\scriptsize{$\pm$ 0.04} & 19.91\scriptsize{$\pm$ 0.08} & \heatgreen{19.83} \\
  & {Attentive}                              & \textbf{90.41\scriptsize{$\pm$ 1.04}} & \underline{74.75\scriptsize{$\pm$ 2.05}} & \textbf{76.85\scriptsize{$\pm$ 3.20}} & \textbf{94.24\scriptsize{$\pm$ 2.03}} & \underline{81.98\scriptsize{$\pm$ 0.46}} & \heatblue[bold]{83.65} &\textbf{39.08\scriptsize{$\pm$ 0.70}} & \underline{20.91\scriptsize{$\pm$ 0.46}} & \textbf{35.85\scriptsize{$\pm$ 0.08}} & \textbf{26.85\scriptsize{$\pm$ 0.13}} & \textbf{46.85\scriptsize{$\pm$ 0.37}} & \underline{61.52\scriptsize{$\pm$ 0.53}} & \textbf{44.21\scriptsize{$\pm$ 0.60}} & \underline{28.02\scriptsize{$\pm$ 0.66}} & \heatblue[underline]{37.74} \\
  \hline
  \multirow{2}{*}{\textbf{\rotatebox[origin=c]{90}{\renewcommand{\arraystretch}{1}
        \begin{tabular}{@{}c@{}}Biolin \\ gual
      \end{tabular}}
    }
  }                                   & {Linear}                                 & 79.20\scriptsize{$\pm$ 1.04} & 54.20\scriptsize{$\pm$ 0.99} & 37.58\scriptsize{$\pm$ 3.30} & 65.83\scriptsize{$\pm$ 5.60} & 72.81\scriptsize{$\pm$ 0.57} & \heatgreen{61.92} &17.63\scriptsize{$\pm$ 0.12} & 6.77\scriptsize{$\pm$ 0.14} & 10.26\scriptsize{$\pm$ 0.12} & 9.90\scriptsize{$\pm$ 0.00} & 25.29\scriptsize{$\pm$ 0.09} & 28.30\scriptsize{$\pm$ 0.29} & 12.43\scriptsize{$\pm$ 0.06} & 12.96\scriptsize{$\pm$ 0.06} & \heatgreen{15.13} \\
  & {Attentive}                              & 83.92\scriptsize{$\pm$ 0.63} & 66.53\scriptsize{$\pm$ 1.03} & \underline{73.15\scriptsize{$\pm$ 1.09}} & 89.21\scriptsize{$\pm$ 1.02} & 78.99\scriptsize{$\pm$ 0.34} & \heatblue{78.36} &33.39\scriptsize{$\pm$ 2.38} & 14.98\scriptsize{$\pm$ 0.08} & 29.07\scriptsize{$\pm$ 0.41} & 23.11\scriptsize{$\pm$ 0.22} & 41.29\scriptsize{$\pm$ 0.74} & 55.42\scriptsize{$\pm$ 0.02} & 31.23\scriptsize{$\pm$ 0.13} & 22.28\scriptsize{$\pm$ 0.26} & \heatblue{31.06} \\
  \hline
  \multirow{2}{*}{\textbf{\rotatebox[origin=c]{90}{\renewcommand{\arraystretch}{1}
        \begin{tabular}{@{}c@{}}Bird \\ AVES
      \end{tabular}}
    }
  }                                   & {Linear}                                 & 63.72\scriptsize{$\pm$ 2.09} & 48.95\scriptsize{$\pm$ 1.06} & 10.35\scriptsize{$\pm$ 0.02} & 41.37\scriptsize{$\pm$ 0.51} & 71.41\scriptsize{$\pm$ 1.18} & \heatgreen{47.16} &10.82\scriptsize{$\pm$ 0.05} & 3.80\scriptsize{$\pm$ 0.02} & 4.50\scriptsize{$\pm$ 0.10} & 7.72\scriptsize{$\pm$ 0.01} & 5.21\scriptsize{$\pm$ 0.14} & 14.27\scriptsize{$\pm$ 0.08} & 3.32\scriptsize{$\pm$ 0.12} & 6.56\scriptsize{$\pm$ 0.07} & \heatgreen{6.48} \\
  & {Attentive}                              & 73.45\scriptsize{$\pm$ 3.34} & 68.47\scriptsize{$\pm$ 1.24} & 51.57\scriptsize{$\pm$ 0.12} & \underline{90.65\scriptsize{$\pm$ 0.00}} & 78.81\scriptsize{$\pm$ 0.46} & \heatblue{72.59} &24.53\scriptsize{$\pm$ 0.86} & 8.56\scriptsize{$\pm$ 0.14} & 20.48\scriptsize{$\pm$ 0.15} & 15.26\scriptsize{$\pm$ 0.20} & 31.54\scriptsize{$\pm$ 1.49} & 39.04\scriptsize{$\pm$ 0.02} & 17.65\scriptsize{$\pm$ 0.27} & 20.60\scriptsize{$\pm$ 0.86} & \heatblue{21.88} \\
  \hline
  \multirow{2}{*}{\textbf{\rotatebox[origin=c]{90}{\renewcommand{\arraystretch}{1}
        \begin{tabular}{@{}c@{}}Bird \\ MAE
      \end{tabular}}
    }
  }                                   & {Linear}                                 & 73.16\scriptsize{$\pm$ 2.92} & 60.03\scriptsize{$\pm$ 1.59} & 53.40\scriptsize{$\pm$ 0.55} & 48.56\scriptsize{$\pm$ 5.60} & 77.17\scriptsize{$\pm$ 0.42} & \heatgreen{62.46} &27.05\scriptsize{$\pm$ 0.68} & 10.83\scriptsize{$\pm$ 0.00} & 20.66\scriptsize{$\pm$ 0.12} & 16.40\scriptsize{$\pm$ 0.09} & 16.37\scriptsize{$\pm$ 0.00} & 43.45\scriptsize{$\pm$ 0.15} & 17.69\scriptsize{$\pm$ 0.07} & 18.26\scriptsize{$\pm$ 0.12} & \heatgreen{20.52} \\
  & {Attentive}                              & \underline{88.94\scriptsize{$\pm$ 1.46}} & \textbf{75.15\scriptsize{$\pm$ 0.64}} & 66.17\scriptsize{$\pm$ 0.25} & 88.85\scriptsize{$\pm$ 0.51} & 80.93\scriptsize{$\pm$ 1.10} & \heatblue[underline]{80.01} &\underline{38.19\scriptsize{$\pm$ 1.35}} & \textbf{26.01\scriptsize{$\pm$ 0.96}} & \underline{35.69\scriptsize{$\pm$ 0.06}} & \underline{26.39\scriptsize{$\pm$ 0.12}} & \underline{45.67\scriptsize{$\pm$ 2.26}} & \textbf{66.26\scriptsize{$\pm$ 1.27}} & \underline{35.58\scriptsize{$\pm$ 0.13}} & \textbf{31.56\scriptsize{$\pm$ 0.07}} & \heatblue[bold]{38.17} \\
  \hline
  \multirow{2}{*}{\textbf{\rotatebox[origin=c]{90}{\renewcommand{\arraystretch}{1}
        \begin{tabular}{@{}c@{}}Conv \\ Next$_{BS}$
      \end{tabular}}
    }
  }                                   & {Linear}                                 & 82.74\scriptsize{$\pm$ 0.63} & \underline{64.75\scriptsize{$\pm$ 1.98}} & 78.26\scriptsize{$\pm$ 0.43} & \underline{88.85\scriptsize{$\pm$ 0.51}} & 76.20\scriptsize{$\pm$ 1.41} & \heatgreen[underline]{78.16} &\underline{38.58\scriptsize{$\pm$ 0.16}} & \textbf{19.92\scriptsize{$\pm$ 0.13}} & 36.19\scriptsize{$\pm$ 0.19} & \textbf{26.35\scriptsize{$\pm$ 0.49}} & \textbf{52.69\scriptsize{$\pm$ 0.08}} & \textbf{66.07\scriptsize{$\pm$ 0.22}} & \underline{39.88\scriptsize{$\pm$ 0.10}} & \underline{32.39\scriptsize{$\pm$ 0.20}} & \heatgreen[bold]{39.07} \\
  & {Restricted}                             & - & - & 82.93\scriptsize{$\pm$ 0.00} & - & - & - &34.17\scriptsize{$\pm$ 0.00} & 17.46\scriptsize{$\pm$ 0.00} & 34.13\scriptsize{$\pm$ 0.00} & 24.79\scriptsize{$\pm$ 0.00} & 48.43\scriptsize{$\pm$ 0.00} & 61.85\scriptsize{$\pm$ 0.00} & 33.97\scriptsize{$\pm$ 0.00} & 29.88\scriptsize{$\pm$ 0.00} & \heatblue{35.79} \\
  \hline
  \multirow{2}{*}{\textbf{\rotatebox[origin=c]{90}{Perch}
    }
  }                                   & {Linear}                                 & 80.09\scriptsize{$\pm$ 0.21} & 53.50\scriptsize{$\pm$ 0.07} & \underline{79.72\scriptsize{$\pm$ 0.16}} & \textbf{89.21\scriptsize{$\pm$ 1.02}} & 74.85\scriptsize{$\pm$ 0.04} & \heatgreen{75.47} &36.26\scriptsize{$\pm$ 0.71} & \underline{19.63\scriptsize{$\pm$ 0.19}} & \textbf{37.87\scriptsize{$\pm$ 0.48}} & 23.88\scriptsize{$\pm$ 0.35} & 49.73\scriptsize{$\pm$ 0.13} & \underline{64.98\scriptsize{$\pm$ 0.11}} & 33.16\scriptsize{$\pm$ 0.49} & 30.72\scriptsize{$\pm$ 0.58} & \heatgreen{37.14} \\
  & {Restricted}                             & - & - & 87.29\scriptsize{$\pm$ 0.00} & - & - & - &30.41\scriptsize{$\pm$ 0.00} & 18.23\scriptsize{$\pm$ 0.00} & 38.09\scriptsize{$\pm$ 0.00} & 26.72\scriptsize{$\pm$ 0.00} & 45.23\scriptsize{$\pm$ 0.00} & 60.67\scriptsize{$\pm$ 0.00} & 28.35\scriptsize{$\pm$ 0.00} & 28.72\scriptsize{$\pm$ 0.00} & \heatblue{35.14} \\
  \hline
  \multirow{2}{*}{\textbf{\rotatebox[origin=c]{90}{\renewcommand{\arraystretch}{1}
        \begin{tabular}{@{}c@{}}Perch \\ 2.0
      \end{tabular}}
    }
  }                                   & {Linear}                                 & 81.12\scriptsize{$\pm$ 1.67} & \textbf{69.43\scriptsize{$\pm$ 0.74}} & \textbf{80.25\scriptsize{$\pm$ 0.04}} & 88.49\scriptsize{$\pm$ 5.09} & \underline{78.40\scriptsize{$\pm$ 0.27}} & \heatgreen[bold]{79.54} &\textbf{43.97\scriptsize{$\pm$ 0.78}} & 18.51\scriptsize{$\pm$ 0.12} & \underline{36.31\scriptsize{$\pm$ 0.23}} & \underline{24.09\scriptsize{$\pm$ 0.12}} & \underline{51.55\scriptsize{$\pm$ 0.12}} & 64.70\scriptsize{$\pm$ 0.08} & \textbf{42.64\scriptsize{$\pm$ 0.02}} & \textbf{32.46\scriptsize{$\pm$ 0.55}} & \heatgreen[underline]{38.61} \\
  & {Restricted}                             & - & - & 76.02\scriptsize{$\pm$ 0.00} & - & - & - &53.77\scriptsize{$\pm$ 0.00} & 23.19\scriptsize{$\pm$ 0.00} & 40.27\scriptsize{$\pm$ 0.00} & 37.97\scriptsize{$\pm$ 0.00} & 53.26\scriptsize{$\pm$ 0.00} & 66.09\scriptsize{$\pm$ 0.00} & 46.84\scriptsize{$\pm$ 0.00} & 34.03\scriptsize{$\pm$ 0.00} & \heatblue{43.09} \\
  \hline
  \multirow{2}{*}{\textbf{\rotatebox[origin=c]{90}{\renewcommand{\arraystretch}{1}
        \begin{tabular}{@{}c@{}}Proto \\ CLR
      \end{tabular}}
    }
  }                                   & {Linear}                                 & 79.06\scriptsize{$\pm$ 0.83} & 63.62\scriptsize{$\pm$ 0.04} & 65.61\scriptsize{$\pm$ 0.23} & 87.41\scriptsize{$\pm$ 2.54} & 77.25\scriptsize{$\pm$ 0.08} & \heatgreen{74.59} &27.80\scriptsize{$\pm$ 0.01} & 13.00\scriptsize{$\pm$ 0.09} & 23.54\scriptsize{$\pm$ 0.00} & 17.77\scriptsize{$\pm$ 0.01} & 27.95\scriptsize{$\pm$ 0.03} & 42.12\scriptsize{$\pm$ 0.00} & 18.80\scriptsize{$\pm$ 0.01} & 19.62\scriptsize{$\pm$ 0.01} & \heatgreen{23.26} \\
  & {Attentive}                              & 75.81\scriptsize{$\pm$ 0.00} & 64.02\scriptsize{$\pm$ 0.18} & 63.88\scriptsize{$\pm$ 0.06} & 88.85\scriptsize{$\pm$ 3.56} & 78.27\scriptsize{$\pm$ 0.53} & \heatblue{74.17} &28.45\scriptsize{$\pm$ 0.92} & 14.83\scriptsize{$\pm$ 0.08} & 26.38\scriptsize{$\pm$ 0.14} & 20.91\scriptsize{$\pm$ 0.23} & 29.53\scriptsize{$\pm$ 1.64} & 49.78\scriptsize{$\pm$ 0.03} & 23.85\scriptsize{$\pm$ 0.21} & 21.63\scriptsize{$\pm$ 0.09} & \heatblue{26.70} \\
  \hline
  \multirow{2}{*}{\textbf{\rotatebox[origin=c]{90}{\renewcommand{\arraystretch}{1}
        \begin{tabular}{@{}c@{}}Surf \\ Perch
      \end{tabular}}
    }
  }                                   & {Linear}                                 & 80.68\scriptsize{$\pm$ 0.21} & 55.75\scriptsize{$\pm$ 0.49} & 59.19\scriptsize{$\pm$ 0.33} & 71.94\scriptsize{$\pm$ 9.16} & 75.15\scriptsize{$\pm$ 0.08} & \heatgreen{68.54} &22.85\scriptsize{$\pm$ 1.34} & 8.98\scriptsize{$\pm$ 0.00} & 21.89\scriptsize{$\pm$ 1.16} & 15.27\scriptsize{$\pm$ 1.11} & 32.48\scriptsize{$\pm$ 0.33} & 35.75\scriptsize{$\pm$ 0.19} & 13.54\scriptsize{$\pm$ 1.05} & 15.73\scriptsize{$\pm$ 2.44} & \heatgreen{20.52} \\
  & {Restricted}                             & - & - & 64.03\scriptsize{$\pm$ 0.00} & - & - & - &23.07\scriptsize{$\pm$ 0.00} & 8.79\scriptsize{$\pm$ 0.00} & 24.44\scriptsize{$\pm$ 0.00} & 20.13\scriptsize{$\pm$ 0.00} & 30.18\scriptsize{$\pm$ 0.00} & 32.22\scriptsize{$\pm$ 0.00} & 13.24\scriptsize{$\pm$ 0.00} & 17.30\scriptsize{$\pm$ 0.00} & \heatblue{20.90} \\
  \hline
  \multirow{2}{*}{\textbf{\rotatebox[origin=c]{90}{\renewcommand{\arraystretch}{1}
        \begin{tabular}{@{}c@{}}ViT \\ INS
      \end{tabular}}
    }
  }                                   & {Linear}                                 & 68.44\scriptsize{$\pm$ 0.00} & 45.50\scriptsize{$\pm$ 0.35} & 14.50\scriptsize{$\pm$ 0.16} & 56.12\scriptsize{$\pm$ 3.05} & 69.85\scriptsize{$\pm$ 0.19} & \heatgreen{50.88} &13.75\scriptsize{$\pm$ 0.22} & 4.66\scriptsize{$\pm$ 0.08} & 7.92\scriptsize{$\pm$ 0.26} & 10.26\scriptsize{$\pm$ 0.09} & 10.34\scriptsize{$\pm$ 0.22} & 20.07\scriptsize{$\pm$ 0.64} & 5.94\scriptsize{$\pm$ 0.09} & 8.43\scriptsize{$\pm$ 0.43} & \heatgreen{9.66} \\
  & {Attentive}                              & 70.35\scriptsize{$\pm$ 3.55} & 61.95\scriptsize{$\pm$ 1.27} & 25.58\scriptsize{$\pm$ 2.50} & 56.12\scriptsize{$\pm$ 9.16} & 73.72\scriptsize{$\pm$ 2.24} & \heatblue{57.54} &17.14\scriptsize{$\pm$ 0.07} & 7.35\scriptsize{$\pm$ 0.01} & 13.20\scriptsize{$\pm$ 0.03} & 11.90\scriptsize{$\pm$ 0.00} & 14.32\scriptsize{$\pm$ 0.01} & 27.74\scriptsize{$\pm$ 0.16} & 9.85\scriptsize{$\pm$ 0.05} & 11.38\scriptsize{$\pm$ 0.08} & \heatblue{13.68} \\
  \bottomrule
\end{tabular}

  }
  \label{tab:results-standard-metrics}
\end{table}

\begin{table}[h]
  \centering
  \caption{Taxonomy distribution (logarithmic scale) of four large datasets—Xeno-Canto (XC), Macaulay Library (MAC), iNaturalist (INA), and Animal Sound Archive (ASA) —across five widely studied biological groups: Birds, Amphibians, Mammals, Insects, and Reptiles~\cite{noauthor_animal_nodate,noauthor_inaturalist_nodate,noauthor_macaulay_nodate,vellinga_xeno-canto_2015}.}
  
\renewcommand{\arraystretch}{1.5} %
\begin{tabular}{l l c c c c c }
    \toprule
    \multirow{2}{*}{\cellcolor{gray!15}\vspace{4mm} \textbf{Datasets}}
                 & \multirow{2}{*}{\cellcolor{gray!15} \textbf{}}
                 & \cellcolor{gray!15}\textbf{Birds}
                 & \cellcolor{gray!15}\textbf{Amphibians}
                 & \cellcolor{gray!15}\textbf{Mammals}
                 & \cellcolor{gray!15}\textbf{Insects}
                 & \cellcolor{gray!15}\textbf{Reptiles}                                                        \\

    \midrule

    \textbf{XC}  & Recordings                                     & 873,376   & 2,486  & 4,098  & 32,082 & 0   \\
    \arrayrulecolor{lightgray}\cline{2-7}
                 & Species                                        & 10,528    & 594    & 529    & 987    & 0   \\
    \hline

    \textbf{MAC} & Recordings                                     & 2,669,609 & 11,542 & 9,515  & 9,060  & 63  \\
    \arrayrulecolor{lightgray}\cline{2-7}
                 & Species                                        & 10,056    & 2,674  & N/A    & N/A    & N/A \\
    \hline

    \textbf{INA} & Recordings                                     & 871,771   & 94,874 & 47,631 & 80,545 & 770 \\
    \arrayrulecolor{lightgray}\cline{2-7}
                 & Species                                        & 6,972     & 1,639  & 923    & 2,166  & 133 \\
    \hline

    \textbf{ASA} & Recordings                                     & 21,285    & 692    & 2,716  & 738    & 1   \\
    \arrayrulecolor{lightgray}\cline{2-7}
                 & Species                                        & N/A       & N/A    & N/A    & N/A    & N/A \\

    \bottomrule
\end{tabular}

  \label{tab:graphic}
\end{table}

\begin{table}[h]
  \centering
  \caption{Spectrogram preprocessing parameter settings for each model. Symbols are used as follows: '\no'~indicates the parameter or method is not applied, '\question'~denotes missing or undocumented information.
  }
  \renewcommand{\arraystretch}{1.5}
\arrayrulecolor{black} %
\resizebox{\textwidth}{!}{%
    \begin{tabular}{l | c c c c c c c c c c c c c}
        \toprule
                                                    & \multicolumn{10}{c}{Bioacoustic foundation models} & \multicolumn{3}{c}{General audio models}                                                                                                                                                      \\
        \cmidrule(lr){2-11} \cmidrule(lr){12-14}
        \cellcolor{gray!15}\textbf{Parameter}       &
        \cellcolor{gray!15}\textbf{Biolingual}      &
        \cellcolor{gray!15}\textbf{BirdMAE}         &
        \cellcolor{gray!15}\textbf{BirdNET}         &
        \cellcolor{gray!15}\textbf{ConvNext$_{BS}$} &
        \cellcolor{gray!15}\textbf{NatureLM-audio}  &
        \cellcolor{gray!15}\textbf{Perch}           &
        \cellcolor{gray!15}\textbf{Perch 2.0}       &
        \cellcolor{gray!15}\textbf{ProtoCLR}        &
        \cellcolor{gray!15}\textbf{SurfPerch}       &
        \cellcolor{gray!15}\textbf{ViT$_{INS}$}     &
        \cellcolor{gray!15}\textbf{BEATs}           &
        \cellcolor{gray!15}\textbf{AudioMAE}        &
        \cellcolor{gray!15}\textbf{EAT}                                                                                                                                                                                                                                                                 \\
        \arrayrulecolor{black}\midrule
        \arrayrulecolor{lightgray}

        n\_fft                                      & 1024                                              & 1024                                     & (2048,1024)      & 1024        & 1024         & 2048        & 1024        & 1024        & 1024        & 1024               & 512        & 1024      & 512        \\ \cline{2-14}
        hop\_length                                 & 1024                                              & 320                                      & (278,280)        & 320         & 160          & 512         & 320         & 320         & 512         & 128                & 160        & 320       & 160        \\ \cline{2-14}
        n\_mels                                     & 64                                                & 128                                      & 96               & 128         & 128          & 96          & 128         & 128         & 128         & 128                & 128        & 128       & 128        \\ \cline{2-14}
        Freq. range (in Hz)                         & 50-14k                                            & 20-16k                                   & (0-3kHz,500-15k) & 0-16k       & 20-8k        & 0-11.025k   & 60-16k      & 50-8k       & 50-16k      & 50-11.025k         & 20-8k      & 20-8k     & 20-8k      \\ \cline{2-14}
        Power                                       & 2                                                 & 2                                        & 2                & 2           & 2            & 2           & 1           & 2           & 2           & 1                  & 2          & 2         & 2          \\ \cline{2-14}
        Sample rate (in Hz)                         & 48k                                               & 32k                                      & 48k              & 32k         & 16k          & 32k         & 32k         & 16k         & 32k         & 22.05k             & 16k        & 16k       & 16k        \\ \cline{2-14}
        Window type                                 & Hann                                              & Hann                                     & Hann             & Hann        & Povey        & Hann        & Hann        & Hann        & Hann        & Hann               & Povey      & Hann      & Hann       \\ \cline{2-14}
        Window size                                 & 1024                                              & \textasciitilde 800                      & 512              & 1024        & 1024         & 2048        & 640         & 1024        & 1024        & 512                & 400        & 400       & 400        \\ \cline{2-14}
        dB scale                                    & \yes                                              & \yes                                     & \yes             & \yes        & \yes         & \question   & \yes        & \yes        & \question   & \yes               & \yes       & \yes      & \yes       \\ \cline{2-14}
        dB cutoff                                   & \question                                         & \no                                      & \no              & 80dB        & \no          & \question   & \question   & \no         & \question   & [-100,0]           & \no        & \no       & \no        \\ \cline{2-14}
        Normalisation                               & Stand.                                            & Stand.                                   & Min-Max          & Stand.      & Stand.       & PCEN        & Log scale   & Stand.      & PCEN        & Rescale to [0,255] & Stand.     & Stand.    & Stand.     \\ \cline{2-14}
        Resolution $(n,t)$                          & $(64,469)$                                        & $(128,500)$                              & $2x(96,516)$     & $(128,500)$ & $(128,1000)$ & $(96,313)$  & $(128,500)$ & $(128,300)$ & $(128,313)$ & $(128,517)$        & (128,1000) & (128,500) & (128,1000) \\
        \arrayrulecolor{black}\bottomrule
    \end{tabular}%
}

  \label{tab:spectrogram-parameters}
\end{table}

\end{document}